\documentclass[prl, twocolumn, superscriptaddress, footinbib, floatfix]{revtex4-2}
\usepackage{amssymb, amsmath, xcolor, graphicx}
\usepackage[colorlinks,citecolor=blue,urlcolor=blue,linkcolor=blue]{hyperref}
\usepackage{wrapfig}
\usepackage{multirow}
\usepackage{slashed}
\usepackage{physics}
\usepackage[switch]{lineno}

\DeclareRobustCommand{\tb}[1]{Table~\ref{tb:#1}}

\newcommand\bets{\begin{table*}}
\newcommand\eets[1]{\label{tb:#1}\end{table*}}
\newcommand{\msbar}{\overline{\mathrm{MS}}}
\def\O{\mathcal{O}}
\def\tilO{\widetilde{\mathcal{O}}}

\def\mn{\rho\sigma}
\def\rs{\rho\sigma}
\def\tf{t_{\rm f}}
\def\ef{\epsilon_{\rm f}}
\def\Nc{N_{\rm c}}

\begin{document}
\title{Lattice-QCD validation of hadron mass and trace-anomaly decomposition sum rules}

\author{Dennis Bollweg}
\affiliation{Scientific Computing Core, Flatiron Institute, New York, New York 10010, USA }

\author{Heng-Tong Ding}
\affiliation{Key Laboratory of Quark \& Lepton Physics (MOE) and Institute of Particle Physics, Central China Normal University, Wuhan 430079, China}

\author{Xiang Gao}
\email{xgao@bnl.gov}
\affiliation{Physics Department, Brookhaven National Laboratory, Upton, New York 11973, USA}

\author{Ran Luo}
\email{ranluo@mails.ccnu.edu.cn}
\affiliation{Key Laboratory of Quark \& Lepton Physics (MOE) and Institute of Particle Physics, Central China Normal University, Wuhan 430079, China}

\author{Swagato Mukherjee}
\affiliation{Physics Department, Brookhaven National Laboratory, Upton, New York 11973, USA}

\date{\today}

\begin{abstract}
We present the first lattice-QCD validation of multiple sum rules associated with quark–gluon decomposition of hadron mass by computing all relevant tensor components of the quark and gluon energy--momentum tensor matrix elements from first principles. We achieve this through nonperturbative renormalization of the QCD energy–momentum tensor, including its trace, in a gradient-flow scheme, followed by continuum extrapolations, two-loop matching to the $\msbar$ scheme, and zero-flow-time extrapolations. These ingredients enable a direct and simultaneous verification, in a common renormalization scheme and scale, of multiple energy-density-based and trace-based mass decomposition sum rules proposed in the literature. We demonstrate the framework for the $\eta_c$ and $J/\psi$ charmonia using three fine lattice spacings with a physical strange-quark and near-physical up- and down-quark masses. We present the first lattice-QCD results for the gravitational form factor $\bar{C}$. We find sizable gluonic contributions to charmonia masses at the hadronic scale, $\sim 15\%$ in the Lorc{\'e} and Metz–Pasquini–Rodini decompositions. The trace-anomaly contribution in the Ji sum rule is $\sim 6\%$, while the gluonic component of the trace anomaly in the Hatta–Rajan–Tanaka sum rule is $\sim 35\%$. The method is general and can be straightforwardly adopted for lattice-QCD calculations of mass and spin decompositions as well as gravitational form factors of other hadrons and nuclei.
\end{abstract}
\maketitle
%
%
\noindent \emph{Introduction.---}
The mass of nearly all visible matter in the Universe arises from asymptotic bound states of quantum chromodynamics (QCD)~\cite{Fritzsch:1972jv,Gross:1973id,Politzer:1973fx}.
Quarks and gluons—almost massless at the Lagrangian level—form massive hadrons through strong, nonperturbative dynamics.
Because of color confinement, partons are never observed in isolation, and how their confined dynamics generate hadron masses remains a fundamental question.

The QCD energy--momentum tensor (EMT) provides the field-theoretic bridge between quark--gluon degrees of freedom and hadronic observables. Its hadronic matrix elements relate the hadron mass to quark and gluon energies, explicit quark-mass terms, and the trace anomaly~\cite{Collins:1976yq,Nielsen:1977sy,Shifman:1978zn,Hatta:2018sqd,Ji:2021pys}.
Gauge-invariant mass decompositions and the associated trace- and energy-density sum rules quantify this connection~\cite{Ji:1994av,Ji:1995sv,Lorce:2017xzd,Lorce:2021xku,Polyakov:2002yz,Lorce:2018egm,Polyakov:2018zvc}. These sum rules state that the hadron mass equals the sum of all quark and gluon contributions, providing a concrete benchmark for how confined partons collectively generate mass.

To test these theoretical relations, two complementary approaches are essential: experimental access to EMT form factors and nonperturbative theoretical determination of EMT matrix elements. Experimentally, the 12~GeV program at Thomas Jefferson National Accelerator Facility and the forthcoming Electron--Ion Collider aim to probe gravitational form factors (GFFs) through deeply virtual exclusive processes, providing leading-twist information on the traceless EMT and its spatial structure~\cite{Burkert:2018bqq,Burkert:2023wzr,Accardi:2012qut,AbdulKhalek:2022hcn,Boer:2011fh,Muller:1994ses,Mueller:1998fv,Ji:1996ek,Radyushkin:1996nd,Ji:1996nm}. However, only limited combinations of EMT components can be reconstructed, and the trace part governing the anomaly contribution remains inaccessible at leading twist. 

A complete picture, therefore, requires first-principles information on all EMT components, including those beyond experimental reach. Lattice QCD offers a gauge-invariant, systematically improvable, nonperturbative framework to address this question. However, the reduction of the continuum rotational to the lattice hypercubic symmetry induces mixing of the EMT trace with power-divergent lower-dimensional operators~\cite{Mandula:1983ut,Gockeler:1996mu}. Additionally, conventional nonperturbative renormalization of lattice EMT using regularization-independent momentum-subtraction (RI/MOM)-type schemes is gauge dependent, infrared sensitive, and statistically noisy. As a result, past studies have largely been restricted to selected traceless components~\cite{Hackett:2023nkr,Hackett:2023rif,Brommel:2005ee,Shanahan:2018pib,Pefkou:2021fni,Loffler:2021afv,Alexandrou:2019ali,Bali:2018zgl,Bhattacharya:2024wtg,Bhattacharya:2023ays,HadStruc:2024rix,Wang:2024lrm,Sun:2020pda}, with the trace-related component inferred using the relevant sum rules~\cite{Yang:2014xsa,Yang:2018nqn,He:2021bof,Hu:2024mas,Hatta:2018sqd,Lorce:2017xzd,Metz:2020vxd}.

We overcome these challenges and provide the first direct lattice-QCD validation of hadron mass decompositions and the trace-anomaly sum rules by computing
each of their constituent components. We employ nonperturbative renormalization of both traceless and trace parts of the EMT in the gradient-flow scheme, followed by continuum limit~\cite{Narayanan:2006rf,Luscher:2009eq,Luscher:2010iy,Luscher:2011bx}, two-loop matching to $\overline{\mathrm{MS}}$ and zero-flow-time extrapolations~\cite{Suzuki:2013gza,Makino:2014taa,Taniguchi:2016ofw,Taniguchi:2017ibr}. This enables a consistent determination of all EMT components in a common scheme and scale, allowing a direct and simultaneous consistency check of multiple energy-density- and trace-based decompositions proposed in the literature~\cite{Ji:1994av,Ji:1995sv,Lorce:2017xzd,Lorce:2021xku,Polyakov:2002yz,Lorce:2018egm,Polyakov:2018zvc,Hatta:2018sqd,Ji:2021pys}. Closely related gradient-flow constructions have also been explored for flowed twist-2 operators entering higher parton distribution function (PDF) moments~\cite{Shindler:2023pdfmoments,Francis:2025rya,Francis:2025pgf}.

%
%
\noindent\emph{$\msbar$-renormalized EMT via gradient flow.---}
The symmetric QCD EMT in Euclidean formulation, related to its Minkowski counterpart by the Wick rotation $[T_{\rs}]_M = i^{\delta_{t\rho} + \delta_{t\sigma}} [T_{\rs}]_E$, is
\begin{equation}
\begin{split}
& T_{\rs}(x) = \mathcal{O}_{1,\rs}(x) - \frac{1}{4} \mathcal{O}_{2,\rs}(x) + \frac{1}{4} \mathcal{O}_{3,\rs}(x) \\
& \qquad \qquad - \frac{1}{2} \mathcal{O}_{4,\rs}(x) - \mathcal{O}_{5,\rs}(x) \,,
\end{split}
\label{eq:Tmunu}
\end{equation}
where,

\begin{equation}
\begin{split}
& \mathcal{O}_{1,\rs} = \frac{2}{g_0^2} \Tr^{\text{c}} \left[ F_{\rho\omega} F_{\sigma\omega} \right] \,, \\
& \mathcal{O}_{2,\rs} = \frac{2}{g_0^2} \delta_{\rs} \Tr^{\text{c}} \left[ F_{\omega\xi} F_{\omega\xi} \right] \,, \\
& \O_{3,\rs} = \sum_f \O_{3,\rs,f} = 
\sum_f \bar{\psi}_f \left( \gamma_\rho \overset{\leftrightarrow}{D}_\sigma 
+ \gamma_\sigma \overset{\leftrightarrow}{D}_\rho \right) \psi_f \,, \\
& \O_{4,\rs} = \delta_{\rs} \sum_f \bar{\psi}_f \overset{\leftrightarrow}{\slashed{D}} \psi_f \,, ~\O_{5,\rs} = 
\delta_{\rs} \sum_f m_f \bar{\psi}_f \psi_f \,.
\end{split}
\label{eq:O123}
\end{equation}
Here, $g_0$ is the bare coupling, $m_f$ the bare quark mass, $\Tr^{\text{c}}$ denotes the color trace, $f$ labels quark flavors, and the covariant derivative is $\overset{\leftrightarrow}{D}_\rho = \overset{\rightarrow}{D}_\rho -\overset{\leftarrow}{D}_\rho$. For analyzing hadron mass decompositions, the EMT is separated into quark and gluon parts~\cite{Belinfante:1962zz,Belitsky:2005qn,Freese:2021jqs},
\begin{equation}
\begin{split}
& T_{\rs} = \sum_{X=q, g} \! T_{X,\rs} \,, 
~\text{with} \quad
T_{q,\rs} = \frac{1}{4} \sum_{\boldsymbol{x}} \mathcal{O}_{3,\rs} (x)  \,, \\
& T_{g,\rs} = \sum_{\boldsymbol{x}} \left[ \mathcal{O}_{1,\rs}(x)  - \frac{1}{4}\mathcal{O}_{2,\rs}(x) \right] \,.
\end{split}
\label{eq:Tg_Tq}
\end{equation}
The total EMT is conserved and, hence, does not require renormalization, but the individual components $T_{X,\rs}$ and $\mathcal{O}_{i,\rs}$ must be renormalized and mix under renormalization. Since renormalization and taking traces do not commute~\cite{Hatta:2018sqd,Lorce:2021xku,Ji:2021qgo}, they must be renormalized first before taking their trace, which is essential for a consistent comparison of different mass decomposition formulations. The QCD equations of motion (EOM) imply $2\langle\mathcal{O}_{5,\rs,f}\rangle = -\langle\mathcal{O}_{4,\rs,f}\rangle$, where angular brackets denote matrix elements in a physical state.

The gradient flow~\cite{Narayanan:2006rf,Luscher:2009eq,Luscher:2010iy,Luscher:2011bx} alleviates the problem of power-divergent mixing of lattice EMT~\cite{Mandula:1983ut,Gockeler:1996mu} by evolving lattice fields along a fictitious flow time $\tf$, which smooths ultraviolet fluctuations and renders composite operators finite at nonzero $\tf$. The flowed operators $\tilO_{i,\rs}(\tf,x)$ are nonperturbatively renormalized in the gradient-flow scheme at flow scale $\tf$, providing a common operator basis for evaluating distinct mass decomposition sum rules on equal footing. Their continuum limits are obtained by extrapolating results at several lattice spacings, $a\to0$. The continuum-extrapolated $\tilO_{i,\rs}(\tf,x)$ are then matched to the $\msbar$ scheme at scale $\mu$ using perturbatively known coefficients $\O^{\msbar}_i(\mu)=M_{ij}(\tf,\mu)\tilO_{j}(\tf)+O(\tf)$, with $i,j=1, 2, 3, 4$, valid for $2\tf\mu^2\sim1$~\cite{Suzuki:2013gza,Makino:2014taa,Taniguchi:2016ofw,Harlander:2018zpi}. The residual $O(\tf)$ corrections are finally removed by extrapolating $\tf\to0$.

%
%
\noindent \emph{Lattice QCD computations.---}
We use three 2+1-flavor highly improved staggered quark (HISQ) gauge ensembles generated by the HotQCD Collaboration~\cite{Follana:2006rc, HotQCD:2014kol, Bazavov:2019www}. The light sea-quark masses correspond to a pion mass of about 160 MeV, and the strange sea-quark mass is tuned to its physical value. The ensembles span three lattice spacings, $a=0.06$, $0.05$, and $0.04$ fm, with respective lattice sizes $64\times48^3$, $64^4$, and $64^4$. For the valence charm quark, we employ the setup in Refs.~\cite{Izubuchi:2019lyk,Gao:2020ito,Gao:2021hvs,Gao:2021xsm,Gao:2021dbh,Gao:2022vyh,Gao:2022iex,Gao:2022uhg,Gao:2023lny,Gao:2023ktu,Ding:2024lfj,Cloet:2024vbv,Ding:2024saz,Gao:2024fbh,Bollweg:2025iol,Gao:2025inf}---a tree-level tadpole-improved Wilson-clover action, where the gauge links entering the Dirac and clover operators are one-step hypercubic (HYP) smeared~\cite{Hasenfratz:2001hp}. 

We compute the $\eta_c$ ($\Gamma=\gamma_5$) and $J/\psi$ ($\Gamma=\gamma_{1,2,3}$) two-point correlation functions, $C^{\mathrm{2pt}}_\Gamma(t) = \sum_{\boldsymbol{x}} \langle \mathcal O_\Gamma(\boldsymbol{x},t) \mathcal{O}^\dagger_\Gamma(\boldsymbol{0},0) \rangle$, using Gaussian-smeared sources and sinks, and meson interpolating operators $\mathcal{O}_\Gamma(\boldsymbol{x},t)=\bar c(\boldsymbol{x},t)\Gamma c(\boldsymbol{x},t)$. The valence charm-quark mass is tuned by requiring the resulting $\eta_c$ and $J/\psi$ masses to reproduce their physical values~\cite{ParticleDataGroup:2024cfk} at the percent level.

Lattice gauge fields are evolved using the Zeuthen-flow discretization~\cite{Ramos:2015baa} and the fermion flow is implemented with a covariant Laplacian consistent with the valence action. Both flows are integrated with a third-order Runge--Kutta scheme~\cite{Luscher:2013cpa}, yielding $O(a^2)$ flow-time discretization effects. Measurements are performed at $\tf=n\,\epsilon_f$ with $\epsilon_f=5.1984\times10^{-4}\,\mathrm{fm}^2$, for integers $n$ spanning the window used in our continuum and $\tf\!\to\!0$ extrapolations. We use clover discretization to construct the flowed operator $\tilde{\mathcal{O}}_{1,\rho\sigma}(\tf)$, and its trace $\tilde{\mathcal{O}}_{2,\rho\sigma}(\tf)$, for the flowed gauge fields. For the flowed operator $\tilde{\mathcal{O}}_{3,\rho\sigma}(\tf)$, and its trace $\tilde{\mathcal{O}}_{4,\rho\sigma}(\tf)$, we employ “ringed’’ fermion fields~\cite{Makino:2014taa} to cancel wave–function renormalization.

Vacuum-subtracted hadronic matrix elements of flowed operators, $\langle H| \tilO_{i,\rs} |H\rangle (\mu)$ with $H=\eta_c,J/\psi$, are extracted from the three-point function $C^{\rs}_{\Gamma}(t,\tau) = \sum_{\boldsymbol{x},\boldsymbol{y}} \langle \mathcal{O}_\Gamma(\boldsymbol{x},t) \tilO_{i,\rs}(\boldsymbol{y},\tau) \mathcal{O}^\dagger_\Gamma(\boldsymbol{0},0)
\rangle - \langle C^{\mathrm{2pt}}_\Gamma(t) \rangle \sum_{\boldsymbol{y}} \langle \tilO_{i,\rs}(\boldsymbol{y},\tau) \rangle$, using source-sink separations $t/a=$ 22, 24, 26, 28, 30, and 32 (see Supplemental Material). Keeping our focus on hadron mass decompositions, we consider only hadrons at rest. The quark-line-connected three-point functions for $\tilO_{3,\rs}$ and $\tilO_{4,\rs}$ are computed using standard sequential-source methods. While computing the disconnected three-point functions for $\tilO_{1,\rs}$ and $\tilO_{2,\rs}$, we use the all-mode averaging (AMA)~\cite{Shintani:2014vja} technique for the corresponding two-point correlation functions to improve statistical precision. The quark-line-disconnected contributions of the charm quark to $\tilO_{3,\rs}$ and $\tilO_{4,\rs}$ are found to be negligible. We omit all light-flavor (up, down, and strange) contributions entering solely through quark-line-disconnected diagrams for the charmonia states. As we shall see later, all sum rules are satisfied without these contributions, indicating that any omitted light-flavor disconnected terms are not resolved at our present level of precision.

We fit the spectral decomposition of the ratio $R^{\rs}_{\Gamma}(t, \tau) = C^{\rs}_{\Gamma}(t,\tau) / C^{\rm 2pt}_{\Gamma}(t)$ at each $\tf$ using a two-exponential ansatz to obtain the EMT matrix elements of $\eta_c$ and $J/\psi$ in the limit $t\rightarrow\infty$. We also employ summation fits, where $R^{\rs}_{\Gamma}(t, \tau)$ are summed over the operator insertion time $\tau$ and extrapolated to $t\rightarrow\infty$ through a linear fit in $1/t$. These complementary methods yield consistent results across ensembles and flow times, providing confidence that excited-state effects due to finite source-sink separations of the three-point functions are under control.

%
%
\begin{figure}[h!]
    \includegraphics[width=0.45\textwidth]{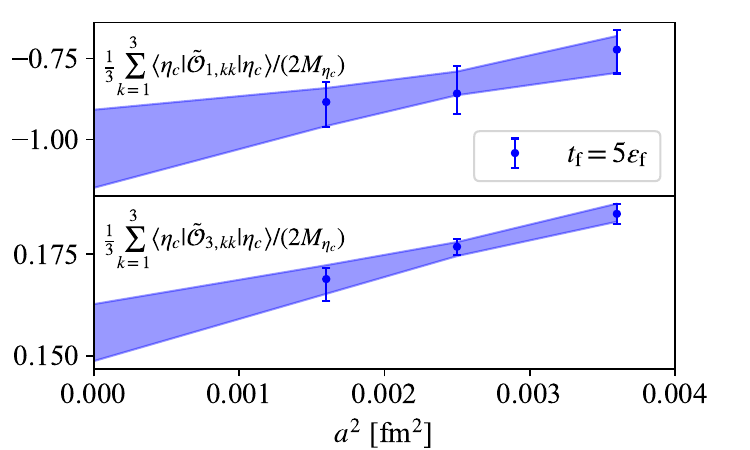}
    \caption{Hadronic matrix elements of gradient-flow-renormalized EMT components are plotted as data points. The shaded bands indicate the continuum extrapolations.}
    \label{fig:continuum}
\end{figure}

\noindent\emph{Hadronic matrix elements of $\msbar$-renormalized EMT.---}
These procedures  provide hadronic matrix elements of component $\tilO_{i,\rs}$, nonperturbatively renormalized in the gradient-flow scheme at multiple flow times $\tf$ and for three lattice spacings. To obtain matrix elements of $\tilO_{i,\rs}$ renormalized in the $\msbar$ scheme at $\mu=2~\mathrm{GeV}$, we perform three sequential steps: (i) continuum, $a\to0$, extrapolations of the gradient-flow–renormalized $\tilO_{i,\rs}$ matrix elements at each $\tf$; (ii) one- and two-loop perturbative scheme conversion of the continuum-extrapolated matrix elements to the $\msbar$ scheme at $\mu=2~\mathrm{GeV}$ for each $\tf$; (iii) zero-flow-time, $\tf\to0$, extrapolations of the resulting $\msbar$-renormalized matrix elements.

At each $\tf$, the $a\to0$ extrapolation is carried out by fitting $\langle H|\tilO_{i,\rs}|H\rangle(\tf,a)=\langle H|\tilO_{i,\rs}|H\rangle(\tf,0)+a^2 X_{i,\rs}(\tf)$ using uncorrelated bootstrap samples from the three ensembles to determine $\langle H|\tilO_{i,\rs}|H\rangle(\tf,0)$. For all operators and $\tf$ values investigated, the linear $a^2$ behavior is well supported by the data. Representative examples are shown in \autoref{fig:continuum} for $\sum_{k=1}^3\langle \eta_c|\tilO_{1,kk}/3|\eta_c\rangle/(2M^2_{\eta_c})$ (upper) and $\sum_{k=1}^3\langle \eta_c|\tilO_{3,kk}/3|\eta_c\rangle/(2M^2_{\eta_c})$ (lower) at $\tf=5\epsilon_f$.

\begin{figure}[h!]
    \includegraphics[width=0.4\textwidth]{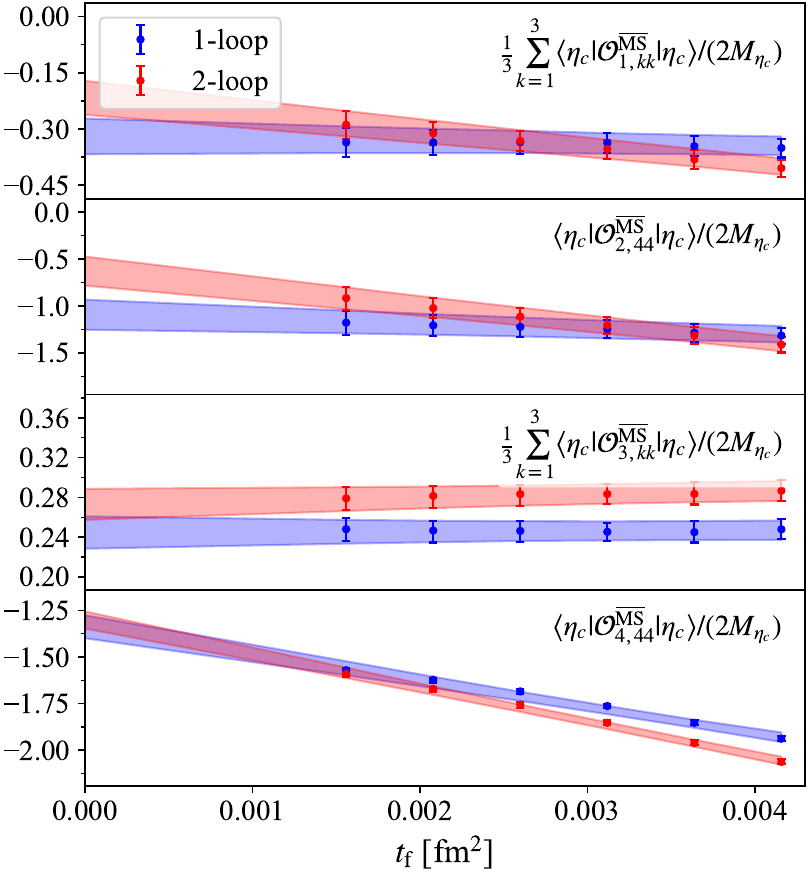}
    \caption{%
    Hadronic matrix elements of $\msbar$-renormalized EMT components at $\mu=2$~GeV, obtained using one-loop (blue) and two-loop (red) perturbative matching, are plotted as data points. The shaded bands indicate the $\tf\to0$ extrapolations.
    }%
    \label{fig:tf_extrapolation}
\end{figure}

The continuum-extrapolated matrix elements are then converted from the gradient-flow scheme to the $\msbar$ scheme at $\mu=2~\mathrm{GeV}$ using $\langle H| \O^{\msbar}_{i} |H\rangle (\mu,\tf) = M_{ij}(\tf,\mu)\, \langle H| \tilO_{j} |H\rangle (\tf,0)$, where $M_{ij}(\tf,\mu)$, with $i,j=1,\dots,4$, are the one- and two-loop perturbative matching coefficients~\cite{Suzuki:2013gza,Makino:2014taa,Taniguchi:2016ofw,Harlander:2018zpi} that incorporate both renormalization and operator mixing. Together with the QCD EoM, this yields $\msbar$-renormalized matrix elements of all five operators in \autoref{eq:O123}. 
To ensure that the scheme conversion remains perturbatively reliable---the $\ln(2\tf\mu^2)$ terms in $M_{ij,\rs}$ remain small at $\mu=2~\mathrm{GeV}$---we restrict ourselves to a small interval $\tf=0.00156-0.00416$ fm$^2$. As shown in \autoref{fig:tf_extrapolation}, within this window the difference between one-loop and two-loop matched results stays small, indicating that the uncertainty due to higher-order perturbative corrections is under control. 

Since gradient flow smooths the gluon and quark fields over an approximate radius $\sqrt{8\tf}$~\cite{Luscher:2009eq,Luscher:2010iy,Luscher:2011bx}, the small-$\tf$ expansion of the flowed operators receives additional contributions from higher-dimensional operators, leading to residual $\tf$ dependence of the matched matrix elements. To remove these contaminations, we perform a $\tf\to0$ extrapolation of the matched matrix elements using
$\langle H| \O^{\msbar}_{i,\rs} |H\rangle (\mu,\tf) = \langle H| \O^{\msbar}_{i,\rs} |H\rangle (\mu) + Y_{i,\rs}\tf$,
with correlated fits to bootstrap samples over the $\tf=0.00156\sim0.00416$ fm$^2$ interval. As illustrated in \autoref{fig:tf_extrapolation}, the data display a clear linear dependence in this range. Smaller flow times are excluded from the fit, since ultraviolet effects there are not yet regulated primarily by the flow, and the simple linear behavior in $\tf$ is not clearly observed.

%
%

%
\begin{table}[h!]
\begin{center}
\begin{tabular}{ c | c c | c c c }
\hline
\multirow{2}{*}{} & \multicolumn{2}{c|}{$\eta_c$} & \multicolumn{3}{c}{$J/\psi$}  \\ [1mm]
& \multicolumn{1}{c}{$A(0)$}   & $\bar{C}(0)$ & \multicolumn{1}{c}{$A(0)$} & \multicolumn{1}{c}{$\bar{C}(0)$} & $\bar{f}(0)$ \\ [2mm]
Gluon (g) & \multicolumn{1}{c}{$0.084(45)$} & 0.058(18) & \multicolumn{1}{c}{$0.077(35)$} & \multicolumn{1}{c}{-0.146(29)}     & -0.012(32)   \\ [2mm] 
Charm (c) & \multicolumn{1}{c}{$0.918(17)$}  & -0.068(4)  & \multicolumn{1}{c}{$0.926(34)$} & \multicolumn{1}{c}{0.176(87)}      & 0.023(5)     \\ [2mm]

Total & \multicolumn{1}{c}{$1.000(46)$} & $-0.010(18)$ & \multicolumn{1}{c}{$0.997(43)$} & \multicolumn{1}{c}{0.030(30)}      & 0.010(32)    \\ [1mm] \hline

\end{tabular}
\end{center}
\vspace{-3mm}
\caption{Gravitational form factors of $\eta_c$ (\autoref{eq:GFF-eta}) and $J/\psi$ (\autoref{eq:GFF-J}) in the $\msbar$ scheme and at the scale $\mu=2$~GeV.}
\label{tab:GFFs_summary}
\end{table}

\noindent\emph{Gravitational form factors.---}
From $\langle H| \O^{\msbar}_{i,\rs} |H\rangle (\mu)$ we construct $\langle H| T^{\msbar}_{X,\rs} |H\rangle (\mu)$ using \autoref{eq:Tg_Tq}, propagating correlated uncertainties using bootstrap samples. 

From here on, we present all physical results in Minkowski space, with Euclidean lattice matrix elements converted accordingly. 

For the pseudoscalar $\eta_c$ meson, the rest-frame matrix elements of $\msbar$-renormalized EMT components~\cite{Pefkou:2021fni,Hackett:2023nkr},
\begin{equation}
\begin{split}
& \langle \eta_c| T^{\msbar}_{X,00} |\eta_c \rangle = 2M_{\eta_c}^2 [A_X(0)+\bar C_X(0)] \,, \\
& \sum_{k=1}^3 \langle \eta_c| T^{\msbar}_{X,kk} |\eta_c \rangle = -6M_{\eta_c}^2 \bar{C}_X(0) \,,
\end{split}
\label{eq:GFF-eta}
\end{equation}
provide two independent relations to extract $\msbar$-renormalized GFFs $A_X(0)$ and $\bar{C}_X(0)$ at vanishing momentum transfer. The $A_X(0)$ gives the partonic momentum fraction and $\bar{C}_X(0)$ encodes the trace contribution.

For the vector $J/\psi$ meson, the rest-frame matrix elements of $\msbar$-renormalized EMT components in the state $J$ polarized along the $\gamma_1$ direction~\cite{Pefkou:2021fni,Polyakov:2019lbq}.
\begin{equation}
\begin{split}
& \langle J| T_{X,00}^{\msbar} |J\rangle = 2M_{J/\psi}^2 [A_{X}(0) -\frac{1}{2} \bar{C}_{X}(0) + \frac{1}{4} \bar{f}_X(0)] \,, \\
& \langle J | T_{X,11}^{\msbar} |J \rangle = 2M_{J/\psi}^2 \left[\frac{1}{2} \bar{C}_{X}(0) + \frac{3}{4} \bar{f}_X(0)\right] \,, \\
& \sum_{k=2,3} \! \langle J| T_{X,kk}^{\msbar} |J \rangle = 4M_{J/\psi}^2 \left[\frac{1}{2} \bar{C}_{X}(0) -\frac{1}{4} \bar{f}_X(0)\right] \,,
\end{split}
\label{eq:GFF-J}
\end{equation}
determine three $\msbar$-renormalized GFFs, $A_{X}$, $\bar{C}_{X}$, and $\bar{f}_X$, at vanishing momentum transfer.

\autoref{tab:GFFs_summary} presents the results of first direct lattice QCD calculations of GFFs $\bar{C}$ and $\bar{f}$. We obtain $\bar{C}_c(0)+\bar{C}_g(0)=0$ and, for the vector meson, $\bar{f}_c(0)+\bar{f}_g(0)=0$ within uncertainties, validating the corresponding trace-anomaly sum rules \footnote{The sign difference of $\bar{C}(0)$ between $\eta_c$ and $J/\psi$ stems from different Minkowski and polarization-normalization conventions used in the standard Lorentz-covariant parametrization of EMT matrix elements for pseudoscalar~\cite{Pefkou:2021fni} and vector~\cite{Polyakov:2019lbq} mesons.}. We also find,  within uncertainties,  the longitudinal momentum sum rule for partons $A_c(0)+A_g(0)=1$ for both $\eta_c$ and $J/\psi$.

%
\noindent\emph{The $\sigma$ term using QCD EOM.---}
The contribution to the hadron mass due to explicit chiral-symmetry breaking for $m_f>0$ is quantified by
\begin{equation}
\sigma_f =
\frac{1}{2 M_H} \langle H| \O^{\msbar}_{5,00,f} |H \rangle = 
- \frac{1}{4 M_H} \langle H| \O^{\msbar}_{4,00,f} |H \rangle .
\label{eq:sigma}
\end{equation}
We utilize the QCD EOM for quark flavor $f$ to eliminate $\O^{\msbar}_{5,\rs,f}$ in favor of $\O^{\msbar}_{4,\rs,f}$. Since $m_f>0$ explicitly breaks the scale invariance of classical QCD, the $\sigma$ term also contributes to the trace of QCD EMT in addition to anomaly generated through quantum corrections. 

An example of $\msbar$-renormalized $\sigma_c$, obtained from $\langle H| \O^{\msbar}_{4,00,c} |H \rangle$, is shown in \autoref{fig:tf_extrapolation}, where it can be seen that the one- and two-loop matching give consistent results in the $\tf \to 0$ limit.

%
%
\begin{figure*}[t!]
    \includegraphics[width=0.45\textwidth]{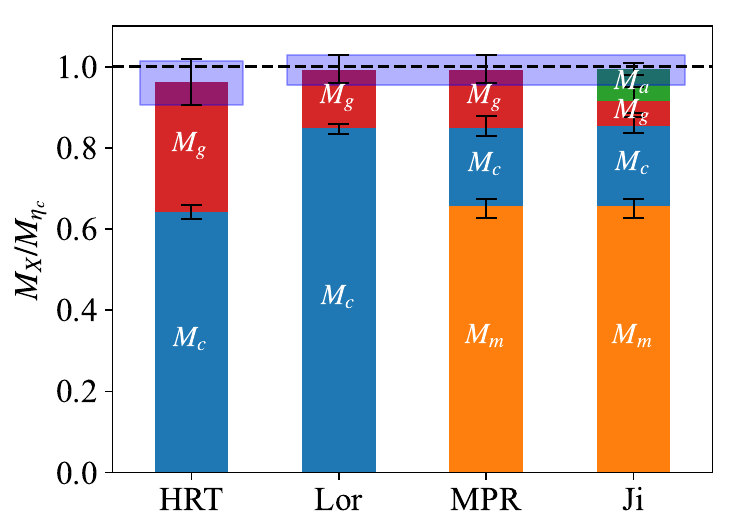}
    \includegraphics[width=0.45\textwidth]{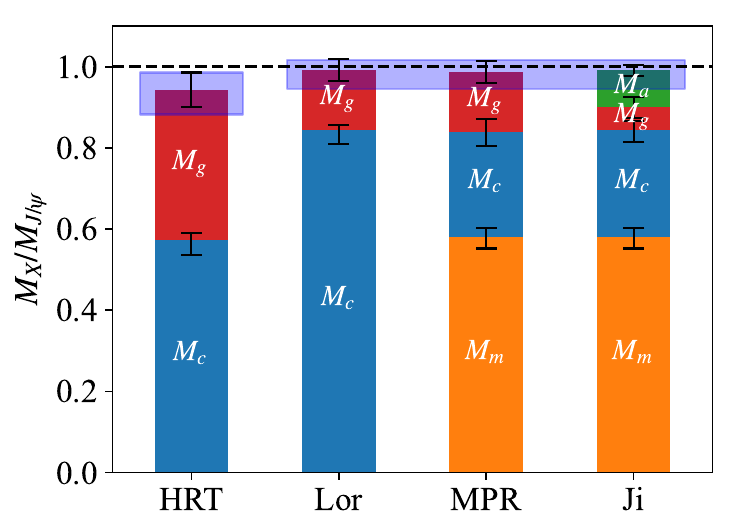}
    \caption{Hatta-Rajan-Tanaka (HRT, \autoref{eq:decomp-HRT}), Lorc{\'e} (Lor, \autoref{eq:decomp-Lor}) , Metz-Pasquini-Rodini (MPR, \autoref{eq:decomp-MPR}), and Ji (Ji, \autoref{eq:decomp-Ji}) decompositions of $\eta_c$ (left panel) and $J/\psi$ (right panel) masses. All contributions are presented at 2~GeV $\msbar$ scale using 2-loop gradient-flow to $\msbar$ matching. The black error bars indicate combined systematic and statistical uncertainty of each individual components. The blue error bands on top indicate combined systematic and statistical uncertainty of the sum over all components in each cases.}
    \label{fig:mass_decomp}
\end{figure*}

\noindent\emph{Hadron mass decompositions.---}
Having determined all necessary $\msbar$-renormalized matrix elements, we directly validate the sum rules underlying several complementary decompositions of the hadron mass. The Hatta-Rajan-Tanaka (HRT)~\cite{Hatta:2018sqd,Tanaka:2018nae,Tanaka:2022wzy,tanaka2026protonmassdecompositionsnnlo} decomposition based on the rest-frame matrix elements of the trace of $\msbar$-renormalized EMT is
\begin{equation}
 M_H = \!\!\! \sum_{X=q,g} \!\!\! M_X^\text{HRT} ,
 ~~\text{with}~~
 M_X^\text{HRT} \! = 
 \frac{\langle H| \left( T^{\msbar}_X \right)^\rho_\rho |H \rangle }{2M_H} .
\label{eq:decomp-HRT}
\end{equation}

The Lorc{\'e}~\cite{Lorce:2017xzd}, Metz-Pasquini-Rodini (MPR)~\cite{Rodini:2020pis,Metz:2020vxd}, and Ji~\cite{Ji:1994av,Ji:1995sv,Ji:2021mtz,Ji:2021qgo} decompositions are all based on the rest-frame matrix element of $T^{\msbar}_{00}$, but organize the quark/gluon contributions in different ways. The Lorc{\'e}'s decomposition in the $\msbar$ scheme reads
\begin{equation}
\begin{split}
 & M_H = \!\!\! \sum_{X=q,g} \!\!\! M^{\text{Lor}}_X , 
 \quad \text{with} \quad 
 M^{\text{Lor}}_X \! = \frac{\langle H| T^{\msbar}_{X,00} |H \rangle }{2M_H} .
\end{split}
\label{eq:decomp-Lor}
\end{equation}
The MPR decomposition explicitly separates out $\sigma_f$, giving
\begin{equation}
\begin{split} 
& M_H = \!\!\!\! \sum_{X=q,g,m } \!\!\!\!\! M_X^{\text{MPR}} ,
\quad \text{with} \quad
M_m^\text{MPR} = \sum_f \sigma_f , \\
& M_q^\text{MPR} = M^\text{Lor}_q - \sum_f \sigma_f , 
\quad 
M_g^\text{MPR} = M^\text{Lor}_g .
\end{split}
\label{eq:decomp-MPR}
\end{equation}
The Ji decomposition distinguishes the contributions from the traceless part of the EMT from the trace anomaly part, yielding
\begin{equation}
\begin{split}
& M_H = \!\!\!\!\! \sum_{X=q,g,m,a} \!\!\!\!\!\! M_X^{\text{Ji}} ,
\quad \text{with} \quad 
M_m^{\text{Ji}} = M_m^\text{MPR} , \\
& M_q^{\text{Ji}} = M_q^{\text{Lor}} - \frac{1}{4} M_q^{\text{HRT}} - \frac{3}{4} M_m^\text{MPR} , \\
& M_g^{\text{Ji}} = M_g^{\text{Lor}} - \frac{1}{4} M_g^{\text{HRT}} , \\
& M_a^{\text{Ji}} = \frac{1}{4} \left[ M_q^{\text{HRT}} + M_g^{\text{HRT}} -M_m^\text{MPR} \right ]\,. \\
\end{split}
\label{eq:decomp-Ji}
\end{equation}

\autoref{fig:mass_decomp} presents HRT, Lorc{\'e}, MPR, and Ji mass decompositions, all at 2~GeV $\msbar$ scale, of $\eta_c$ (left panel) and $J/\psi$ (right panel). For the quark sector, we include only the charm flavor and find all mass rules are satisfied within uncertainty, demonstrating the mutual consistency of these distinct decomposition schemes when evaluated in the same renormalization framework. This suggests that omitted light-flavor disconnected contributions are smaller than our current total precision, although a dedicated calculation of these terms is left to future work. The dominant contributions to $M_{\eta_c}=2984.1~\mathrm{MeV}$ and $M_{J/\psi}=3096.9~\mathrm{MeV}$ come from the charm quark: $M_c^{\text{HRT}} = 1918^{+53}_{-54}~\mathrm{MeV}$, $M_c^{\text{Lor}} = 2536^{+25}_{-48}~\mathrm{MeV}$, and $M_c^{\text{Ji}} + M_m^{\text{Ji}}= 2542^{+26}_{-40}~\mathrm{MeV}$ for $\eta_c$, and $M_c^{\text{HRT}} = 1776^{+51}_{-117}~\mathrm{MeV}$, $M_c^{\text{Lor}} = 2617^{+35}_{-110}~\mathrm{MeV}$, and $M_c^{\text{Ji}} + M_m^{\text{Ji}}= 2614^{+40}_{-78}~\mathrm{MeV}$ for $J/\psi$. Gluons also provide sizable contributions: $M_g^{\text{HRT}} = 953^{+168}_{-167}~\mathrm{MeV}$, $M_g^{\text{Lor}} = 426^{+111}_{-100}~\mathrm{MeV}$, and $M_g^{\text{Ji}} = 188^{+102}_{-91}~\mathrm{MeV}$ for $\eta_c$, and $M_g^{\text{HRT}} = 1144^{+134}_{-129}~\mathrm{MeV}$, $M_g^{\text{Lor}} = 455^{+82}_{-85}~\mathrm{MeV}$, and $M_g^{\text{Ji}} = 179^{+78}_{-82}~\mathrm{MeV}$ for $J/\psi$. These first-principles QCD calculations demonstrate QCD confinement not only provides substantial gluon-energy contributions but also generates non-negligible trace-anomaly contributions---$M_a^{\text{Ji}} = 231^{+45}_{-44}~\mathrm{MeV}$ for $\eta_c$ and $M_a^{\text{Ji}} = 277^{+43}_{-45}~\mathrm{MeV}$ for $J/\psi$---to the hadron masses, even for deeply bound $c\bar{c}$ states.

%
%
\noindent\emph{Conclusions.---}
In this Letter, we have presented the first direct lattice-QCD validation of sum rules associated with quark–gluon decomposition of hadron mass. By constructing the QCD energy–momentum tensor nonperturbatively in a gradient-flow scheme, we achieve a formulation that controls operator mixing and avoids power-divergent subtractions, enabling a well-defined continuum limit extrapolation, followed by two-loop matching to the $\msbar$ scheme and zero-flow-time extrapolations. This establishes a common renormalization framework in which multiple mass decomposition sum rules can be verified simultaneously.

Applying this framework to the $\eta_c$ and $J/\psi$ charmonia, we demonstrate the consistency of energy-density-based and trace-based decompositions and quantify substantial contributions from gluonic energy and the QCD trace anomaly to heavy quark bound-state masses. The first lattice determination of the gravitational form factor $\bar{C}$ further illustrates the scope of the approach. Residual effects from finite volume, the partial quenching from the omission of a dynamical charm quark in the sea, and neglected light-flavor disconnected contributions are expected to be subleading at the present level of precision, but should be quantified more systematically in future higher-precision studies.

Beyond charmonium, the methodology developed here provides a general foundation for first-principles studies of hadron mass, spin structure, and gravitational form factors. In particular, it enables access to components of the energy–momentum tensor, such as trace and higher-twist contributions, that are challenging to constrain experimentally, thereby providing essential nonperturbative input complementary to ongoing and future programs at the Thomas Jefferson National Accelerator Facility and the future Electron–Ion Collider.

%
%
Work used GRID~\cite{boyle2015gridgenerationdataparallel}, GPT (Grid Python Toolkit)~\cite{GPT}, and RunDec~\cite{Chetyrkin:2000yt} software packages.

\noindent\emph{Acknowledgments.---}
This material is based upon work supported by the U.S.~Department of Energy (DOE), Office of Science, Office of Nuclear Physics through Contract No.~DE-SC0012704, and within the frameworks of Scientific Discovery through Advanced Computing (SciDAC) award ``Fundamental Nuclear Physics at the Exascale and Beyond'' and the Topical Collaboration in Nuclear Theory ``Heavy-Flavor Theory (HEFTY) for QCD Matter.'' This work is supported partly by the National Natural Science Foundation of China under Grants No.~12325508, No.~12293064, and No.~12293060 as well as the National Key Research and Development Program of China under Contract No.~2022YFA1604900 and the Fundamental Research Funds for the Central Universities, Central China Normal University under Grants No.~30101250314 and No.~30106250152.

This research used awards of computer time provided by the Oak Ridge Leadership Computing Facility, which is a DOE Office of Science User Facility supported under Contract No.~DE-AC05-00OR22725; the National Energy Research Scientific Computing Center, a DOE Office of Science User Facility supported by the Office of Science of the U.S.~Department of Energy under Contract No.~DE-AC02-05CH11231 using NERSC Award No.~NP-ERCAP0036132; and the Nuclear Science Computing Center at Central China Normal University ($\mathrm{NSC}^{3}$).

\onecolumngrid
\begin{center}
\textbf{\large Supplemental Material}
\end{center}
\appendix
%
%
\section{Renormalized EMT Operators via Gradient Flow}\label{sec:flow}

To compute EMT matrix elements on the lattice, one must address operator renormalization. Gluonic operators, in particular, suffer from power divergences and complicated operator mixing. The gradient flow method provides a systematic solution~\cite{Narayanan:2006rf,Luscher:2009eq,Luscher:2010iy,Luscher:2011bx}: by evolving fields along a fictitious flow time $\tf$, short-distance fluctuations are suppressed, rendering flowed operators finite at nonzero $\tf$ without requiring further renormalization, and enabling a well-defined continuum limit by taking $a\to0$ at fixed physical $\tf$.

For the gauge field $B_\mu(\tf,x)$, the flow equation is:
\begin{align}
\partial_{\tf} B_\mu(\tf,x) &= D_\nu F_{\nu\mu}(\tf,x), \qquad B_\mu(0,x) = A_\mu(x), 
\end{align}
and for the fermion field $\chi(\tf,x)$,
\begin{align}
\partial_{\tf} \chi(\tf,x) &= \Delta \chi(\tf,x), \qquad \chi(0,x) = \psi(x).
\end{align}
where $D_\nu$ is the gauge-covariant derivative and $\Delta$ is the covariant Laplacian. The flowed fields are smooth and free from power-divergent mixing at nonzero $\tf$, making them suitable for defining composite operators without further renormalization.

We construct the flowed versions of the relevant EMT operators using the flowed gauge and fermion fields, subtracting their vacuum expectation values (VEV) to eliminate contributions from the vacuum.

For the gluonic sector, the flowed operators are given by,
\begin{equation}
    \begin{aligned}
        &\tilO_{1,\mn}(\tf,x)\equiv2{\rm Tr}^{\rm c}\left[F_{\rho\omega}(\tf,x)F_{\sigma\omega}(\tf,x)\right],\\
        &\tilO_{2,\mn}(\tf,x)\equiv2\delta_{\mn}{\rm Tr}^{\rm c}\left[F_{\rho\sigma}(\tf,x)F_{\rho\sigma}(\tf,x)\right],
    \end{aligned}\label{eq:tilO12}
\end{equation}
where the field strength tensor is constructed from the clover definition of the flowed link fields. 

For the quark sector, we use ringed fermion fields $\mathring{\chi}$ and $\mathring{\bar{\chi}}$~\cite{Makino:2014taa}, which cancel the wavefunction renormalization by construction:
\begin{equation}
\begin{aligned}
    &\mathring{\chi}_f(\tf,x)\equiv\sqrt{\frac{-6}{(4\pi\tf)^2\langle\bar{\chi}_f(\tf,x)\overset{\leftrightarrow}{\slashed{D}}\chi_f(\tf,x)\rangle}}\chi_f(\tf,x),\\
    &\mathring{\bar{\chi}}_f(\tf,x)\equiv\sqrt{\frac{-6}{(4\pi\tf)^2\langle\bar{\chi}_f(\tf,x)\overset{\leftrightarrow}{\slashed{D}}\chi_f(\tf,x)\rangle}}\bar{\chi}_f(\tf,x).
\end{aligned}\label{eq:ring}
\end{equation}
Using these, we define the flowed fermionic operators as,
\begin{align}
    \begin{aligned}
        &\tilO_{3,\mn}(\tf,x)\equiv\sum_f\mathring{\bar{\chi}}_f(\tf,x)\left(\gamma_{\mu}\overset{\leftrightarrow}{D}_{\nu}+\gamma_{\nu}\overset{\leftrightarrow}{D}_{\mu}\right)\mathring{\chi}_f(\tf,x),\\
        &\tilO_{4,\mn}(\tf,x)\equiv\delta_{\mn}\sum_f\mathring{\bar{\chi}}_f(\tf,x)\overset{\leftrightarrow}{\slashed{D}}\mathring{\chi}_f(\tf,x)\, .
    \end{aligned}\label{eq:tilO34}
\end{align}
These flowed operators are finite and related to the renormalized EMT in the $\overline{\text{MS}}$ scheme through a small-flow-time matching procedure $\langle H| \O^{\msbar}_{i} |H\rangle (\mu,\tf) = M_{ij}(\tf,\mu)\, \langle H| \tilO_{j} |H\rangle (\tf,0)$ with $i,j=1, 2, 3, 4$, as described in the main text~\cite{Suzuki:2013gza,Makino:2014taa,Taniguchi:2016ofw,Harlander:2018zpi}. The relation is valid up to $O(\tf)$ corrections, which are later removed by an extrapolation to $\tf \to 0$. 

%
%
\section{Details of lattice setup}\label{sec:app_lat}

\begin{table}[]
\begin{tabular}{|c|c|c|c|c|c|c|}
\hline
$a$ {[}fm{]} & $N_s$ & $N_t$               & $m_s^{\rm sea}/m_l^{\rm sea}$ & $M_{\pi}^{\rm sea}$ {[}MeV{]} & $c_{\rm sw}$ & $am^{\rm val}_c$ \\ \hline
0.06         & 48    & \multirow{3}{*}{64} & \multirow{3}{*}{20}           & \multirow{3}{*}{160}          & 1.0336       & 0.306            \\ \cline{1-2} \cline{6-7} 
0.05         & 64    &                     &                               &                               & 1.030934     & 0.236            \\ \cline{1-2} \cline{6-7} 
0.04         & 64    &                     &                               &                               & 1.02868      & 0.167            \\ \hline
\end{tabular}
\caption{Summary of gauge ensemble parameters.}
\label{tb:tab_ensembles}
\end{table}
\begin{table}[]
\begin{tabular}{|c|c|c|c|c|c|c|}
\hline
inserted                             & $a$ {[}fm{]} & $N_{\rm conf}$ & $N_{\rm src}$ sloppy & $N_{\rm src}$ exact & $\sigma_{\rm sm}/a$ & $N_{\rm sm}$ \\ \hline
\multirow{3}{*}{$\tilO_1$,$\tilO_2$} & 0.06         & 556            & \multirow{3}{*}{512} & \multirow{3}{*}{16} & 0                   & 0            \\ \cline{2-3} \cline{6-7} 
                                     & 0.05         & 580            &                      &                     & 3                   & 30           \\ \cline{2-3} \cline{6-7} 
                                     & 0.04         & 599            &                      &                     & 4                   & 40           \\ \hline
\multirow{3}{*}{$\tilO_3$,$\tilO_4$} & 0.06         & 50             & \multirow{3}{*}{0}   & \multirow{3}{*}{10} & 4                   & 40           \\ \cline{2-3} \cline{6-7} 
                                     & 0.05         & 50             &                      &                     & 4.8                 & 48           \\ \cline{2-3} \cline{6-7} 
                                     & 0.04         & 50             &                      &                     & 6                   & 60           \\ \hline
\end{tabular}
\caption{The measurement details are summarized, including the number of configurations ($N_{\rm conf}$), number of sources per configuration ($N_{\rm src}$), Gaussian smearing parameters—smearing width $\sigma_{\rm sm}$ and number of steps $N_{\rm sm}$. For $N_{\rm src}$, “sloppy” sources refer to single-precision with a tolerance $10^{-8}$, while “exact” sources use double precision with a tolerance $10^{-18}$.}
\label{tb:tab_3pt}
\end{table}

This section provides details on the construction of two-point and three-point correlation functions, the application of gradient flow, and the computation of flowed operator matrix elements in our study.

We compute both connected and disconnected three-point (3pt) functions involving flowed quark and gluon EMT operators as defined in \autoref{eq:tilO12} and \autoref{eq:tilO34}. For the gluon sector, which enters entirely through disconnected diagrams, we employ the all-mode averaging (AMA) technique~\cite{Shintani:2014vja} to improve statistical precision. Each configuration includes 512 ``sloppy" sources computed in single precision with residual tolerance $10^{-8}$, and 16 ``exact" sources computed in double precision with tolerance $10^{-18}$. The number of configurations and measurement details are summarized in Table~\ref{tb:tab_3pt}.

In the quark sector, we evaluate only connected three-point functions, which are expected to dominate for charm quarks. The corresponding three-point correlators are constructed in the standard way using the sequential-source technique for connected insertions. For each ensemble, we analyze 50 gauge configurations with 10 exact source positions per configuration. We explicitly verify that charm-quark disconnected contributions are negligible compared with both the connected charm and the gluon contributions, as shown in \autoref{fig:test_dis_charm} and discussed later. We therefore omit contributions from light and strange quarks. Our final results demonstrate that the retained connected charm and disconnected gluon contributions are sufficient to reproduce the charmonium mass spectrum within uncertainties. 

To enhance the overlap with the charmonium ground states, we apply Gaussian smearing to both the source and sink quark fields on the finer lattices. This is implemented via gauge-covariant Wuppertal smearing, which approximates a Gaussian profile through repeated applications of the spatial covariant Laplacian:
\begin{align}
\begin{aligned}
\psi(x) &\rightarrow \left(1 + \frac{\sigma_{\rm sm}^2}{4N_{\rm sm}} \Delta \right)^{N_{\rm sm}} \psi(x), \\
\Delta\psi(x) &\equiv \sum_{\mu=1}^{3} \big[ U_\mu(x)\psi(x+\hat{\mu}) + U^\dagger_\mu(x-\hat{\mu})\psi(x-\hat{\mu}) - 2\psi(x) \big],
\end{aligned}
\end{align}
where $\Delta$ denotes the spatial gauge-covariant Laplacian. The smearing parameters $(\sigma_{\rm sm}, N_{\rm sm})$ used for the smeared ensembles are listed in \tb{tab_3pt}. To further suppress excited-state contamination, we perform the three-point function measurements with large source-sink separations, $t \in [22a, 32a]$. 

We apply gradient flow to systematically renormalize the quark and gluon fields in a nonperturbative manner, as discussed in main text. Specifically, gluon operators are constructed from the flowed gauge fields at finite flow time $\tf$, while quark operators are built from the flowed valence propagators using fermion flow techniques \cite{Luscher:2013cpa}. To obtain the correct EMT in the continuum limit, it is necessary to extrapolate $\tf \to 0$. Therefore, we perform measurements at multiple values of $\tf$ (in physical units) for all three ensembles. Measurements are performed in units of $\epsilon_{\rm f} = 5.1984 \times 10^{-4},\mathrm{fm}^2$, while the fields are evolved using integration step sizes between $\epsilon_{\rm f}/40$ and $\epsilon_{\rm f}/2$.

For operators involving ringed fermion fields, we compute the necessary vacuum expectation values including $\langle \bar{\chi}_f(\tf,x) \overset{\leftrightarrow}{\slashed{D}} \chi_f(\tf,x) \rangle$, as described in \autoref{eq:ring}. This requires estimating all-to-all propagators. To this end, we employ $Z_2$ noise vectors with 10 random sources per configuration. Our tests confirm that for charm quarks, this setup ensures systematic uncertainty below $10^{-4}$, which is subdominant relative to our typical statistical uncertainties of $10^{-3}$ or larger.

%
%
\section{Two-point functions and charmonium mass}\label{sec:2pt}

\begin{figure}
    \includegraphics[width=0.45\textwidth]{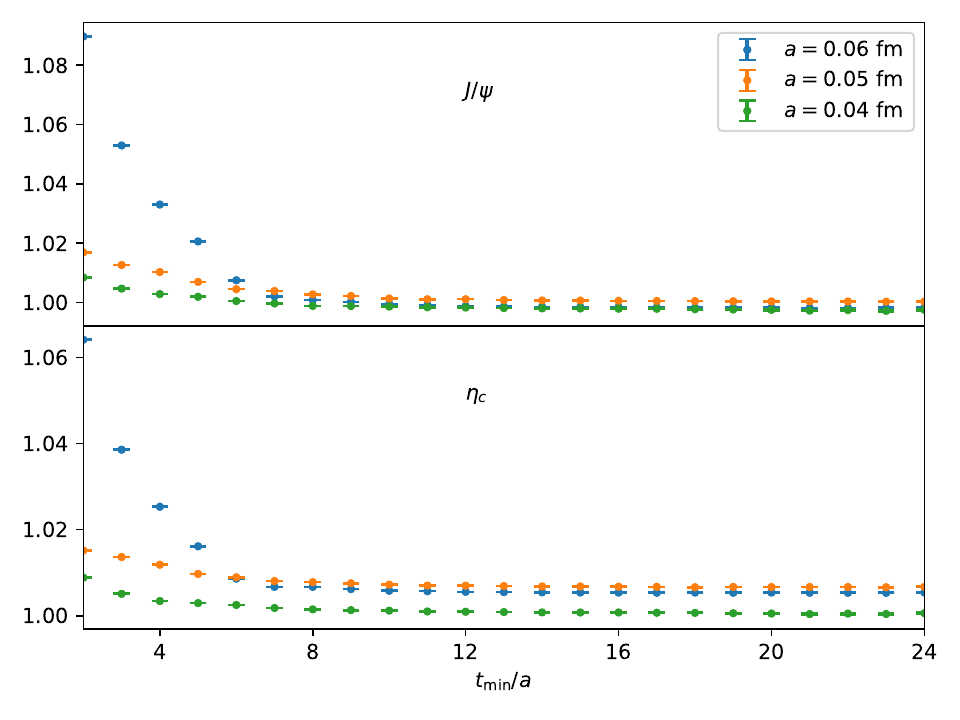}
    \caption{$E_0/E_0^{\rm PDG}$ \cite{ParticleDataGroup:2024cfk} of $\eta_c$ and $J/\psi$ determined from 2-state fit of the 2pt functions are shown for three gauge ensembles, as a function of $t_{\rm min}$. The fit range is [$t_{\rm min},32a$].}
    \label{fig:mass}
\end{figure}

Extracting hadronic matrix elements of the EMT requires knowledge of the ground-state contribution to the relevant three-point functions. To isolate the ground-state signal, we analyze meson two-point (2pt) correlation functions, which provide access to the spectrum and amplitudes of low-lying states. The Euclidean two-point function is defined as:
\begin{equation}
\begin{aligned}
C^{\rm 2pt}_{\Gamma}(t)\equiv&\sum_{\boldsymbol{x}}\langle\bar{c}(x)\Gamma c(x)\bar{c}(x_0)\Gamma c(x_0)\rangle=-\sum_{\boldsymbol{x}}{\rm Tr}\left[S_c(x_0|x)\Gamma S_c(x|x_0) \Gamma\right]\,,
\end{aligned}
\label{eq:2pt_x_x0}
\end{equation}
where $S_c(x|x_0)$ denotes the charm quark propagator from source $x_0\equiv(\boldsymbol{x}_0,0)$ to sink $x\equiv(\boldsymbol{x},t)$, and the trace is taken over spin and color indices. The gamma matrix $\Gamma$ specifies the meson interpolator: $\Gamma = \gamma_5$ corresponds to the pseudoscalar $\eta_c$ channel, while $\Gamma = \gamma_\lambda$ corresponds to the vector $J/\psi$ channel with polarization along the $\lambda$-th spatial direction. We average over the $\lambda=1,2,3$ for $J/\psi$ to reduce statistical noise and improve rotational symmetry. The spatial sum over $\boldsymbol{x}$ projects onto zero-momentum states. In this study, we neglect the quark-line disconnected contributions to the charmonium 2pt functions.

To quantitatively isolate the ground-state and subtract excited-state contamination, we perform correlated two-state fits to the 2pt functions using the ansatz,
\begin{equation}
C^{\rm 2pt}(t) = \sum_{n=0}^{n_{\rm max}} |Z_n|^2 \left[e^{-E_n t}+e^{-E_n (N_ta-t)}\right],
\label{eq:2ptfit}
\end{equation}
where $E_n$ and $Z_n$ are the energy and overlap amplitude of the $n$-th state, and $N_t$ is the temporal extent of the lattice.  We set $n_{\rm max}=1$ to account for the leading excited state. 

\autoref{fig:mass} shows the ratio between our fitted ground-state energies and the corresponding physical meson masses reported by the Particle Data Group (PDG)~\cite{ParticleDataGroup:2024cfk}, i.e., $E_0/E_0^{\rm PDG}$, for the $\eta_c$ (lower panel) and $J/\psi$ (upper panel). These results are obtained from two-state fits to the 2pt correlators on all three ensembles, using fit ranges $[t_{\rm min}, 32a]$ with varying $t_{\rm min}$. The ratios are remarkably close to unity, with deviations below 1\%, indicating that our bare charm-quark mass is accurately tuned to reproduce the physical charmonium spectrum.

The fitted values of $E_0$ and $E_1$ from each bootstrap sample are used in the subsequent analysis of three-point functions to isolate ground-state matrix elements of the energy-momentum tensor.

%
%
\section{Three-point functions of flowed quark EMT operators}\label{sec:3ptquark}

The matrix elements of the EMT are extracted from three-point (3pt) correlation functions, where the EMT operator is inserted between interpolating operators for charmonium states. Only the operators $\tilde{\mathcal{O}}_{3,\rho\sigma}$ and $\tilde{\mathcal{O}}_{4,\rho\sigma}$ as defined in \autoref{eq:tilO34} are required to reconstruct the physical matrix elements of the quark EMT. Since $\tilde{\mathcal{O}}_{4,\rho\sigma}$ is the trace of $\tilde{\mathcal{O}}_{3,\rho\sigma}$, it suffices to extract only the matrix elements of $\tilde{\mathcal{O}}_{3,\rho\sigma}$. The 3pt function is defined as,
\begin{align}\label{eq:3ptq}
    C^{\rho\sigma}_{q,\Gamma}(t,\tau)\equiv\sum_{\boldsymbol{x}}\langle\bar{c}(x)\Gamma c(x) \tilO_{3,\mn}(\tau)\bar{c}(x_0)\Gamma c(x_0)\rangle,
\end{align}
where the operator insertion is defined by,
\begin{align}
    \tilO_{3,\mn}(\tau)=\sum_{\boldsymbol{y}}\tilO_{3,\mn}(y).
\end{align}
Here, $x_0\equiv(\boldsymbol{x}_0,0)$, $y\equiv(\boldsymbol{y},\tau)$, and $x\equiv(\boldsymbol{x},t)$ denote the coordinates of the source, operator inserted, and sink, respectively. The temporal variables $t$ and $\tau$ correspond to the source-sink separation and the operator insertion time. The gamma matrix $\Gamma$ correspond to the meson interpolators, with choices depending on the meson channel, as discussed previously. The corresponding three-point correlators are constructed in the standard way using the sequential-source technique for connected insertions.

The covariant derivative $\overset{\leftrightarrow}{D}_\mu$ in the EMT operator is discretized on the lattice in a symmetric form:
\begin{equation}
    \begin{aligned}
    &\overset{\leftrightarrow}{D}\equiv\overset{\rightarrow}{D}-\overset{\leftarrow}{D},\\
    &\overset{\rightarrow}{D}_{\mu}c(y)\equiv\frac{1}{2}\left[
    U_{\mu}(y)c(y+\hat{\mu}) - U_{\mu}(y-\hat{\mu})^{\dagger}c(y-\hat{\mu})\right],\\
    &\bar{c}(y)\overset{\leftarrow}{D}_{\mu}\equiv\frac{1}{2}\left[
    \bar{c}(y+\hat{\mu})U_{\mu}(y)^{\dagger} - \bar{c}(y-\hat{\mu})U_{\mu}(y-\hat{\mu})
    \right].
\end{aligned}\label{eq:deriv}
\end{equation}

To achieve nonperturbative renormalization and eliminate short-distance singularities, we construct the EMT operator using flowed fields. Specifically, we replace the local operator $\bar{c}(y)\gamma_{\mu} \overset{\leftrightarrow}{D}_{\nu}c(y)$ by its flowed counterpart:
\begin{equation}
\bar{c}(\tf,y)\gamma_{\mu} \overset{\leftrightarrow}{D}_{\nu}(\tf) c(\tf,y),
\end{equation}
where both the quark fields and the gauge links entering the derivative operator are defined at positive flow time $\tf$.

The flowed fermion fields are defined by the fermion flow equations~\cite{Luscher:2013cpa}. In practice, we implement the fermion flow at the level of Wick contractions by convolving the unflowed charm-quark propagators with the fermion flow kernel,
\begin{equation}
\begin{aligned}
\left\langle {\bar{c}(\tf,y)\, c(x)} \right\rangle
&= \sum_z K(\tf, y; 0, z)\, S_c(z|x), \\
\left\langle {c(x)\, \bar{c}(\tf,y)} \right\rangle
&= \sum_z S_c(x|z)\, K^\dagger(\tf, y; 0, z),
\end{aligned}
\label{eq:flow_contraction}
\end{equation}
where $S_c$ is the unflowed charm-quark propagator and $K(\tf, y; 0, z)$ is the fermion flow kernel. The kernel is proportional to the identity in Dirac space and acts as a gauge-covariant smearing function in color and position space. In the $\tf\to0$ limit it reduces to a delta function,
\begin{equation}
\lim_{\tf \to 0} K(\tf, y; 0, z) = \delta_{yz}.
\end{equation}
These relations allow us to construct flowed three-point functions entirely from the unflowed propagators together with the known flow kernel, so that no additional flowed inversions are required.

%
%
\section{Three-point functions of flowed gluon EMT operators}\label{sec:3ptgluon}

The matrix elements of the gluon EMT are extracted from three-point correlation functions with gluonic operator insertions. Unlike the quark case, where connected diagrams dominate, the gluon EMT contributes only through disconnected diagrams. The gluon EMT is constructed from two operators, $\tilde{\mathcal{O}}_{1,\rho\sigma}$ and $\tilde{\mathcal{O}}_{2,\rho\sigma}$, as defined in \autoref{eq:tilO12}. As noted earlier, $\tilde{\mathcal{O}}_{2,\rho\sigma}$ is proportional to the trace of $\tilde{\mathcal{O}}_{1,\rho\sigma}$, so we only need to evaluate matrix elements of $\tilde{\mathcal{O}}_{1,\rho\sigma}$ to reconstruct the full gluon EMT.

The gluonic three-point correlation function is defined as,
\begin{equation}\label{eq:3ptg}
\begin{aligned}
    C^{\rs}_{g,\Gamma}(t,\tau) = \sum_{\boldsymbol{x},\boldsymbol{y}} \langle \mathcal{O}_\Gamma(\boldsymbol{x},t) \tilO_{1,\rs}(\boldsymbol{y},\tau) \mathcal{O}^\dagger_\Gamma(\boldsymbol{0},0)
\rangle - \langle C^{\mathrm{2pt}}_\Gamma(t) \rangle \sum_{\boldsymbol{y}} \langle \tilO_{1,\rs}(\boldsymbol{y},\tau) \rangle,
\end{aligned}
\end{equation}
where $C^{\mathrm{2pt}}_{\Gamma}$ is defined in \autoref{eq:2pt_x_x0}, and the operator insertion is summed over spatial positions at time slice $\tau$. The subtraction removes vacuum contributions.

The gluon field strength tensor $F_{\rho\sigma}(x)$ is constructed using a clover-type discretization:
\begin{equation}
    \begin{aligned}
        &U_{\mn}(x)=U_{\rho}(x)U_{\sigma}(x+\hat{\rho})U_{\rho}^{\dagger}(x+\hat{\sigma})U_{\sigma}^{\dagger}(x),\\
        &P_{\mn}(x)=U_{\mn}(x)+U_{\sigma,-\rho}(x)+U_{-\rho,-\sigma}(x)+U_{-\sigma,\rho}(x),\\
        &F_{\mn}'(x)=\frac{1}{8}\left[P_{\mn}(x)-P_{\mn}^{\dagger}(x)\right],\\
        &F_{\mn}(x)=-i\left[F_{\mn}'(x)-\frac{{\rm Tr}^{\rm c}F_{\mn}'(x)}{\Nc}I\right]\,,
    \end{aligned}
\end{equation}
where $\mathrm{Tr}^{\mathrm{c}}$ denotes the trace in the color space. To construct the flowed operator $F_{\rho\sigma}(\tf,x)$ at positive flow time $\tf$, all gauge links $U_\rho(x)$ are replaced by their flowed counterparts generated through the gradient flow.

\section{Fitting strategies for ground-state matrix elements extraction from three-point functions}\label{sec:app_fit_strategies}
To extract the matrix elements of the quark or gluon EMT, we analyze the ratio of 3pt to 2pt correlation functions constructed using flowed EMT operators. This ratio is defined as
\begin{equation}
R^{\rho\sigma}_{X,\Gamma}(t, \tau) = \frac{C^{\rho\sigma}_{X,\Gamma}(t,\tau)}{C^{\rm 2pt}_{\Gamma}(t)},
\end{equation}
where $C^{\rho\sigma}_{X,\Gamma}(t,\tau)$ and $C^{\rm 2pt}_{\Gamma}(t)$ denote the three- and two-point functions respectively with $X=q,g$. $\Gamma$ specifies the Dirac structure of the interpolating operator for the external meson state.

The three-point function admits the spectral decomposition:
\begin{equation}
\label{eq:3ptfit}
C^{\rho\sigma}_{X,\Gamma}(t,\tau) = \sum_{m,n} Z_m^* Z_n \frac{\langle m | \tilO_{i,\mn} | n \rangle}{2M} e^{-\tau E_n} e^{-(t - \tau) E_m}.
\end{equation}
Here, $M$ represent the meson mass, while $Z_n$ and $E_n$ are the overlap amplitudes and energy levels of the meson states, which are identical to those appearing in the two-point function decomposition. $\langle 0|\tilO_{i,\mn}|0\rangle/2M$ is the target matrix element in the ground-state limit.

To reliably isolate ground-state contributions at finite time separations, we employ both two-state and summation fitting strategies. All fits are fully correlated, utilizing covariance matrices estimated via bootstrap resampling. The fit window for the operator insertion time $\tau$ is restricted to the interval $\tau \in [a N_{\rm skip},, t - a N_{\rm skip}]$, where the skip factor $N_{\rm skip}$ is chosen as
\begin{align}
    N_{\rm skip}={\rm int}(\sqrt{8\tf}/a)+N_{\rm side},
\end{align}
where $\tf$ denotes the gradient flow time and $N_{\rm side}\in[6,12]$ depends on the lattice spacing and fit strategy. This choice reflects the fact that gradient flow smears the operator over a radius $\sqrt{8\tf}$, which limits the temporal region where the flowed operator can reliably probe hadronic structure. A larger $\tf$ reduces the usable fit range and simultaneously worsens the condition number of the covariance matrix. Therefore, when $\tf$ is small, more $(t,\tau)$ data points are retained to improve the fit signal, while for large $\tf$, more points are discarded near the boundaries to ensure numerical stability.

To further reduce correlations among nearby $(t,\tau)$ values, we introduce a stride in both $t$ and $\tau$ selection. For example, a stride of 2 uses separation $t/a=28,30,32$ for quark and $t/a=26,28,30$ for gluon, and insertion times $\tau/a = N_{\rm skip}, N_{\rm skip}+2, N_{\rm skip}+4, ...$ We perform two variants of the two-state fit, using stride-2 and stride-4 samples, denoted as $\mathrm{Fit_{str2}}$ and $\mathrm{Fit_{str4}}$, respectively.

In the spectral decomposition of Eq.~\eqref{eq:3ptfit}, we truncate the sum to the lowest two states ($m,n = 0,1$) and neglect the $m=n=1$ term, which is exponentially suppressed for large $t$. The resulting two-state ansatz for the ratio takes the form
\begin{equation}\label{eq:2s_ansatz_recap}
\begin{aligned}
    &R^{\rho\sigma}_{X,\Gamma}(t, \tau)= \frac{1}{2M}\frac{\langle 0|\tilO_{i,\mn}|0\rangle - \frac{{\rm Re}(Z_1^*Z_0\langle 0|\tilO_{i,\mn}|1\rangle)}{|Z_0|^2} \left[ e^{-\Delta E \tau} + e^{-\Delta E (t - \tau)} \right]}{1 + (|Z_1|^2/|Z_0|^2) e^{-\Delta E t}}.
\end{aligned}
\end{equation}
Since the external interpolating operators are the same in both two-point and three-point functions, we first analyze two-point functions to determine $Z_n$ and $E_n$ of each bootstrap sample, and then use them as fixed inputs in the analysis of the three-point function ratios.

We also apply the summation fit by adding the ratios over $\tau$:
\begin{equation}
R^{\rho\sigma, \text{sum}}_{X,\Gamma}(t) = \sum_{\tau = a N_{\rm skip}}^{t - a N_{\rm skip}} R^{\rho\sigma}_{X,\Gamma}(t, \tau).
\end{equation}
In large-$t$, this approximates
\begin{equation}
R^{\rho\sigma, \text{sum}}_{X,\Gamma}(t) = \frac{\langle 0|\tilO_{i,\mn}|0\rangle}{2M} \cdot t + B,
\end{equation}
where the slope directly yields the ground-state matrix element. We denote this method by $\rm SUM$.

We perform 600 bootstrap samples for each parameter set ($\Gamma$, $a$, $\tf$, $\tilO_{i,\mn}$), and apply all three fitting strategies ($\rm Fit_{str2}$, $\rm Fit_{str4}$, and $\rm SUM$), resulting in 1800 fits per observable. All results are collected into an extended bootstrap ensemble to capture both statistical and fit-related systematic uncertainties.

%
%
\section{Extraction of EMT matrix elements for the $\eta_c$ meson}

\begin{figure*}
    \centerline{
    \includegraphics[width=0.32\linewidth]{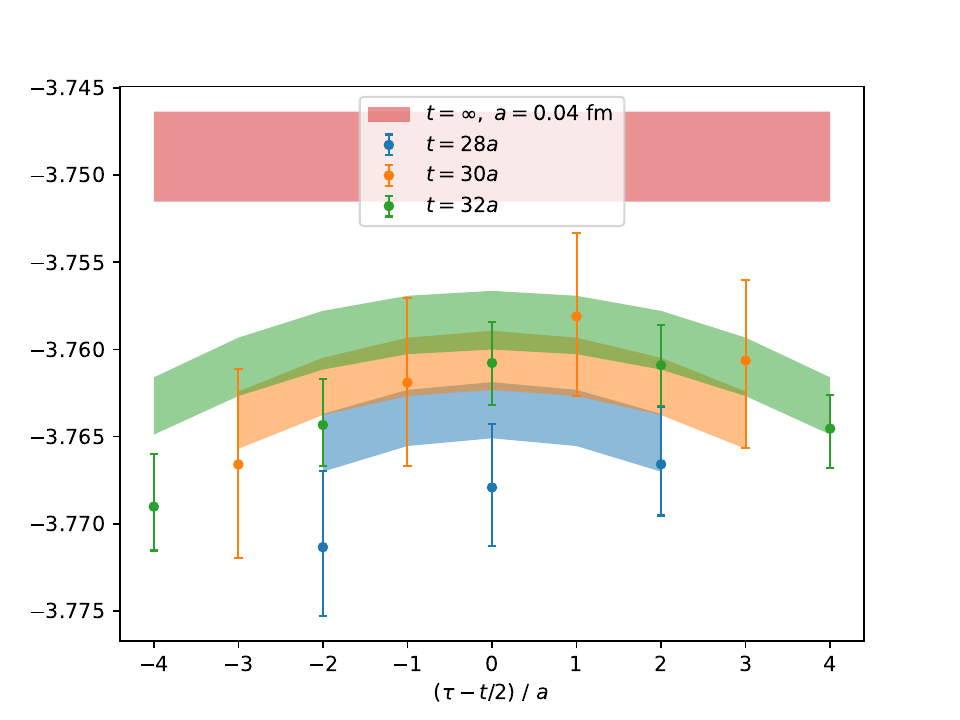}
    \includegraphics[width=0.32\linewidth]{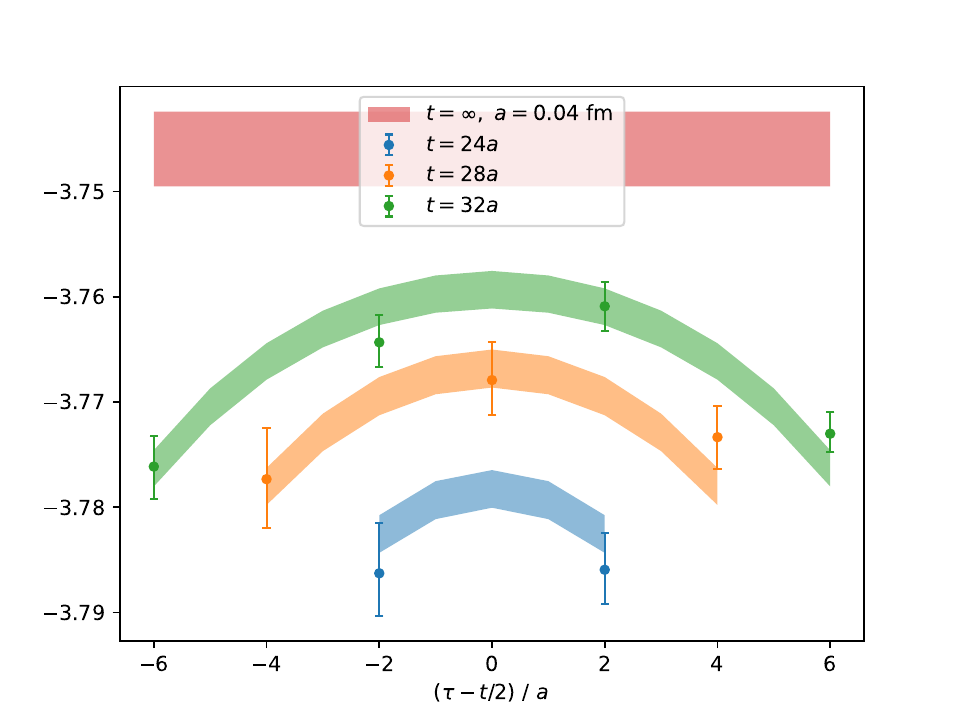}
    \includegraphics[width=0.32\linewidth]{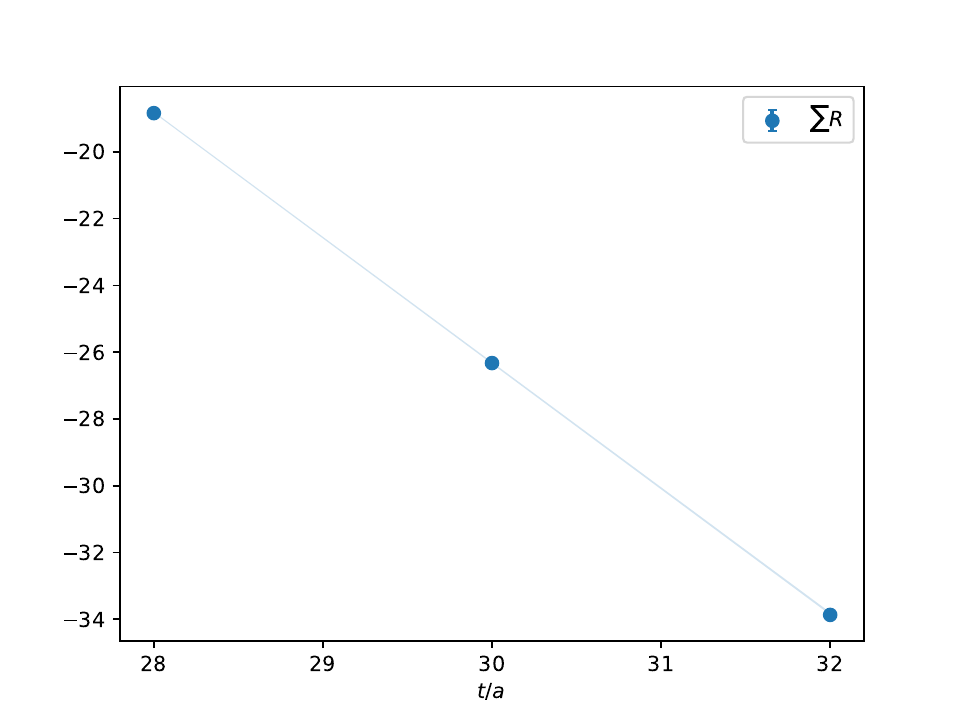}
    }
    \centerline{
    \includegraphics[width=0.32\linewidth]{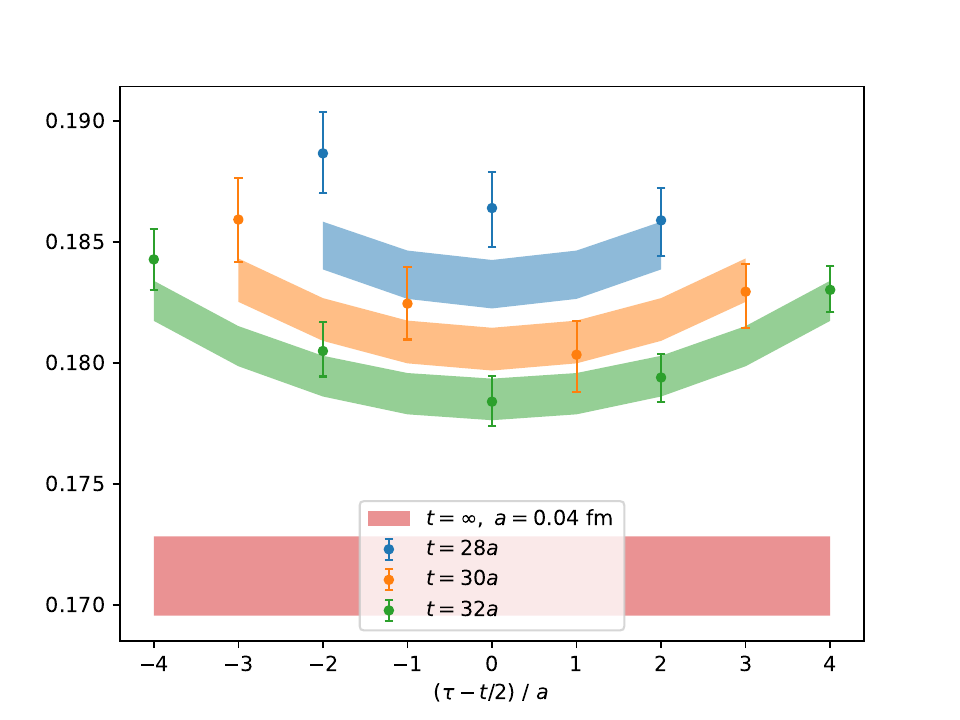}
    \includegraphics[width=0.32\linewidth]{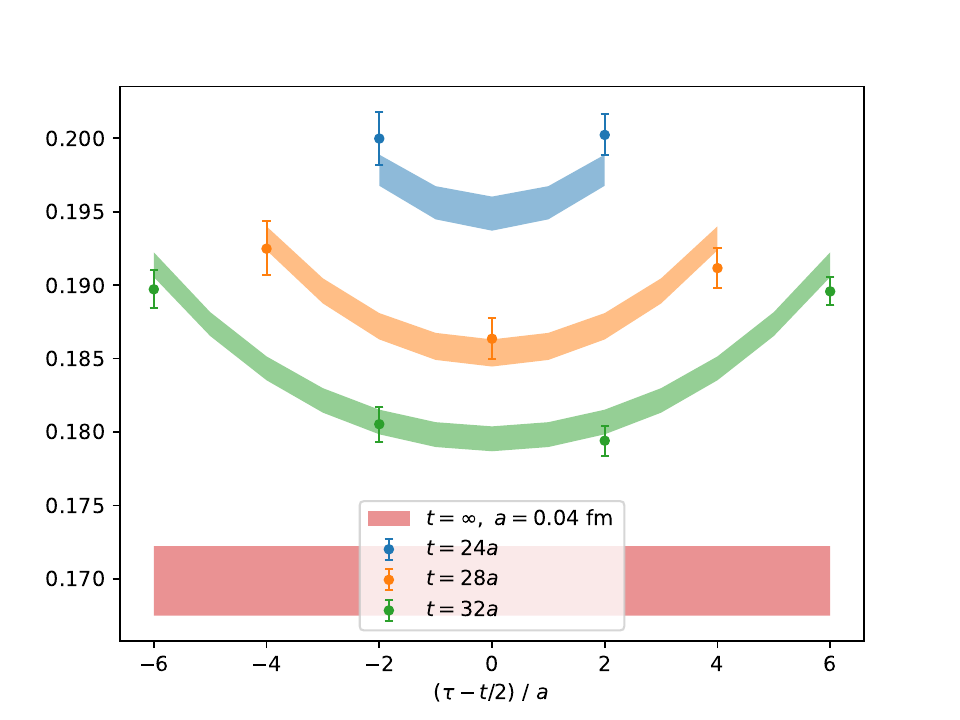}
    \includegraphics[width=0.32\linewidth]{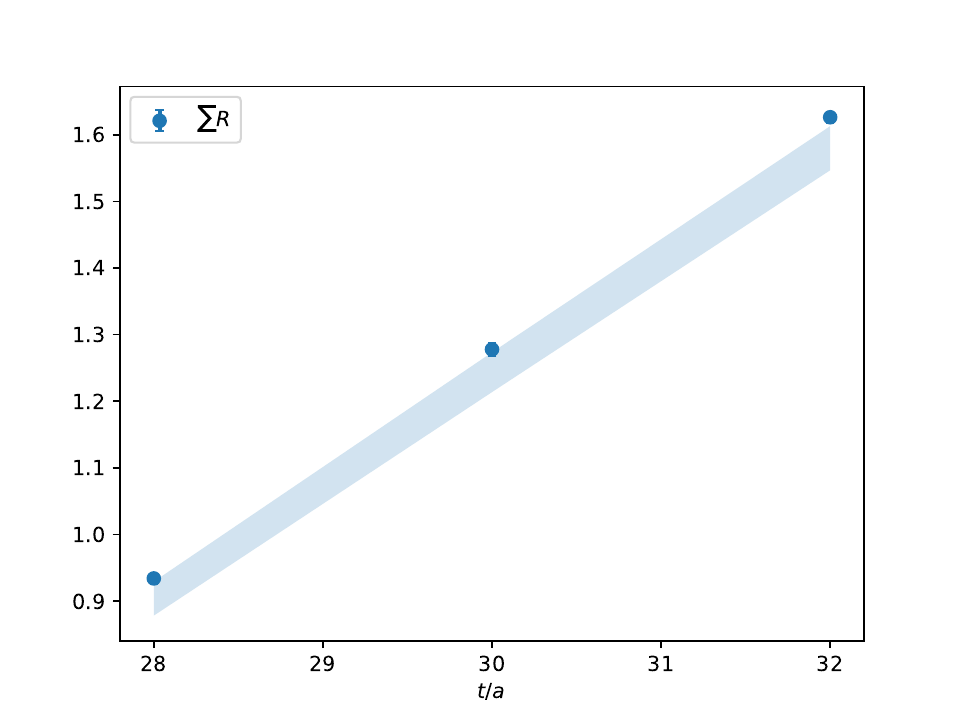}
    }
    \caption{Left and middle panels: ratios of quark three-point to two-point correlation functions for the $\eta_c$ meson of $a=0.04$ fm lattice at flow time $\tf=5\ef$, plotted as functions of $t - \tau/2$. The curved bands represent reconstructions from the two-state fits $\rm Fit_{str2}$ and $\rm Fit_{str4}$, while the horizontal bands indicate the extracted ground-state matrix elements. Right panels: Summed ratios as functions of the source-sink separation $t$, with curved bands representing results from the summation fit SUM. The upper and lower rows show results for temporal components $R^{44}_{q,\gamma_5}$ and spatial components $\sum_{k=1}^3R^{kk}_{q,\gamma_5}/3$ respectively.}
    \label{fig:etacRatioq}
\end{figure*}

\begin{figure*}
    \centerline{
    \includegraphics[width=0.32\linewidth]{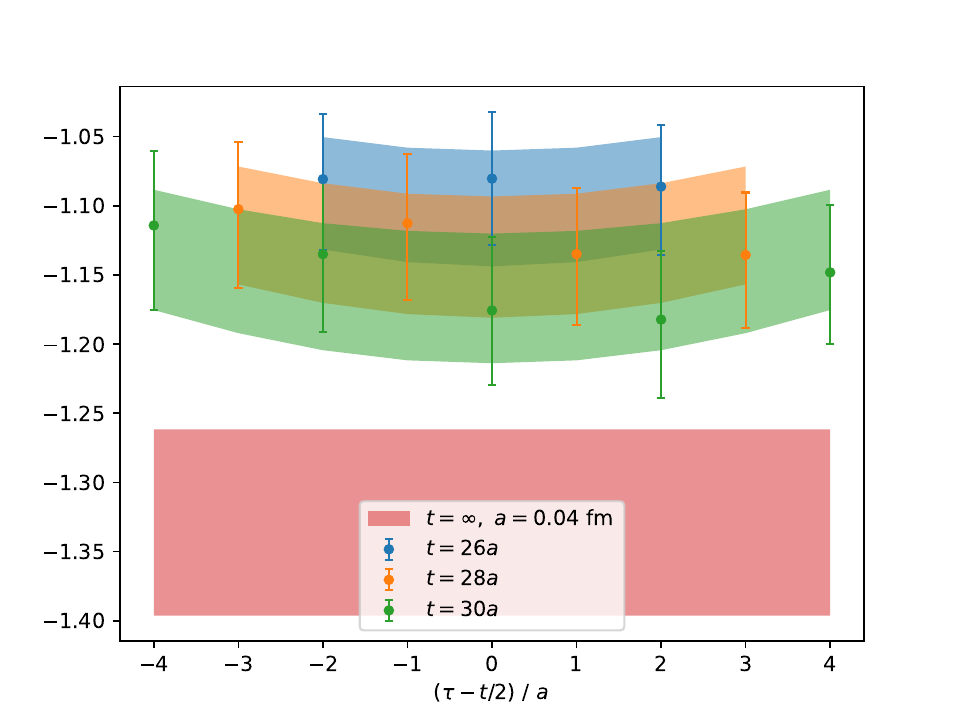}
    \includegraphics[width=0.32\linewidth]{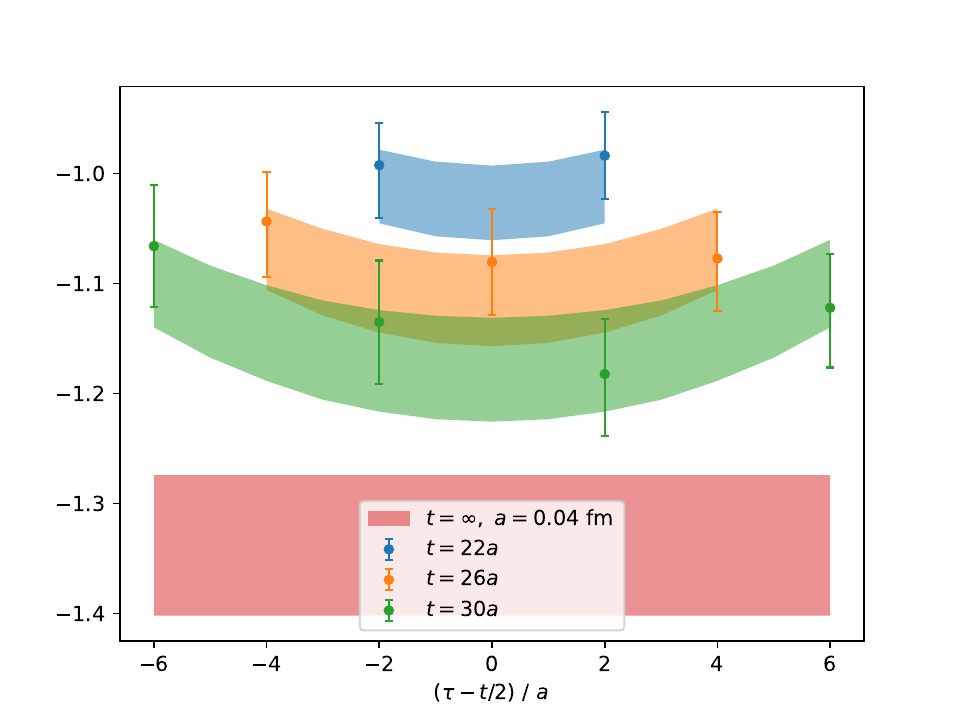}
    \includegraphics[width=0.32\linewidth]{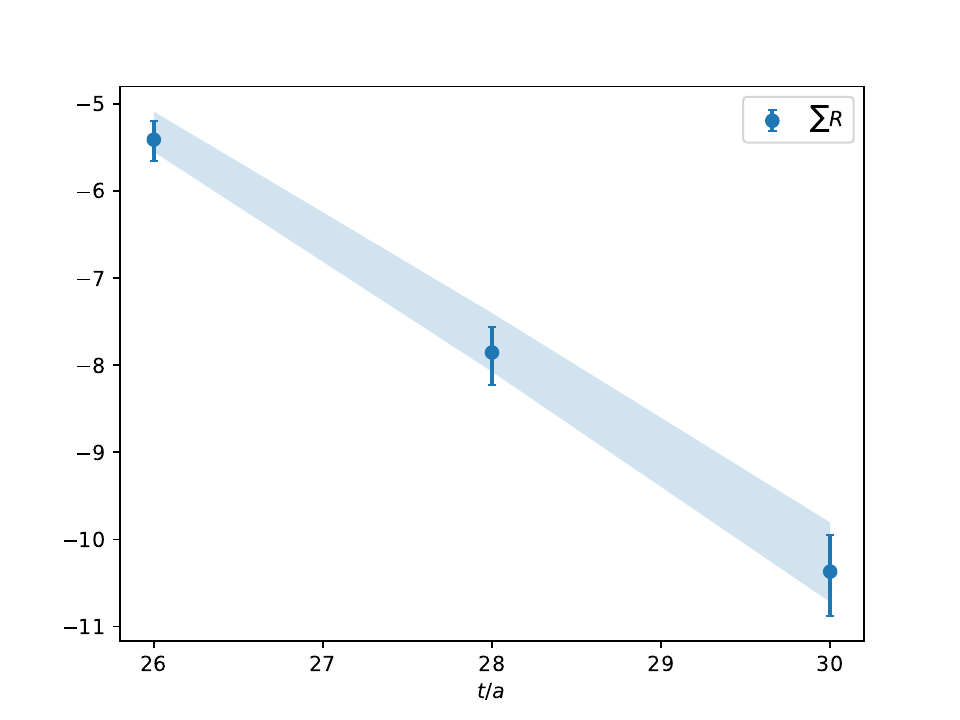}
    }
    \centerline{
    \includegraphics[width=0.32\linewidth]{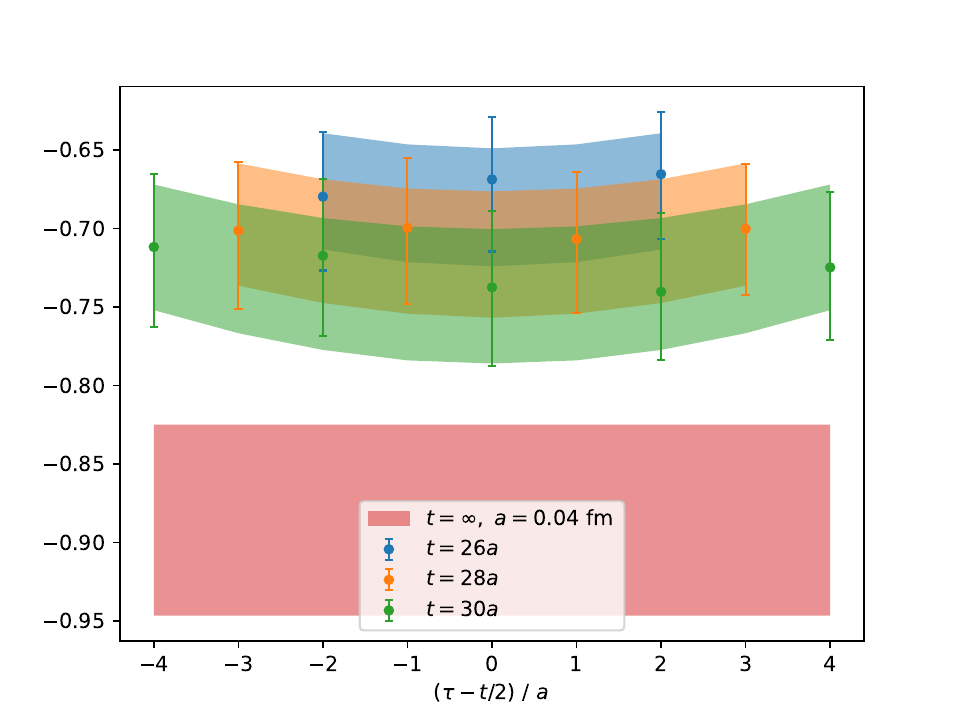}
    \includegraphics[width=0.32\linewidth]{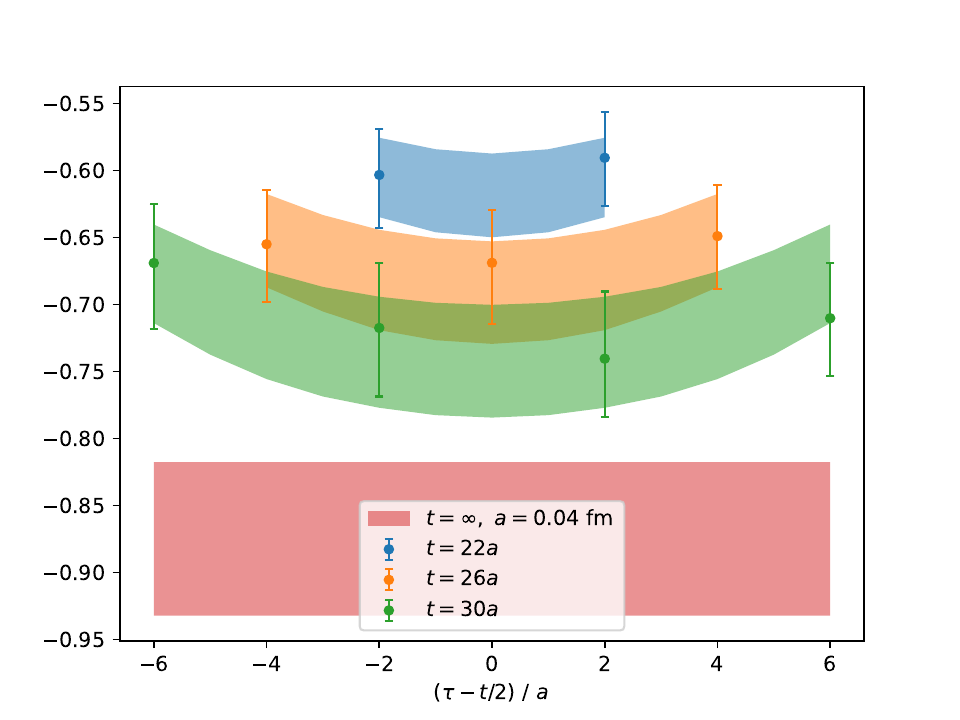}
    \includegraphics[width=0.32\linewidth]{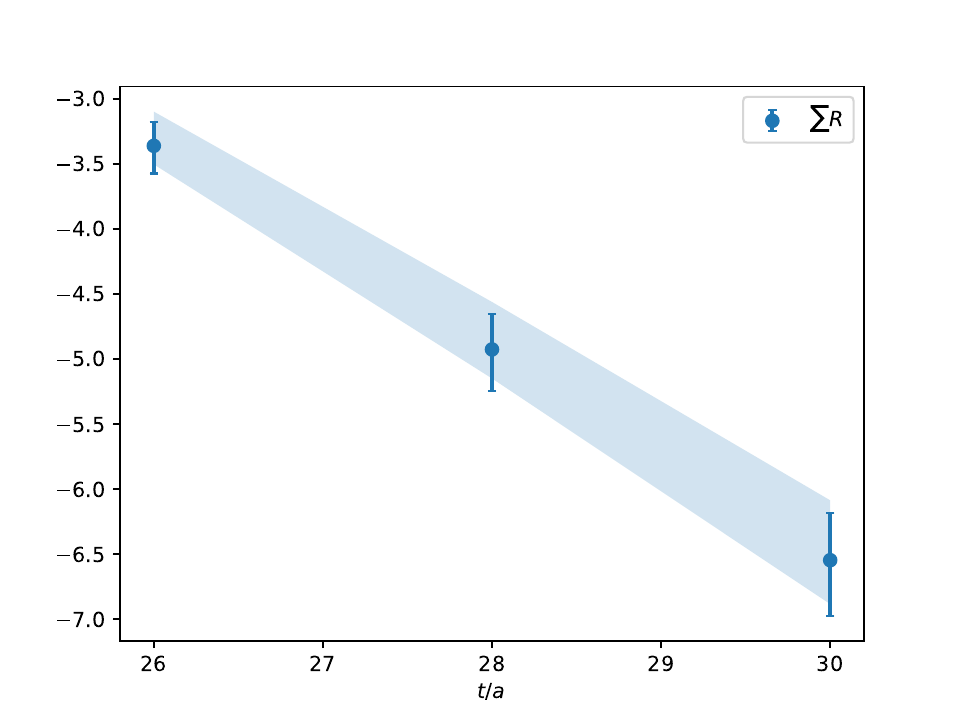}
    }
    \caption{Similar figures as \autoref{fig:etacRatioq} but for gluonic matrix elements of $\eta_c$: $R^{44}_{g,\gamma_5}$ (upper row) and $\sum_{k=1}^3R^{kk}_{g,\gamma_5}/3$ (lower row).}
    \label{fig:etacRatiog}
\end{figure*}

\begin{figure*}
    \includegraphics[width=0.4\linewidth]{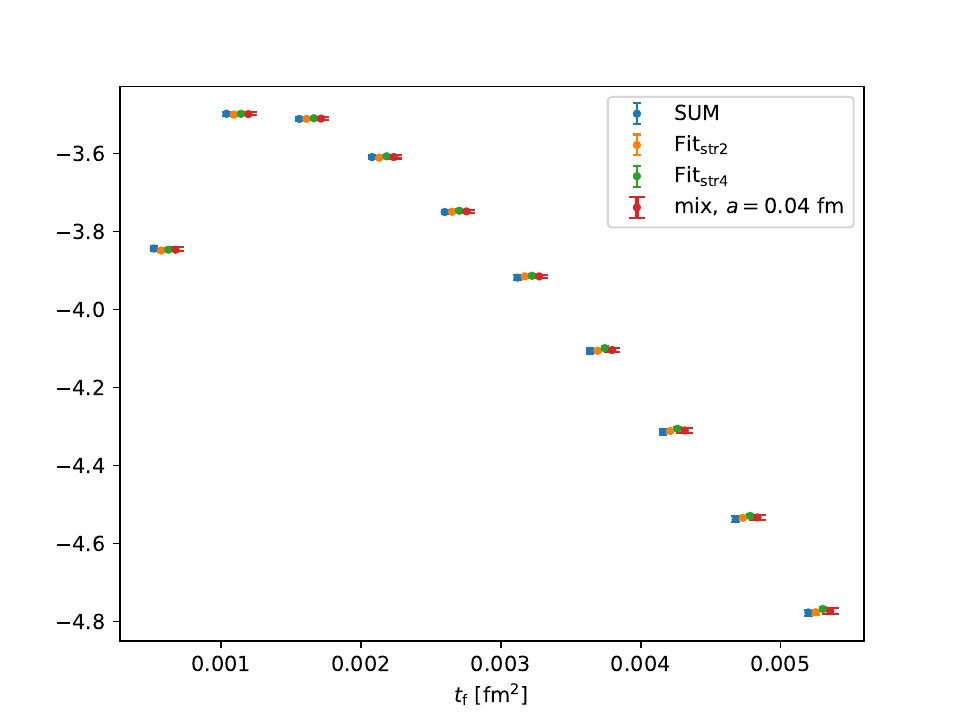}
    \includegraphics[width=0.4\linewidth]{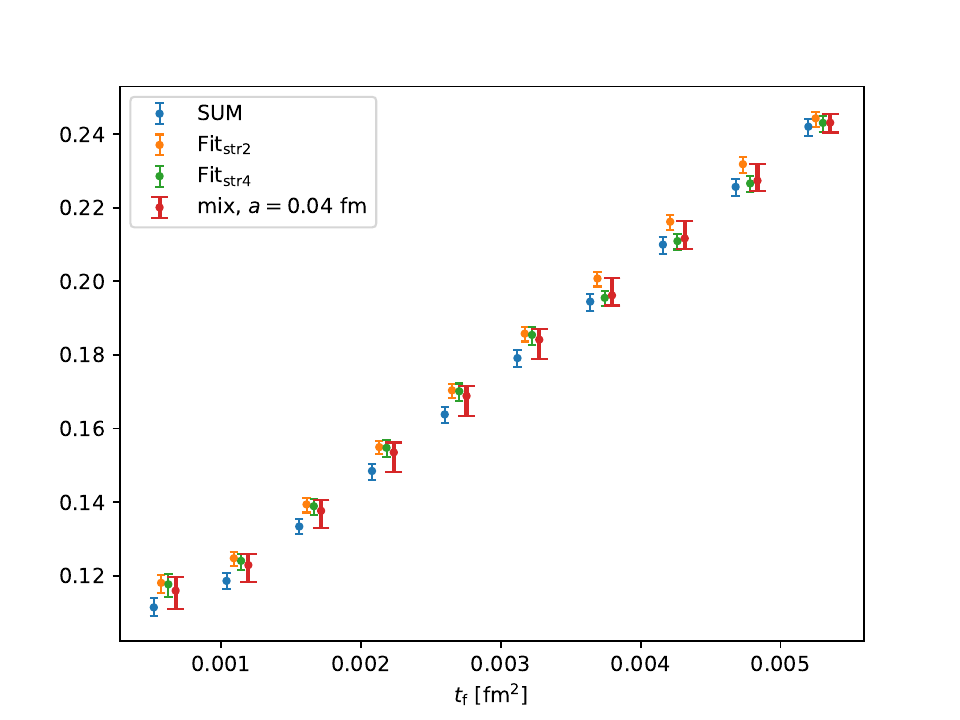}
    \includegraphics[width=0.4\linewidth]{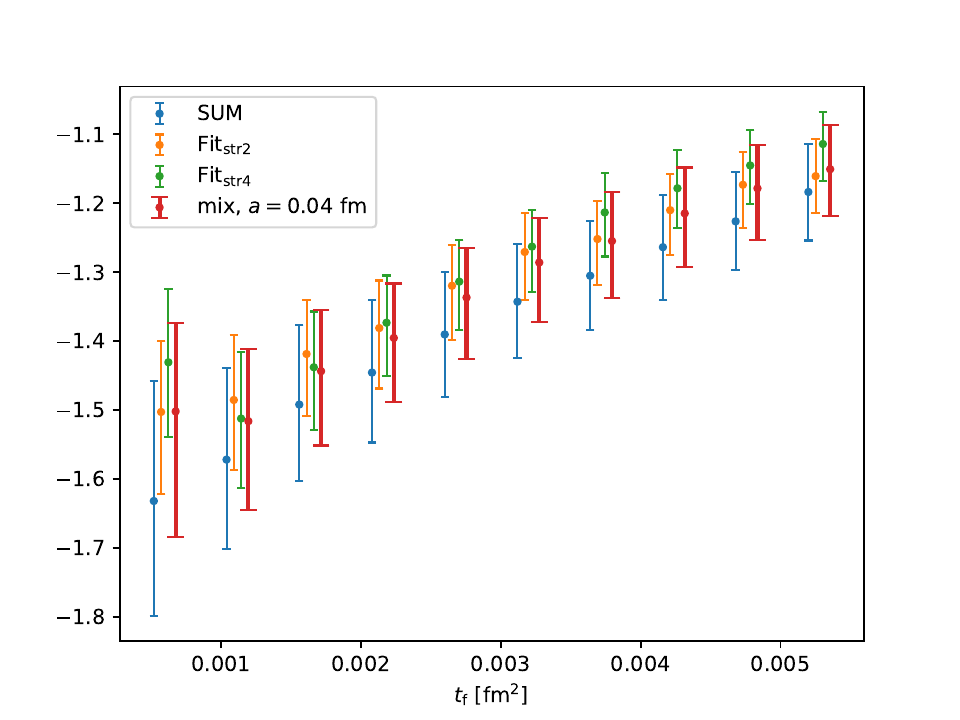}
    \includegraphics[width=0.4\linewidth]{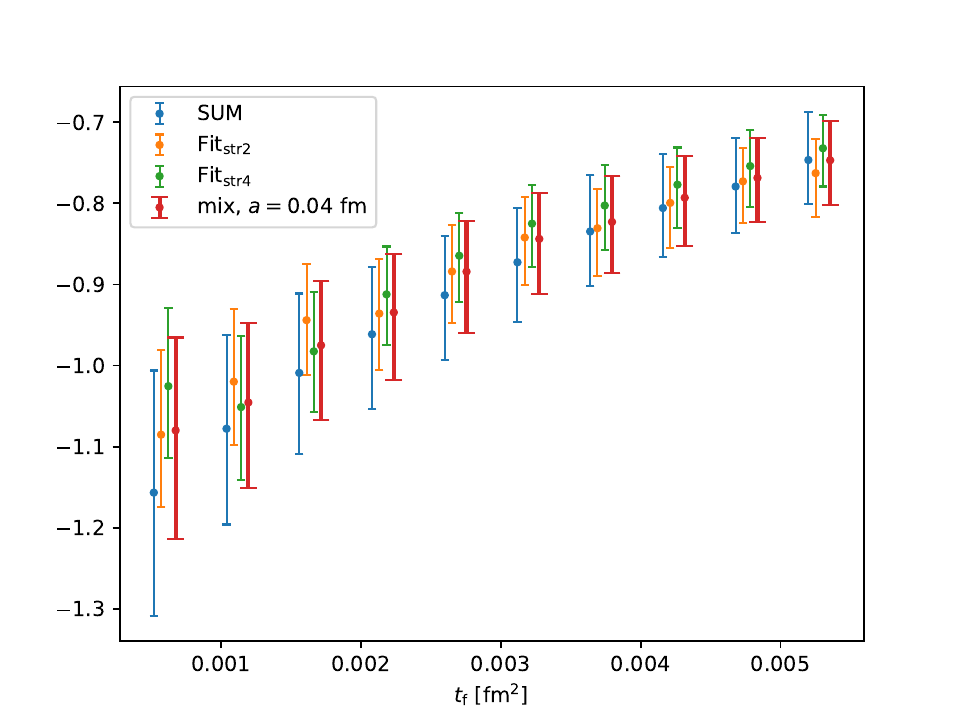}
    \caption{Extracted EMT matrix elements for $\eta_c$ from the summation fit SUM and two two-state fits $\rm Fit_{str2}$ and $\rm Fit_{str4}$ are shown as a function of flow time $\tf$. The averaged results are also shown (denoted by mix). The upper panels show quark matrix elements $\frac{1}{2M}\langle \eta_c|\tilO_{3,44}|\eta_c\rangle$ (left) and $\frac{1}{2M}\sum_{k=1}^3\langle \eta_c|\tilO_{3,kk}/3|\eta_c\rangle$ (right). The low panels show gluon matrix elements $\frac{1}{2M}\langle \eta_c|\tilO_{1,44}|\eta_c\rangle$ (left) and $\frac{1}{2M}\sum_{k=1}^3\langle \eta_c|\tilO_{1,kk}/3|\eta_c\rangle$ (right).}
    \label{fig:etacME}
\end{figure*}

For the pseudoscalar $\eta_c$, we choose $\Gamma=\gamma_5$ in the meson interpolator for the three-point functions defined in \autoref{eq:3ptq} and \autoref{eq:3ptg}. In the asymptotic limit $t\gg\tau\gg0$, the temporal and spatial components of the ratios approach
\begin{equation}
\begin{aligned}
\frac{1}{2M}\langle \eta_c|\tilO_{3,44}|\eta_c\rangle &= \lim_{t \gg \tau \gg 0} R^{44}_{q,\gamma_5} , \\
\frac{1}{3}\sum_{k=1}^3\frac{1}{2M}\langle \eta_c|\tilO_{3,kk}|\eta_c\rangle &= \lim_{t \gg \tau \gg 0} \frac{1}{3}\sum_{k=1}^3 R^{kk}_{q,\gamma_5},
\end{aligned}
\end{equation}
for the quark sector, and
\begin{equation}
\begin{aligned}
\frac{1}{2M}\langle \eta_c|\tilO_{1,44}|\eta_c\rangle &= \lim_{t \gg \tau \gg 0} R^{44}_{g,\gamma_5}, \\
\frac{1}{3}\sum_{k=1}^3\frac{1}{2M}\langle \eta_c|\tilO_{1,kk}|\eta_c\rangle &= \lim_{t \gg \tau \gg 0} \frac{1}{3}\sum_{k=1}^3 R^{kk}_{g,\gamma_5},
\end{aligned}
\end{equation}
for the gluon sector, where $k=1,2,3$ label spatial directions. In practice, we average the spatial components in the rest frame to improve statistical precision.

\autoref{fig:etacRatioq} and \autoref{fig:etacRatiog} show the extraction of the $\eta_c$ quark and gluon EMT matrix elements on the $a=0.04$~fm ensemble at $\tf=5\ef$ respectively, for the temporal components $R^{44}_{X,\gamma_5}$ (upper rows) and the averaged spatial components $\sum_{k=1}^3 R^{kk}_{X,\gamma_5}/3$ (lower rows), with $X=q,g$.
In the left and middle panels, the three-point to two-point ratios are plotted as functions of $t-\tau/2$. The reconstructed curves from the two-state fits, $\rm Fit_{str2}$ and $\rm Fit_{str4}$, track the data well over the full range of source--sink separations, indicating good control of excited-state contamination; the corresponding ground-state matrix elements are shown as horizontal bands.
The right panels display the summed ratios versus $t$, where the $\rm SUM$ bands exhibit a clear linear behavior, consistent with the expected suppression of excited-state effects.

The extracted ground-state matrix elements from $\rm Fit_{str2}$, $\rm Fit_{str4}$, and $\rm SUM$ are summarized in \autoref{fig:etacME} as functions of the flow time $\tf$ for the same $a=0.04$~fm ensemble (upper panels for quarks and lower panels for gluons). The three strategies yield mutually consistent results within statistical uncertainties across the $\tf$ range, demonstrating the stability of the extraction with respect to both the excited-state treatment and the fit method. We also show the averaged results obtained by combining all $3\times600=1800$ bootstrap samples (labeled ``mix''), which are used in the subsequent analysis.

%
%
\section{Extraction of EMT matrix elements for the $J/\psi$ meson}

\begin{figure*}
    \centerline{
    \includegraphics[width=0.32\linewidth]{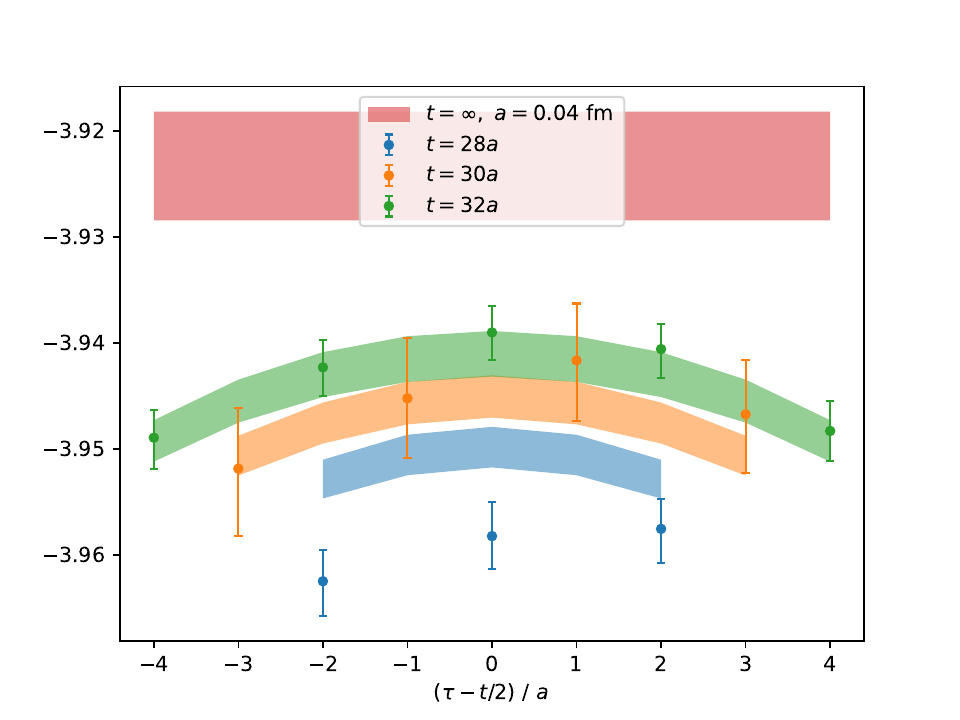}
    \includegraphics[width=0.32\linewidth]{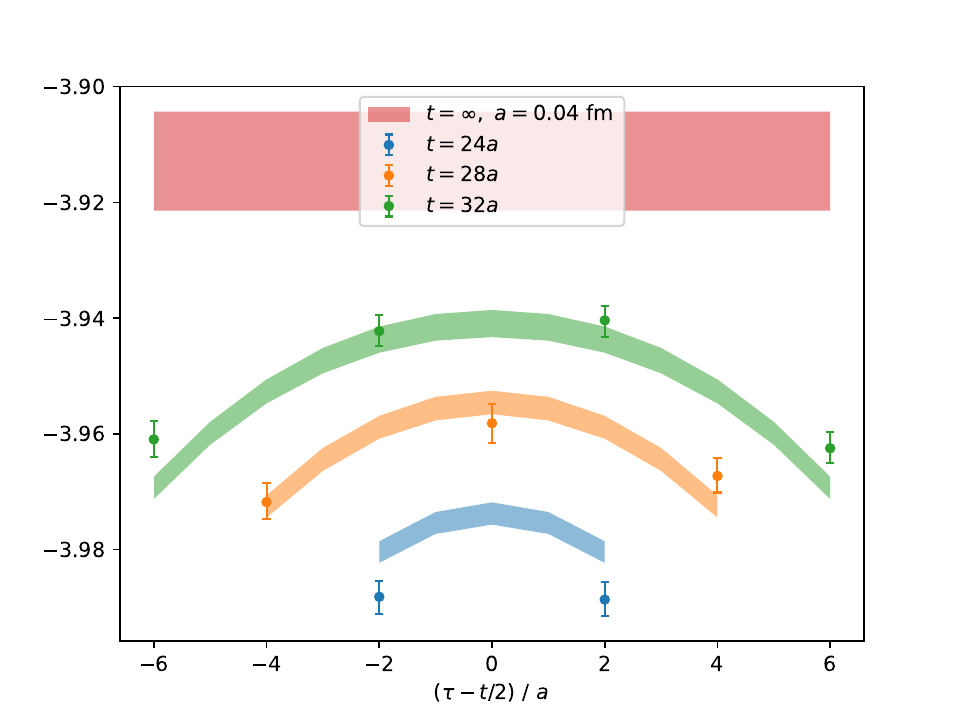}
    \includegraphics[width=0.32\linewidth]{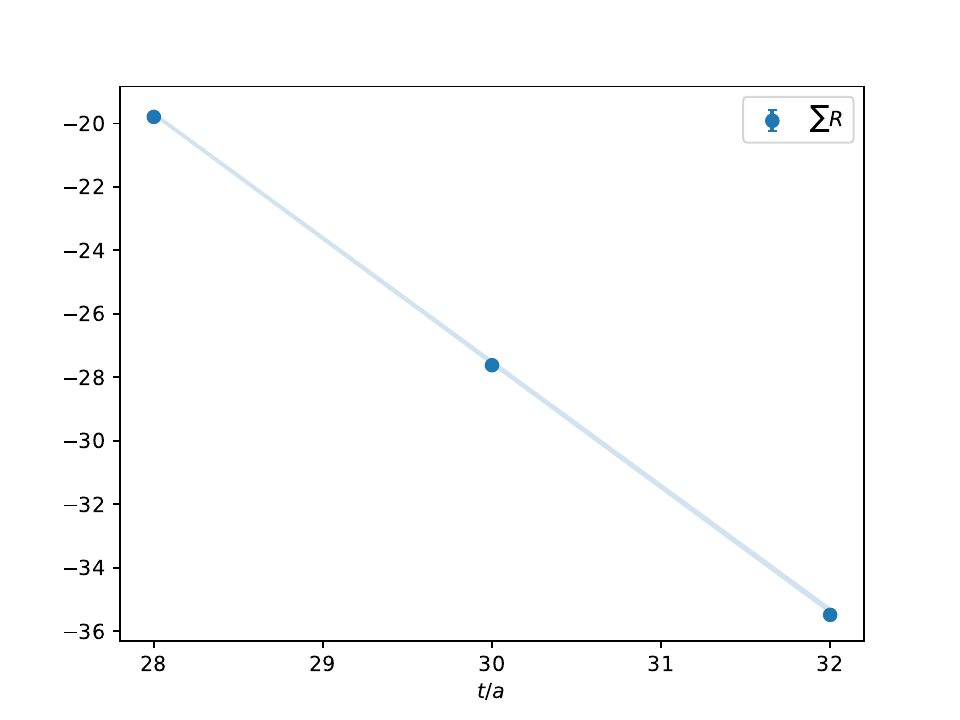}
    }
    \centerline{
    \includegraphics[width=0.32\linewidth]{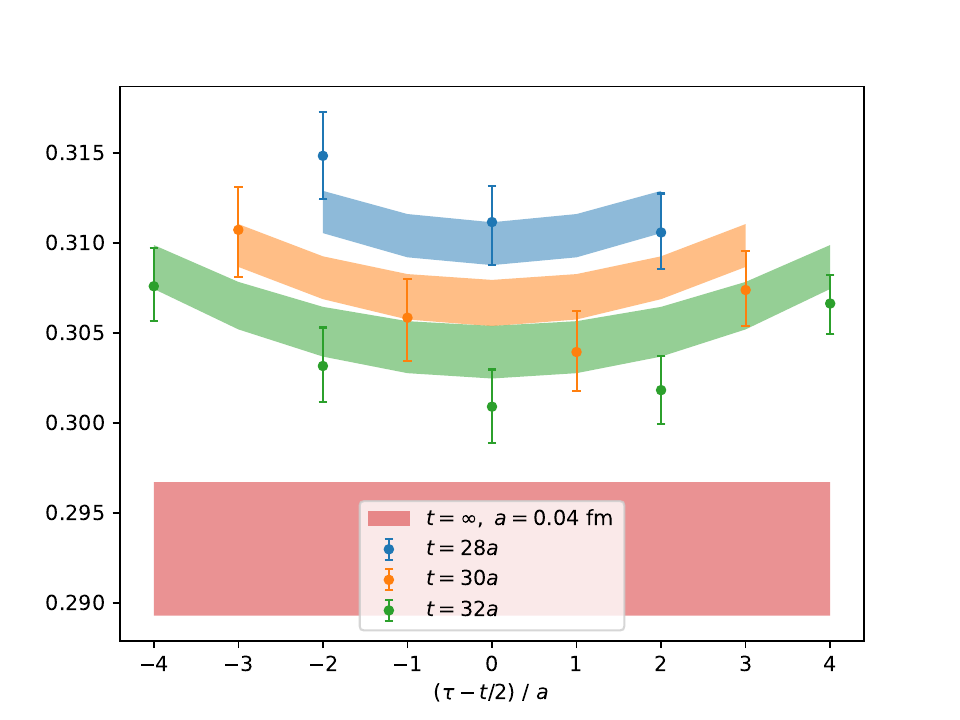}
    \includegraphics[width=0.32\linewidth]{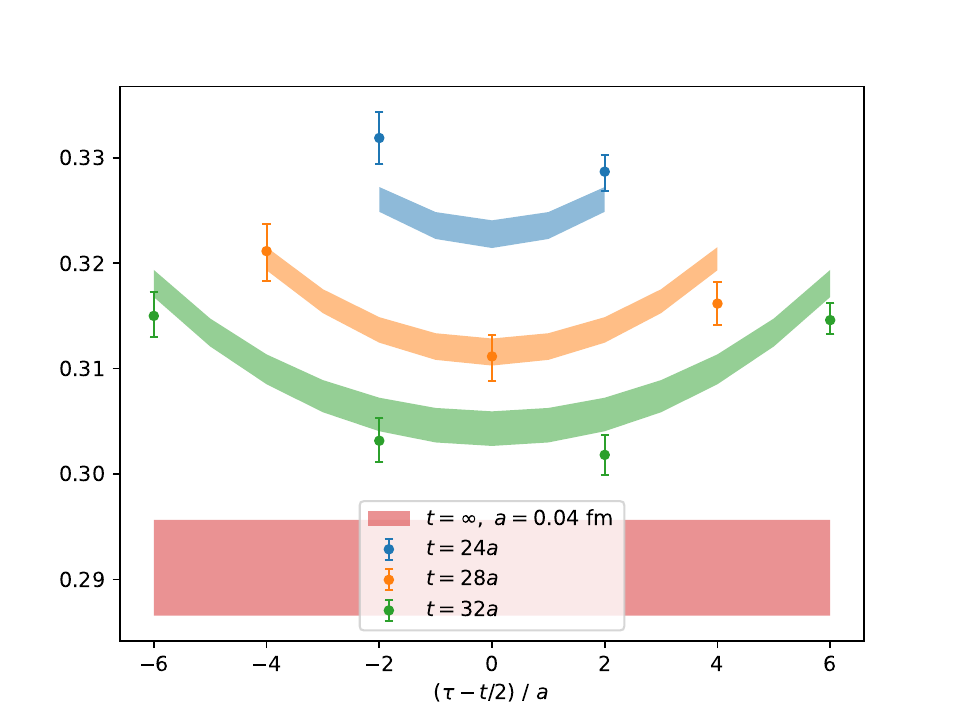}
    \includegraphics[width=0.32\linewidth]{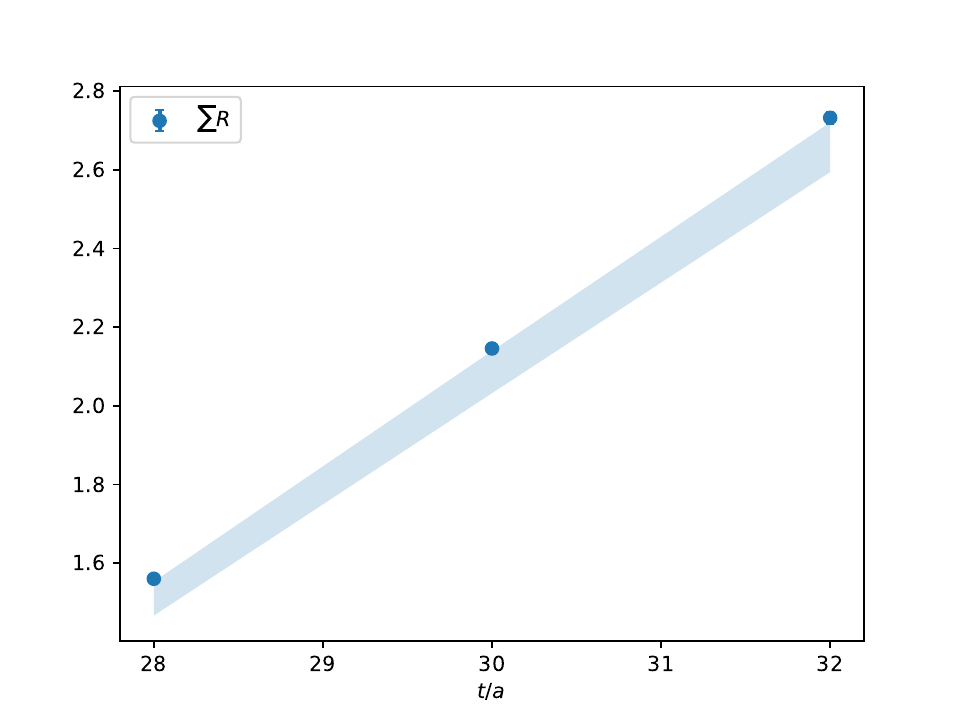}
    }
    \centerline{
    \includegraphics[width=0.32\linewidth]{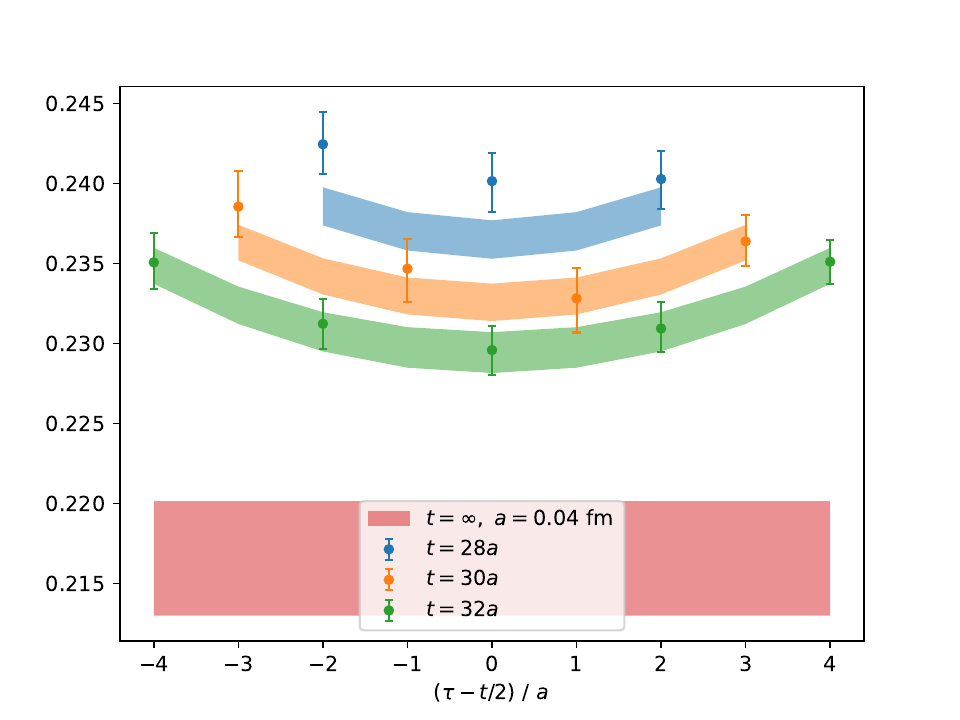}
    \includegraphics[width=0.32\linewidth]{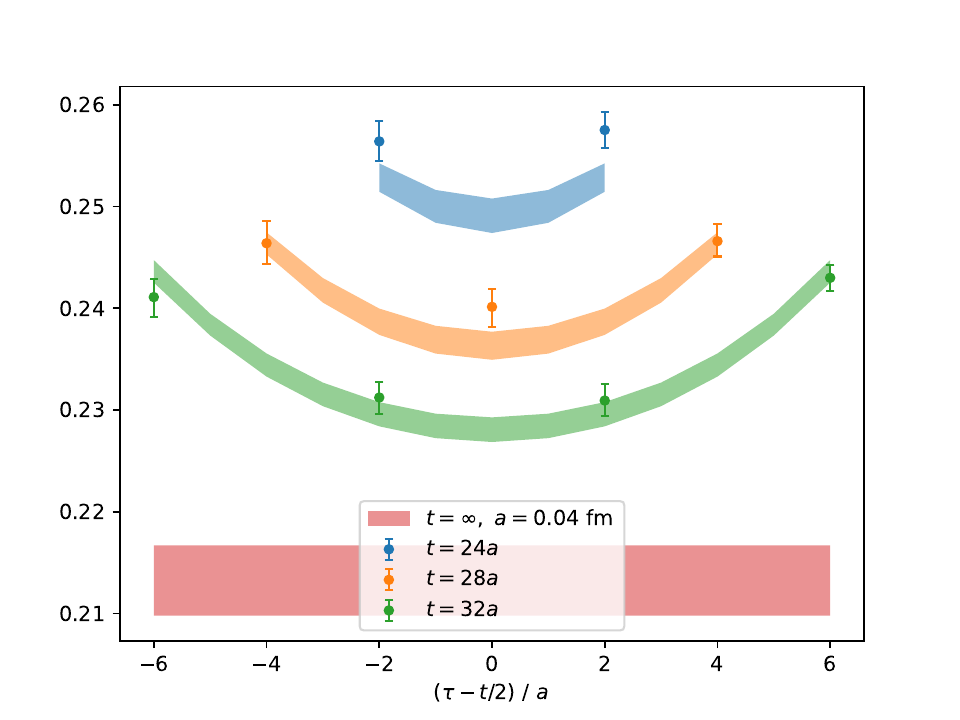}
    \includegraphics[width=0.32\linewidth]{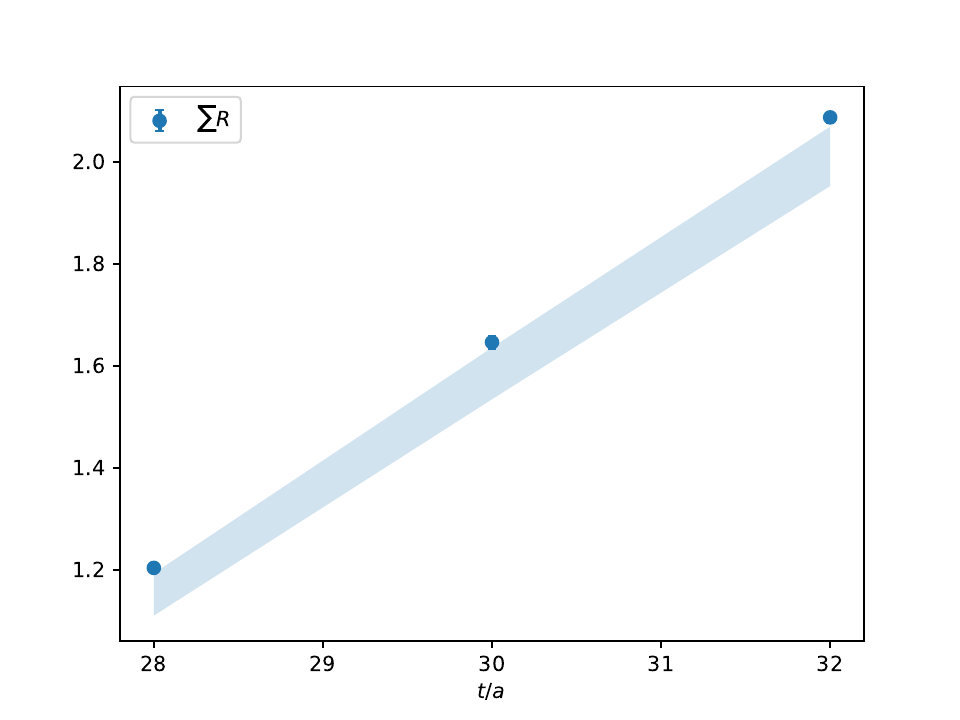}
    }
    \caption{Left and middle panels: Quark three-point to two-point function ratios for the $J/\psi$ meson as functions of $t - \tau/2$, for the temporal ($R^{44}_{q,\gamma_1}$, first row), longitudinal ($R^{11}_{q,\gamma_1}$, second row), and transverse ($(R^{22}_{q,\gamma_1} + R^{33}_{q,\gamma_1})/2$, third row) components. Curved bands correspond to the two-state fits $\rm Fit_{str2}$ and $\rm Fit_{str4}$; horizontal bands indicate the extracted matrix elements. Right panels: Summed ratios plotted as functions of $t$, with straight bands from the summation fit $\rm SUM$.}
    \label{fig:jpsiRatioq}
\end{figure*}

\begin{figure*}
    \centerline{
    \includegraphics[width=0.32\linewidth]{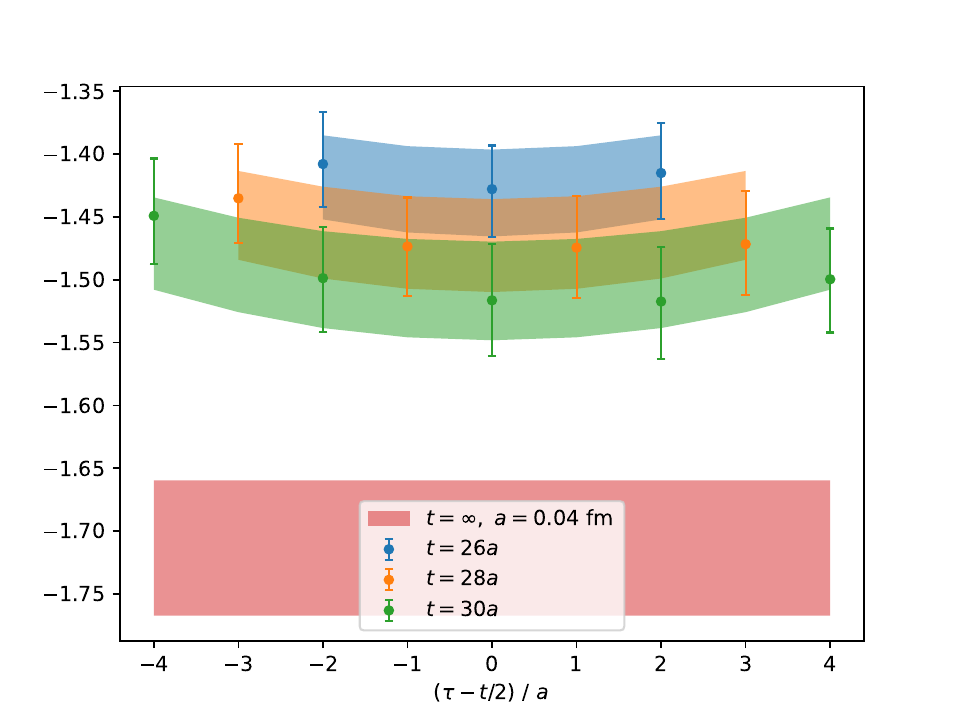}
    \includegraphics[width=0.32\linewidth]{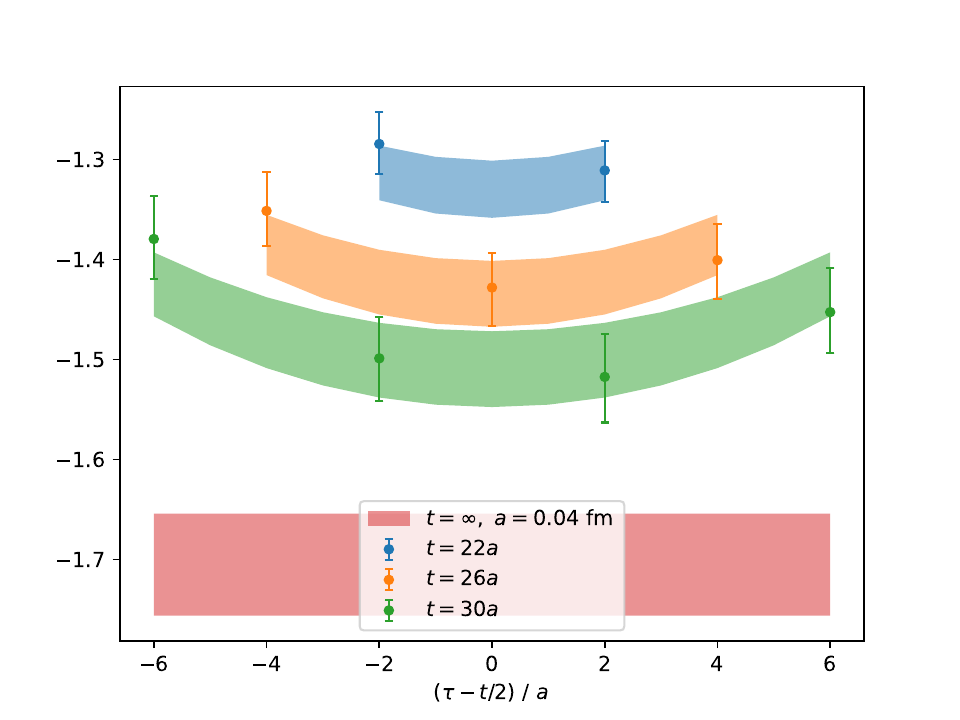}
    \includegraphics[width=0.32\linewidth]{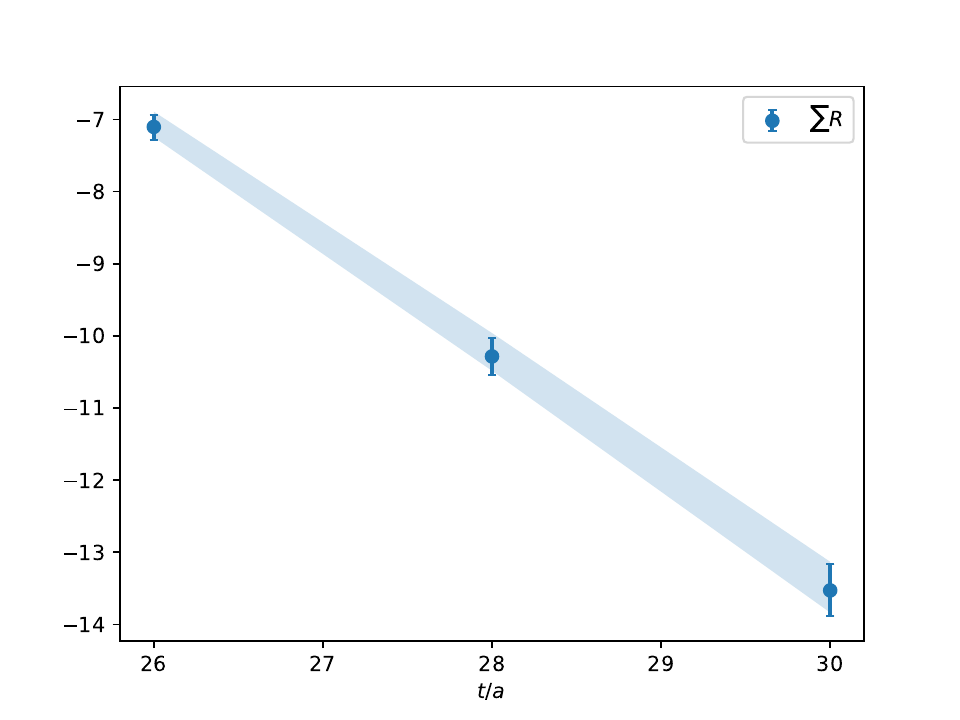}
    }
    \centerline{
    \includegraphics[width=0.32\linewidth]{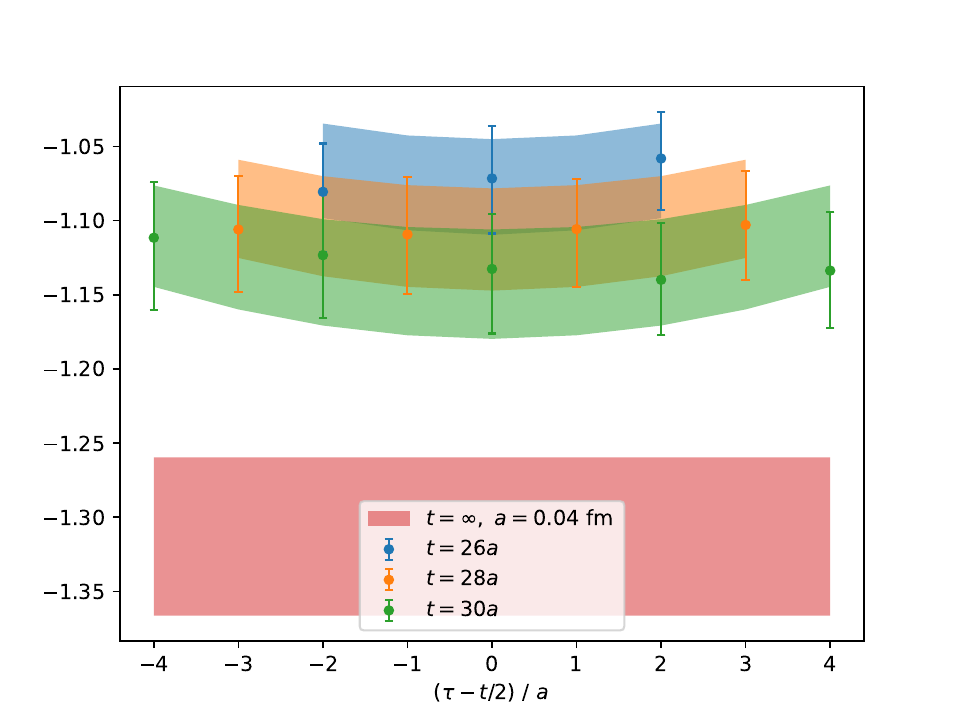}
    \includegraphics[width=0.32\linewidth]{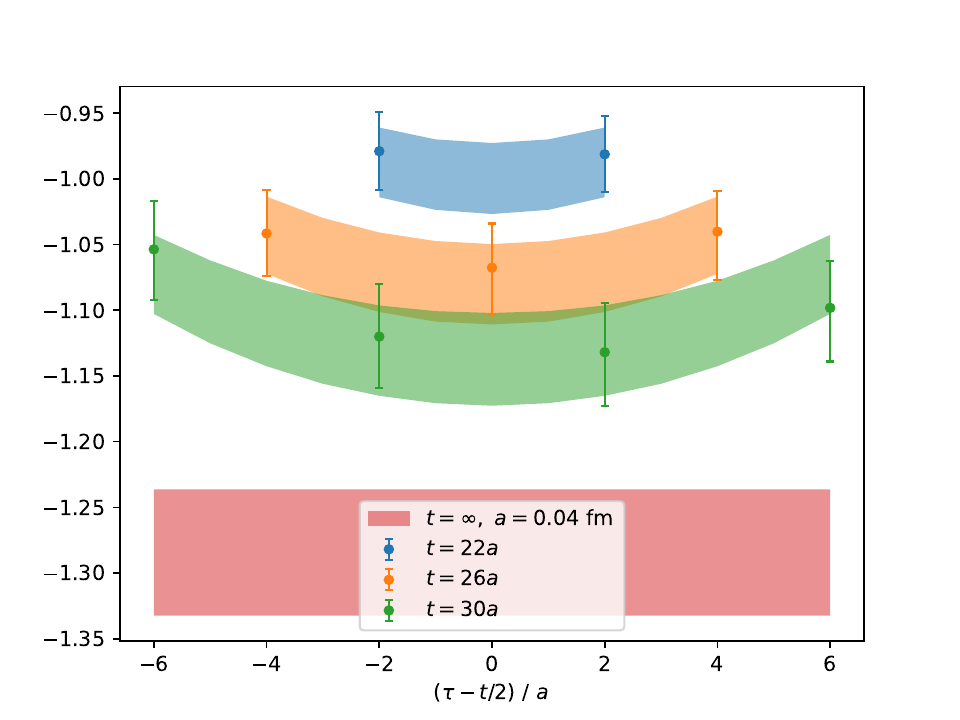}
    \includegraphics[width=0.32\linewidth]{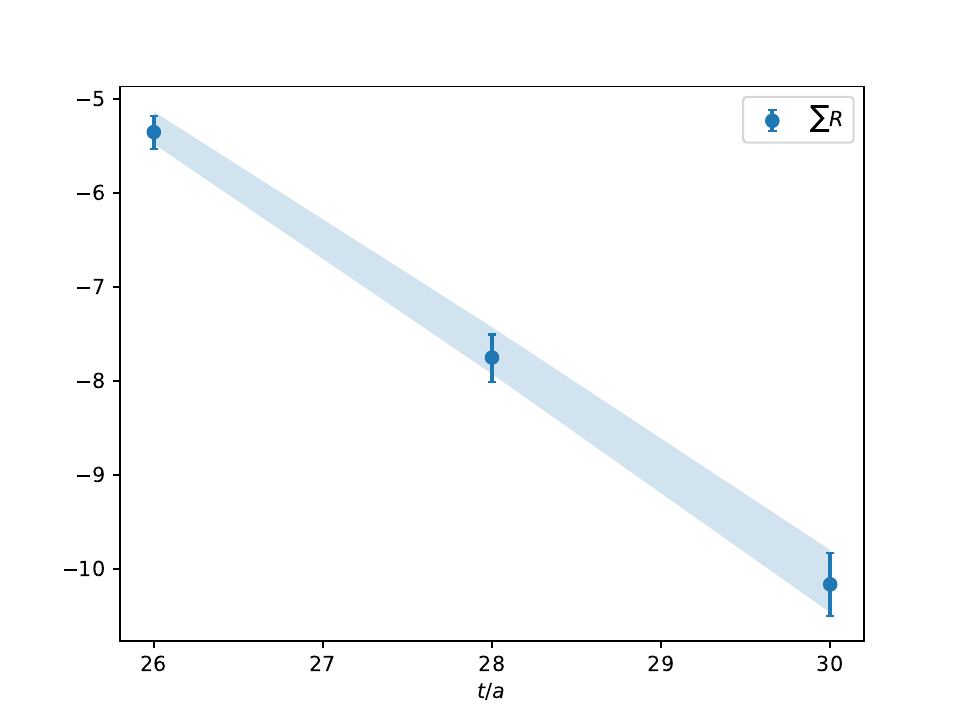}
    }
    \centerline{
    \includegraphics[width=0.32\linewidth]{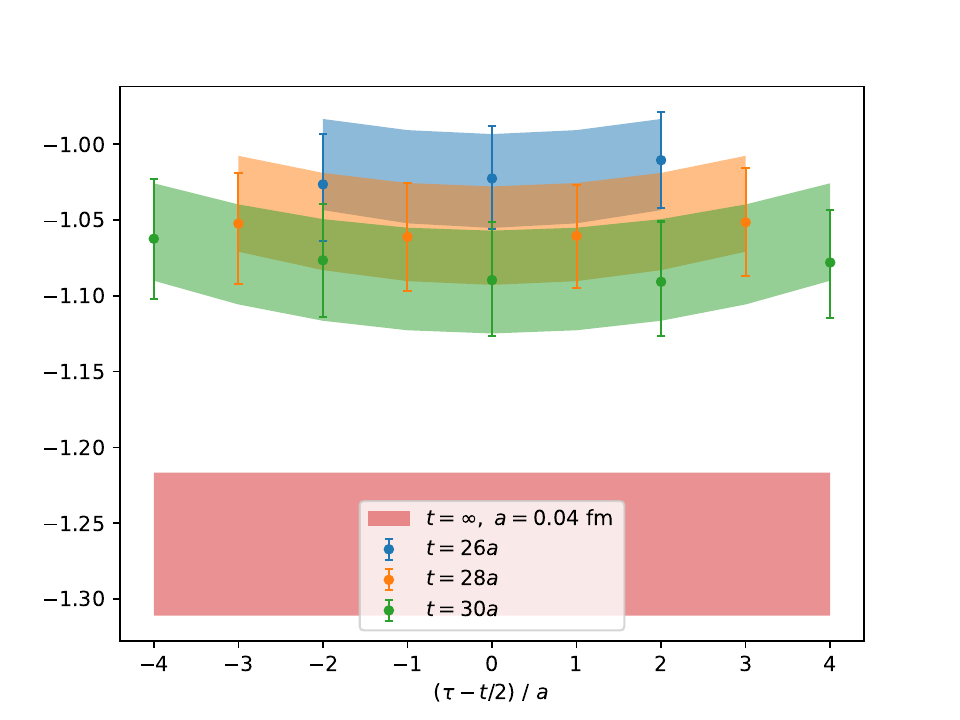}
    \includegraphics[width=0.32\linewidth]{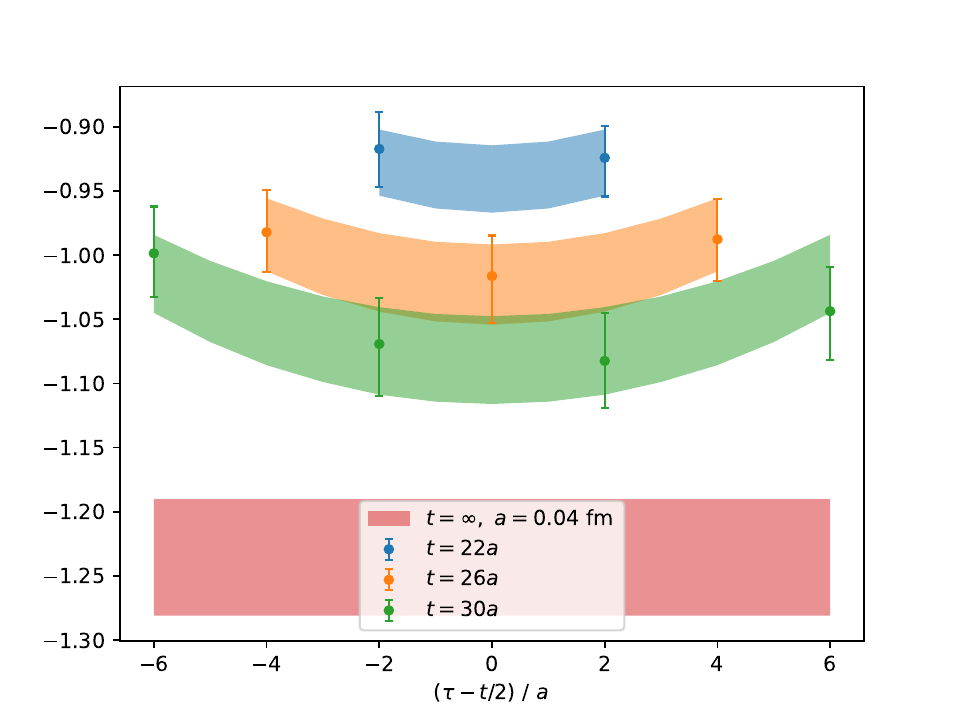}
    \includegraphics[width=0.32\linewidth]{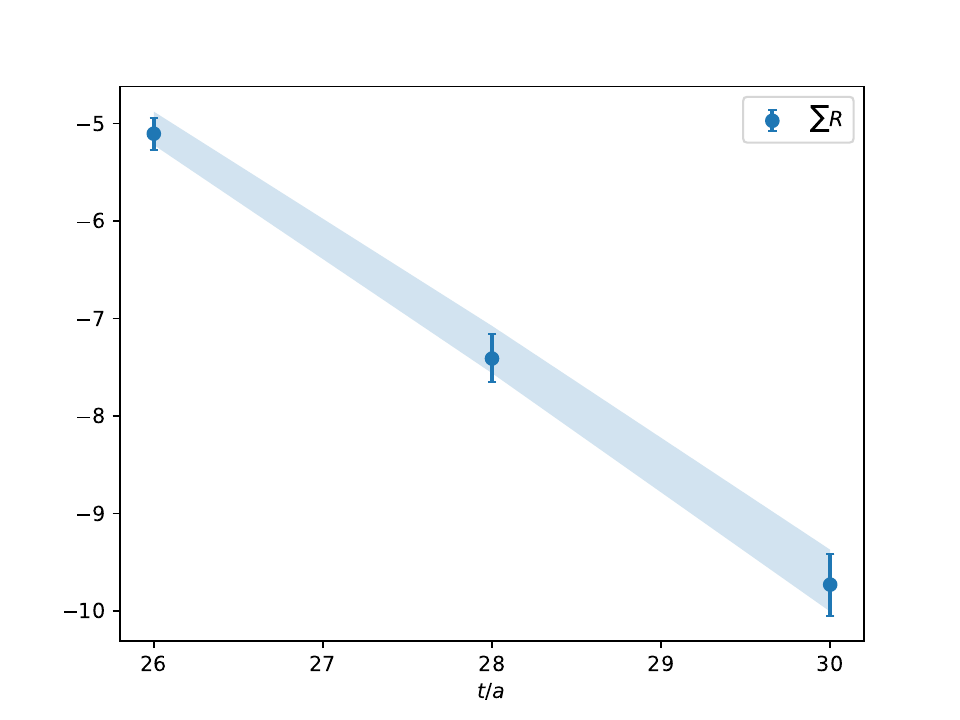}
    }
    \caption{Similar figures as \autoref{fig:jpsiRatioq} but for gluonic matrix elements of $J/\psi$: $\frac{1}{3}\sum_{\lambda=1}^3R^{44}_{g,\gamma_\lambda}$ (first row), $\frac{1}{3}\sum_{\lambda=1}^3R^{\lambda\lambda}_{g,\gamma_\lambda}$ (second row) and $\frac{1}{6}\sum_{\lambda=1}^3 \sum_{k\neq \lambda} R^{kk}_{g,\gamma_\lambda}$ (third row).}
    \label{fig:jpsiRatiog}
\end{figure*}

\begin{figure*}
    \includegraphics[width=0.32\linewidth]{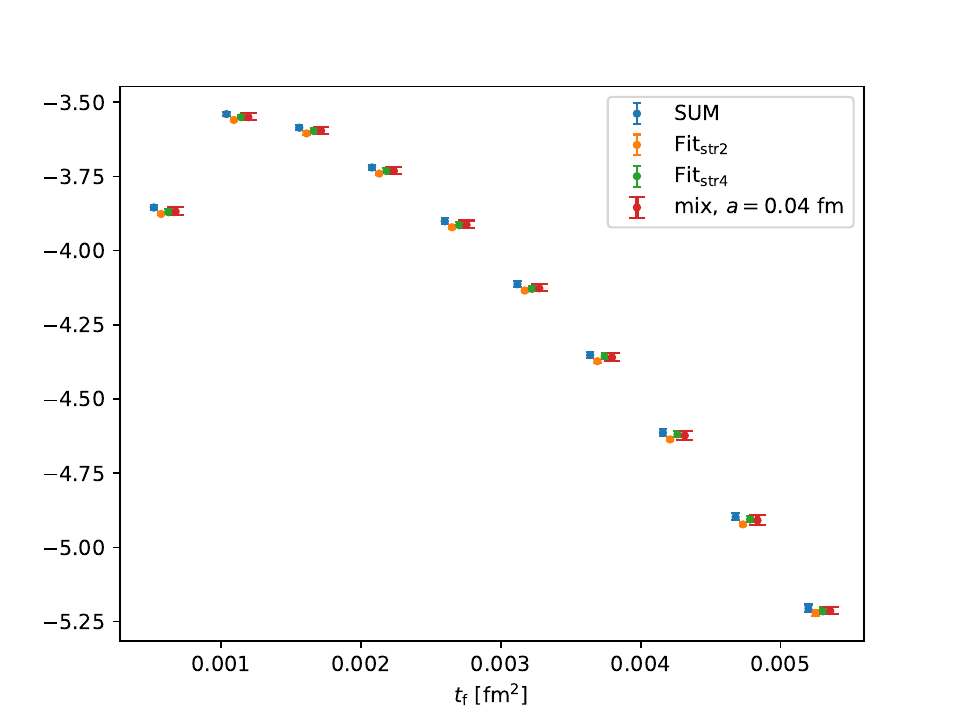}
    \includegraphics[width=0.32\linewidth]{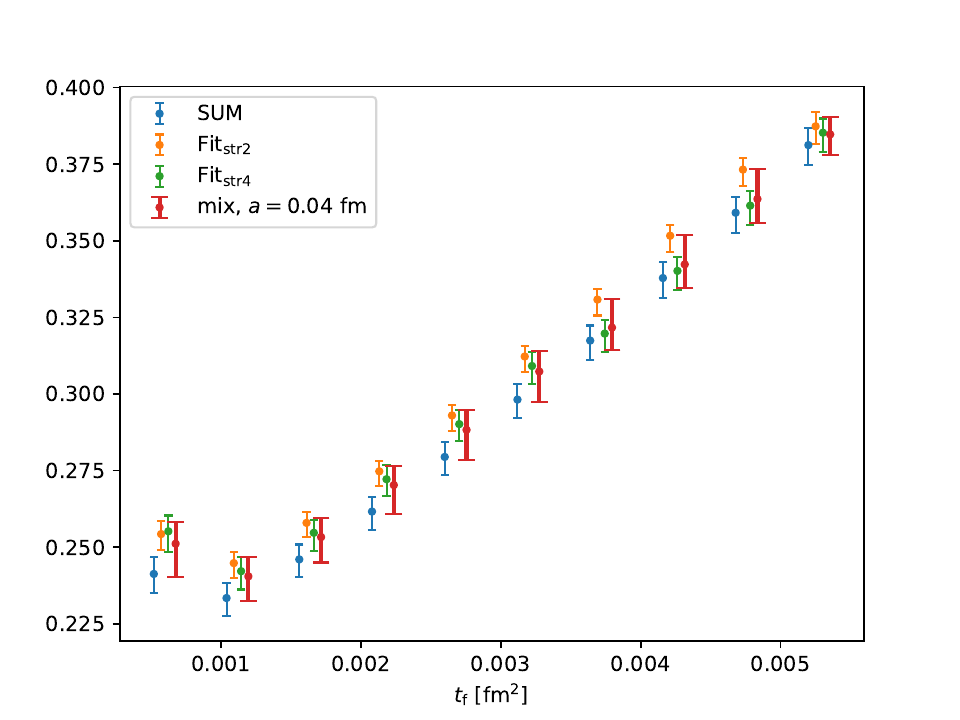}
    \includegraphics[width=0.32\linewidth]{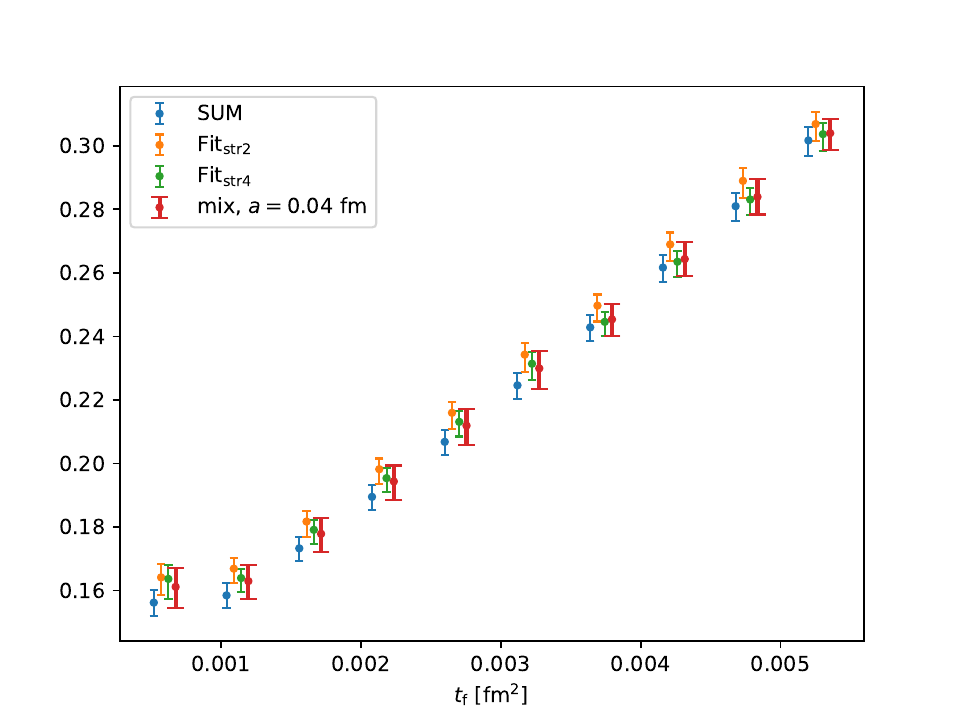}
    \includegraphics[width=0.32\linewidth]{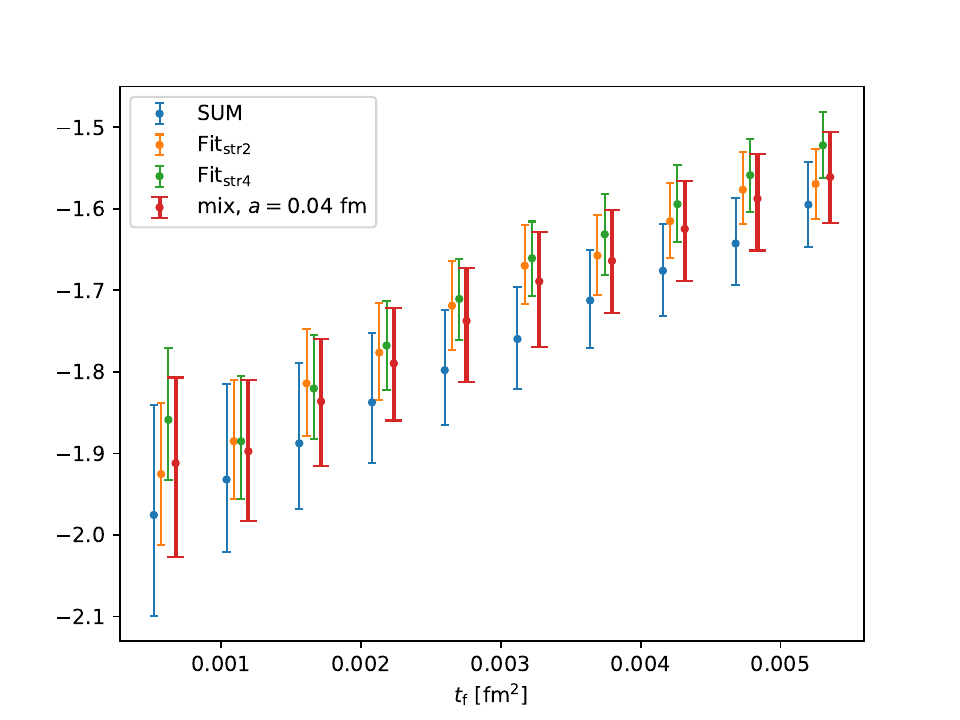}
    \includegraphics[width=0.32\linewidth]{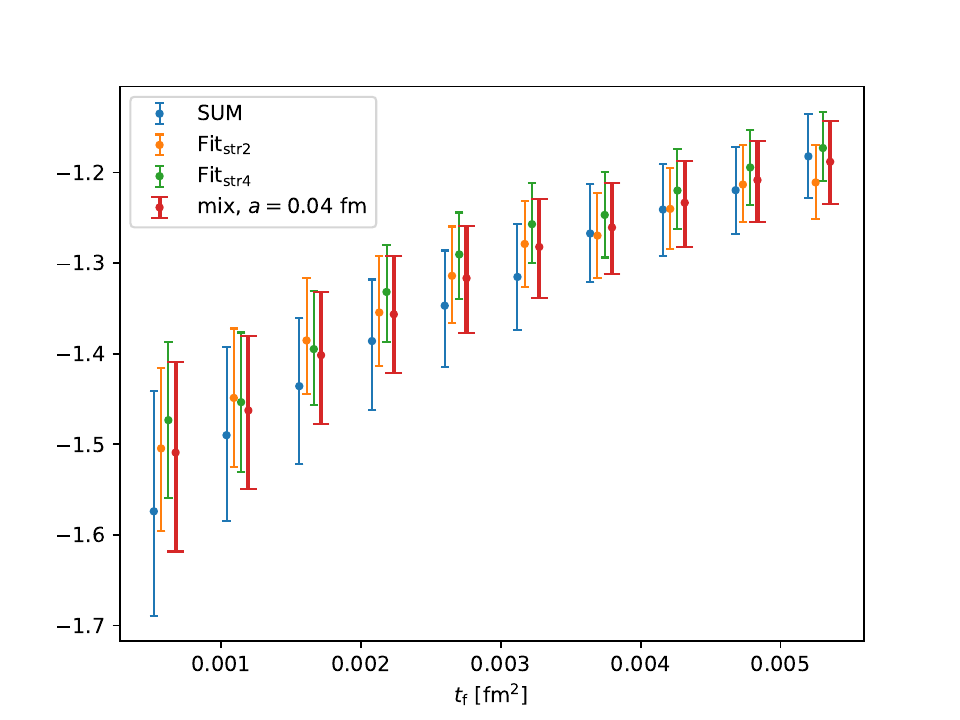}
    \includegraphics[width=0.32\linewidth]{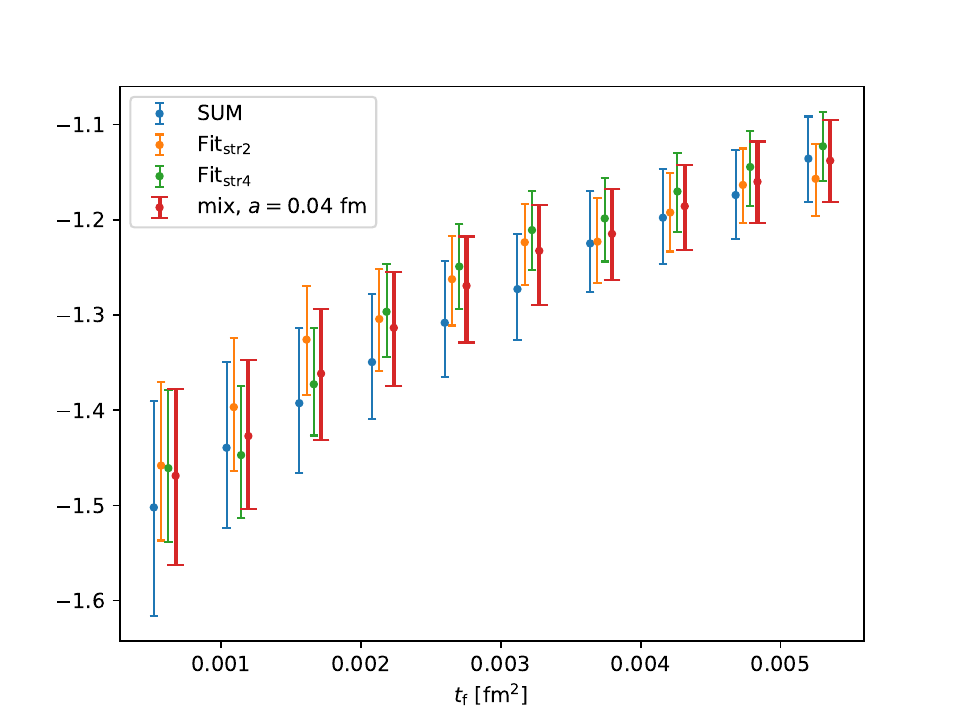}
    \caption{Extracted EMT matrix elements for $J/\psi$ from the summation fit SUM and two two-state fits $\rm Fit_{str2}$ and $\rm Fit_{str4}$ are shown as a function of flow time $\tf$. The averaged results are also shown (denoted by mix). The upper panels show quark matrix elements $\frac{1}{2M}\langle J/\psi,\gamma_1|\tilO_{3,44}|J/\psi,\gamma_1\rangle$ (left), $\frac{1}{2M}\langle J/\psi,\gamma_1|\tilO_{3,11}|J/\psi,\gamma_1\rangle$ (middle) and $\frac{1}{2M}\langle J/\psi,\gamma_1|(\tilO_{3,22}+\tilO_{3,33})/2|J/\psi,\gamma_1\rangle$ (right). The low panels show gluon matrix elements $\frac{1}{2M}\langle J/\psi,\gamma_1|\tilO_{1,44}|J/\psi,\gamma_1\rangle$ (left), $\frac{1}{2M}\langle J/\psi,\gamma_1|\tilO_{1,11}|J/\psi,\gamma_1\rangle$ (middle) and $\frac{1}{2M}\langle J/\psi,\gamma_1|(\tilO_{1,22}+\tilO_{1,33})/2|J/\psi,\gamma_1\rangle$ (right).}
    \label{fig:jpsiMX}
\end{figure*}

For the vector $J/\psi$ channel we consider polarization along the $x$ direction and use the interpolating operator $\Gamma=\gamma_1$. In the rest frame, the relevant quark EMT matrix elements are obtained from the ground-state limits of the ratios,
\begin{equation}
\begin{aligned}
\frac{1}{2M}\langle J/\psi,\gamma_1|\tilO_{3,44}|J/\psi,\gamma_1\rangle &= \lim_{t \gg \tau \gg 0} R^{44}_{q,\gamma_1}, \\
\frac{1}{2M}\langle J/\psi,\gamma_1|\tilO_{3,11}|J/\psi,\gamma_1\rangle &= \lim_{t \gg \tau \gg 0} R^{11}_{q,\gamma_1}, \\
\frac{1}{2M}\left\langle J/\psi,\gamma_1\left|\frac{\tilO_{3,22}+\tilO_{3,33}}{2}\right|J/\psi,\gamma_1\right\rangle
&= \lim_{t \gg \tau \gg 0} \frac{R^{22}_{q,\gamma_1}+R^{33}_{q,\gamma_1}}{2}\,,
\end{aligned}
\end{equation}
where the indices $k=2,3$ are transverse to the polarization direction. As in the pseudoscalar case, we average over the degenerate transverse components to improve statistical precision.

\autoref{fig:jpsiRatioq} summarizes the quark-channel extraction for the $a=0.04$~fm ensemble at $\tf=5\ef$. The left and middle columns show the ratios $R^{44}_{q,\gamma_1}$ (upper row), $R^{11}_{q,\gamma_1}$ (middle row), and $(R^{22}_{q,\gamma_1}+R^{33}_{q,\gamma_1})/2$ (lower row) as functions of $t-\tau/2$, together with the reconstructed curves from the two-state fits using $\rm Fit_{str2}$ and $\rm Fit_{str4}$. The fit reconstructions track the data well across the full range of source--sink separations, indicating robust isolation of the ground-state signal in all three channels. The right column displays the summed ratios versus $t$ and the corresponding summation fits; the observed linear behavior provides an independent check that excited-state effects are suppressed. In \autoref{fig:jpsiMX}, we compare the extracted ground-state matrix elements as functions of $\tf$ for $\rm Fit_{str2}$, $\rm Fit_{str4}$ and $\rm SUM$ in the upper panels, which agree within uncertainties.

For the gluon EMT matrix elements we average over the three polarization directions $\lambda=1,2,3$ with $\Gamma=\gamma_\lambda$, and use rotational symmetry to improve statistics. The corresponding relations in the rest frame are
\begin{equation}
\begin{aligned}
\frac{1}{2M}\langle J/\psi,\gamma_1|\tilO_{1,44}|J/\psi,\gamma_1\rangle
&= \lim_{t \gg \tau \gg 0}\,\frac{1}{3}\sum_{\lambda=1}^3 R^{44}_{g,\gamma_\lambda}, \\
\frac{1}{2M}\langle J/\psi,\gamma_1|\tilO_{1,11}|J/\psi,\gamma_1\rangle
&= \lim_{t \gg \tau \gg 0}\,\frac{1}{3}\sum_{\lambda=1}^3 R^{\lambda\lambda}_{g,\gamma_\lambda}, \\
\frac{1}{2M}\left\langle J/\psi,\gamma_1\left|\frac{\tilO_{1,22}+\tilO_{1,33}}{2}\right|J/\psi,\gamma_1\right\rangle
&= \lim_{t \gg \tau \gg 0}\,\frac{1}{6}\sum_{\lambda=1}^3 \sum_{k\neq \lambda} R^{kk}_{g,\gamma_\lambda},
\end{aligned}
\end{equation}
where $k\neq\lambda$ denotes EMT components transverse to the polarization direction. Since $\lambda=1,2,3$ are equivalent, on the left-hand side we display the $\gamma_1$ case for brevity.

\autoref{fig:jpsiRatiog} presents the corresponding gluon-channel analysis. The left and middle columns show the ratios and the reconstructed curves from the two-state fits, the right panels display the summed ratios and summation fits. The lower panels of \autoref{fig:jpsiMX} compile the resulting gluon matrix elements versus $\tf$ from $\rm Fit_{str2}$, $\rm Fit_{str4}$, and $\rm SUM$. The consistency among these extraction strategies provides a cross-check of excited-state control and systematic stability in the gluon channel.

%
%
\section{Continuum extrapolation}
\begin{figure*}
    \centerline{
    \includegraphics[width=0.33\textwidth]{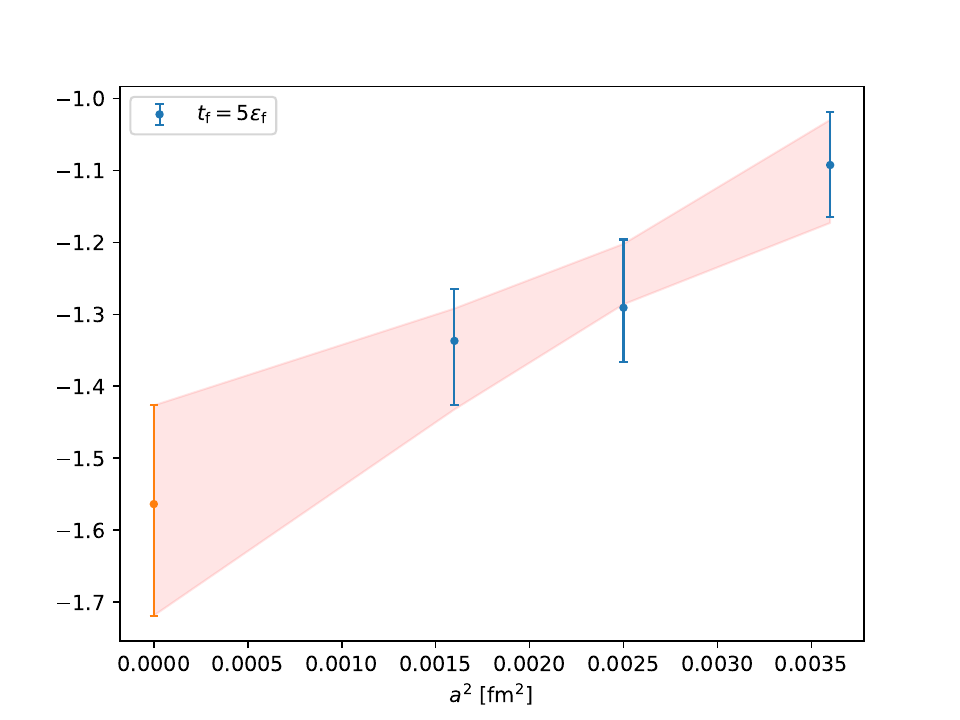}
    \includegraphics[width=0.33\textwidth]{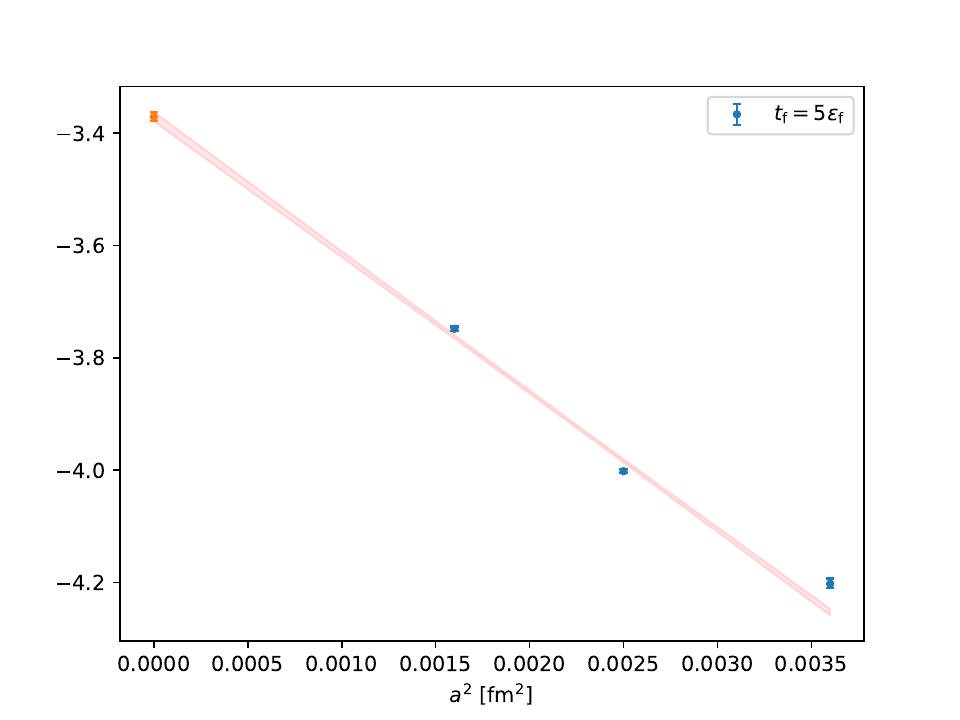}
    \includegraphics[width=0.33\textwidth]{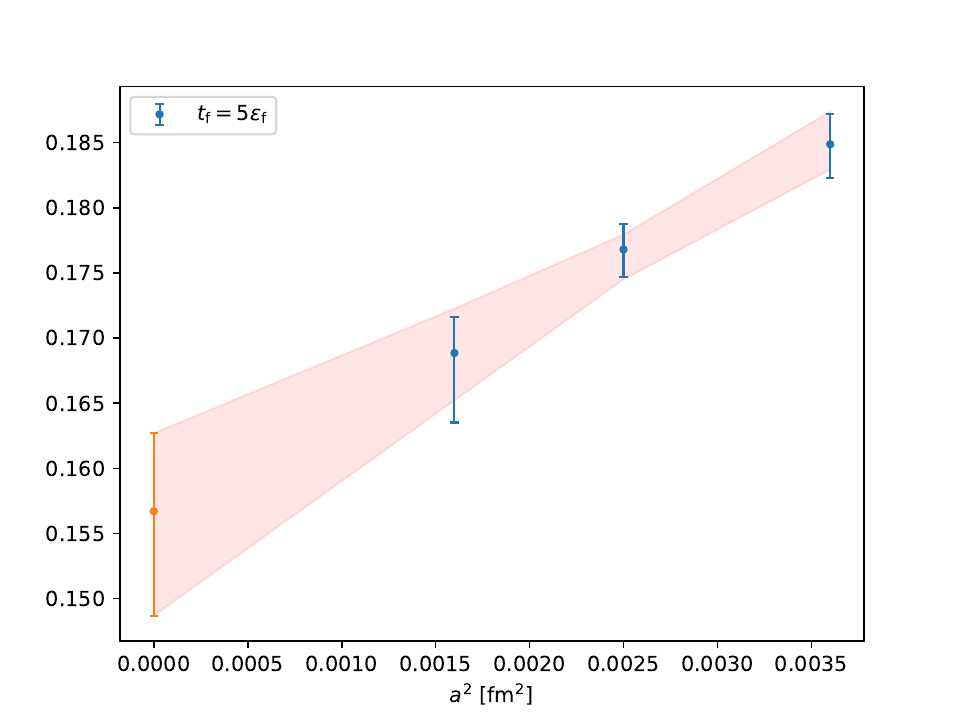}
    }
    \centerline{
    \includegraphics[width=0.33\textwidth]{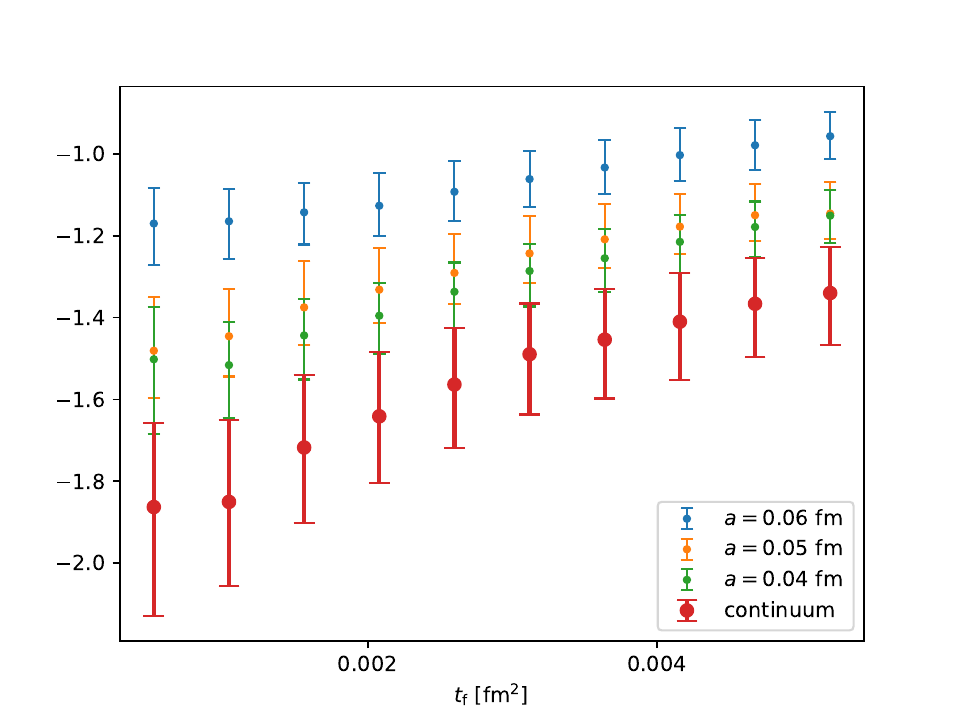}
    \includegraphics[width=0.33\textwidth]{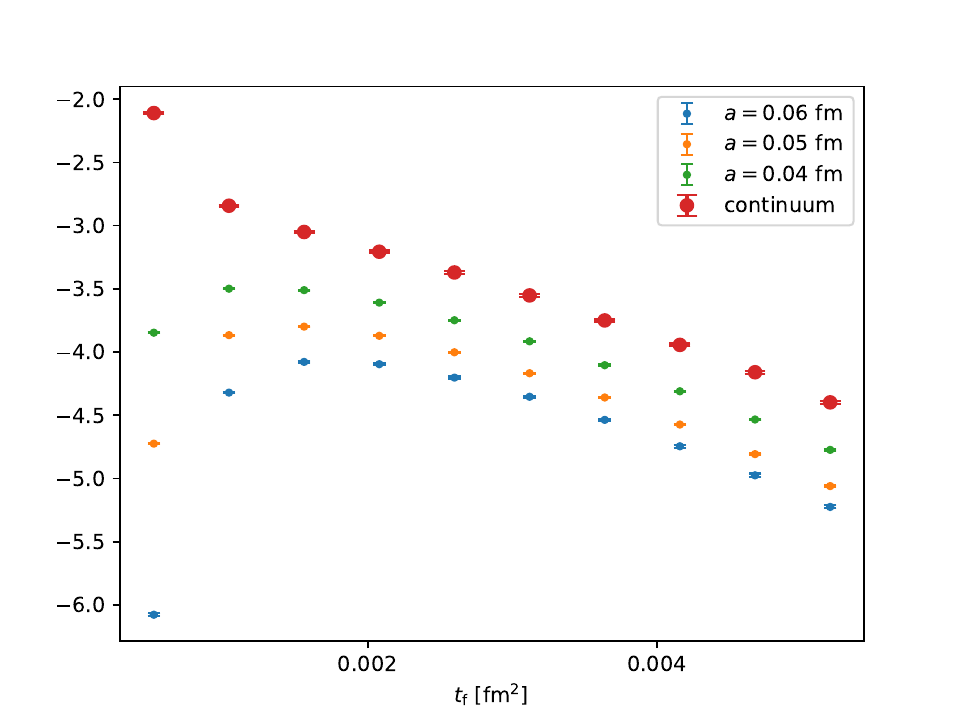}
    \includegraphics[width=0.33\textwidth]{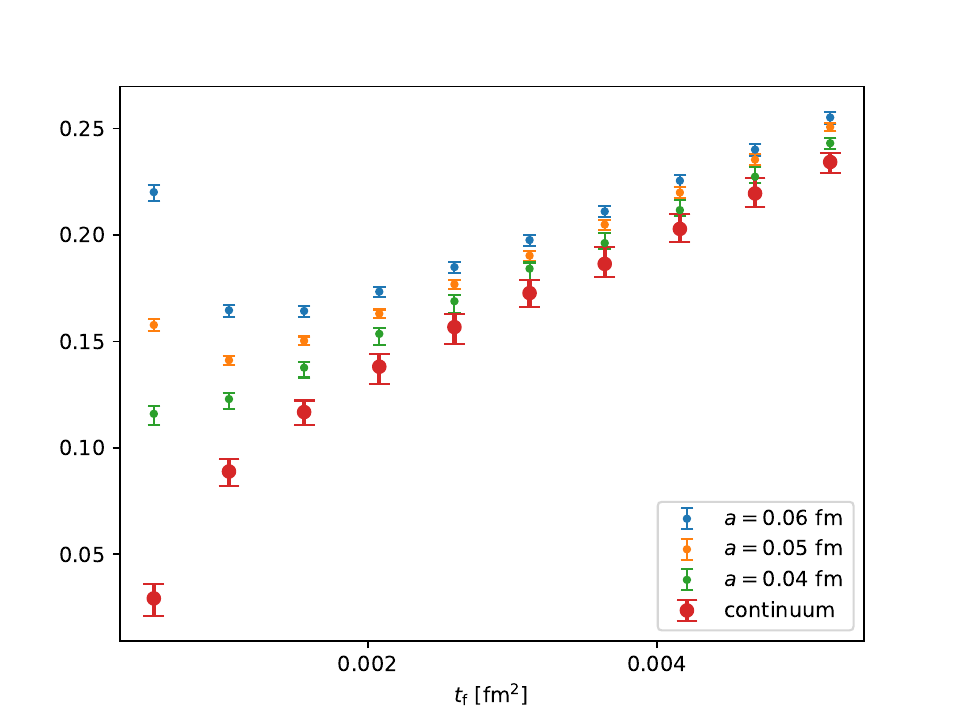}
    }
    \caption{Continuum extrapolation of selected gluon and quark EMT matrix elements in the $\eta_c$ channel.
    Left: Temporal component of the gluon EMT, $\frac{1}{2M}\langle \eta_c|\tilO_{1,44}|\eta_c\rangle$.
    Middle: Temporal component of the quark EMT, $\frac{1}{2M}\langle \eta_c|\tilO_{3,44}|\eta_c\rangle$.
    Right: Spatial component of the quark EMT, $\frac{1}{2M}\langle \eta_c|\tilO_{3,kk}/3|\eta_c\rangle$.
    Top row displays the continuum extrapolation at flow time $\tf=5\ef$.
    Bottom row shows the matrix elements at three lattice spacings and the continuum values as functions of flow time $\tf$.
    }
    \label{fig:continuum_etac}
\end{figure*}

\begin{figure*}
    \centerline{
    \includegraphics[width=0.33\textwidth]{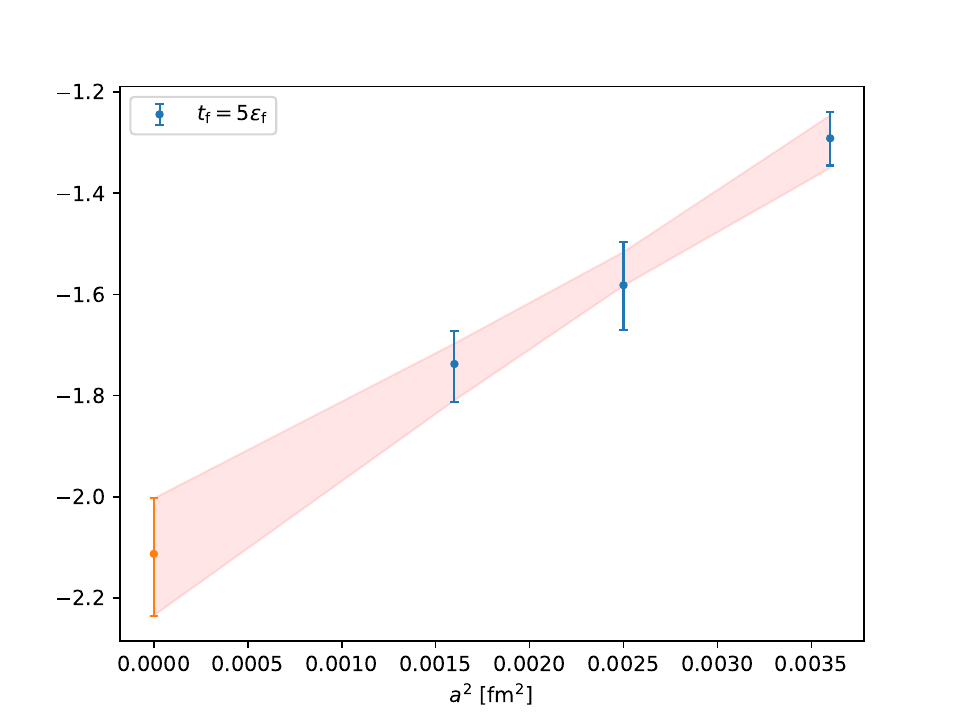}
    \includegraphics[width=0.33\textwidth]{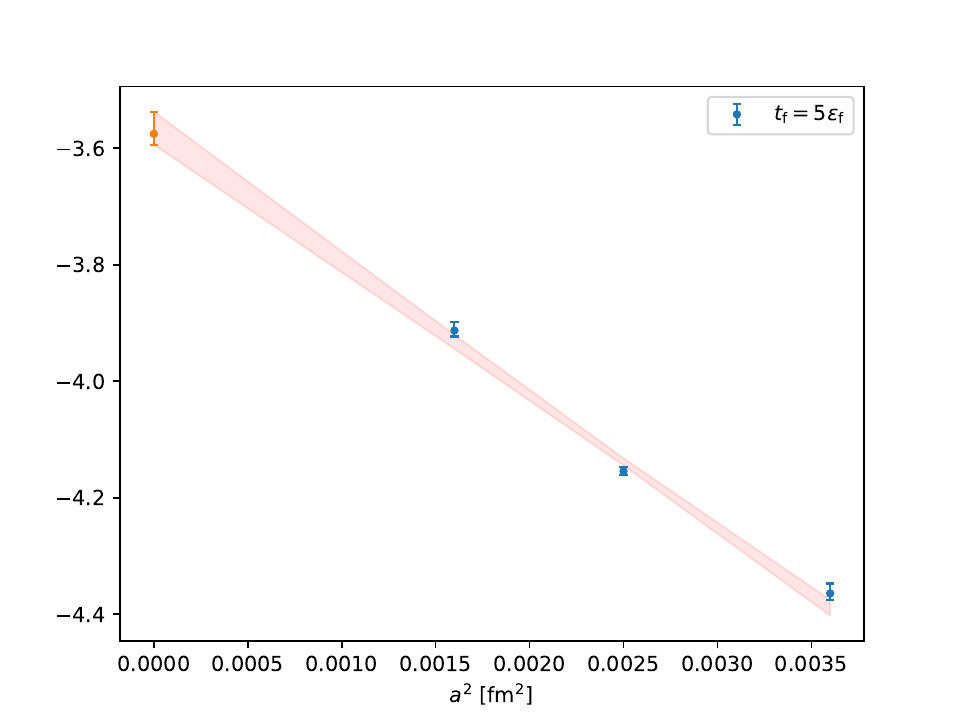}
    \includegraphics[width=0.33\textwidth]{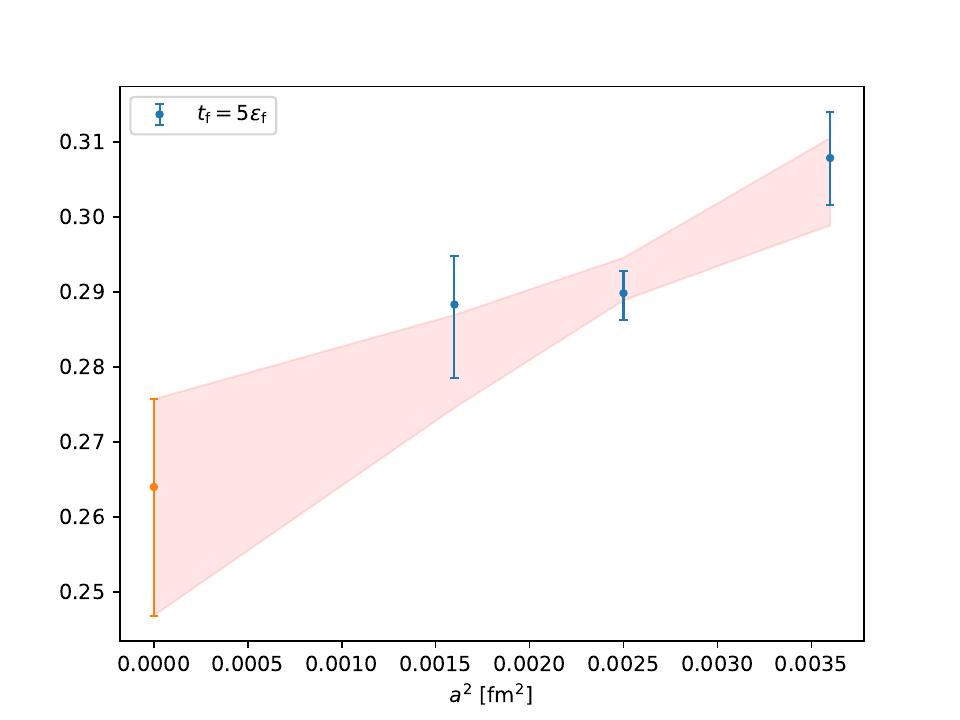}
    }
    \centerline{
    \includegraphics[width=0.33\textwidth]{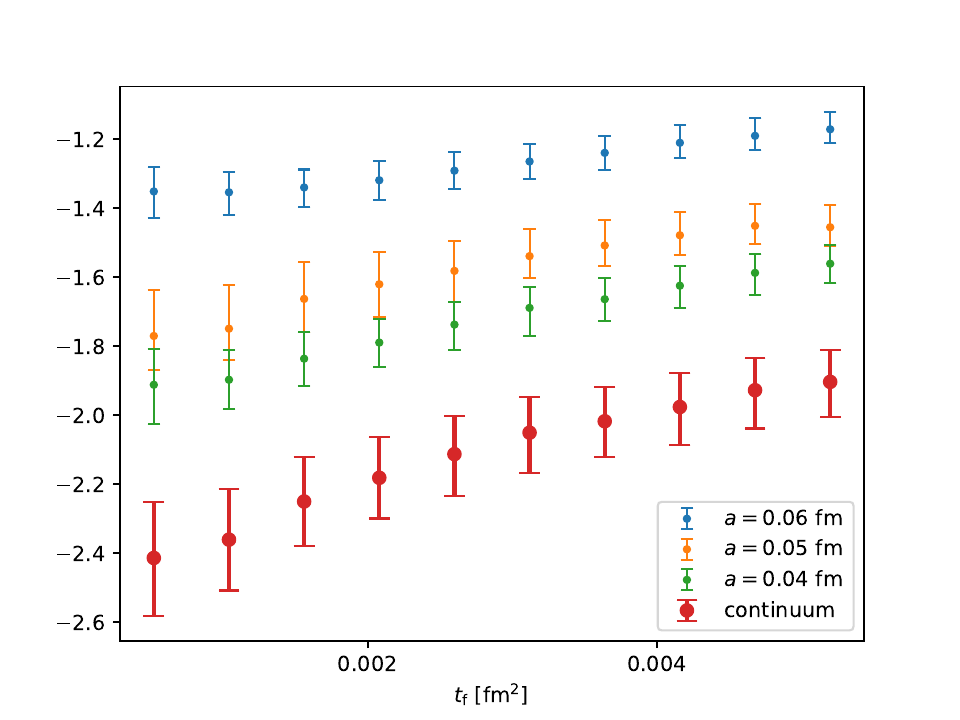}
    \includegraphics[width=0.33\textwidth]{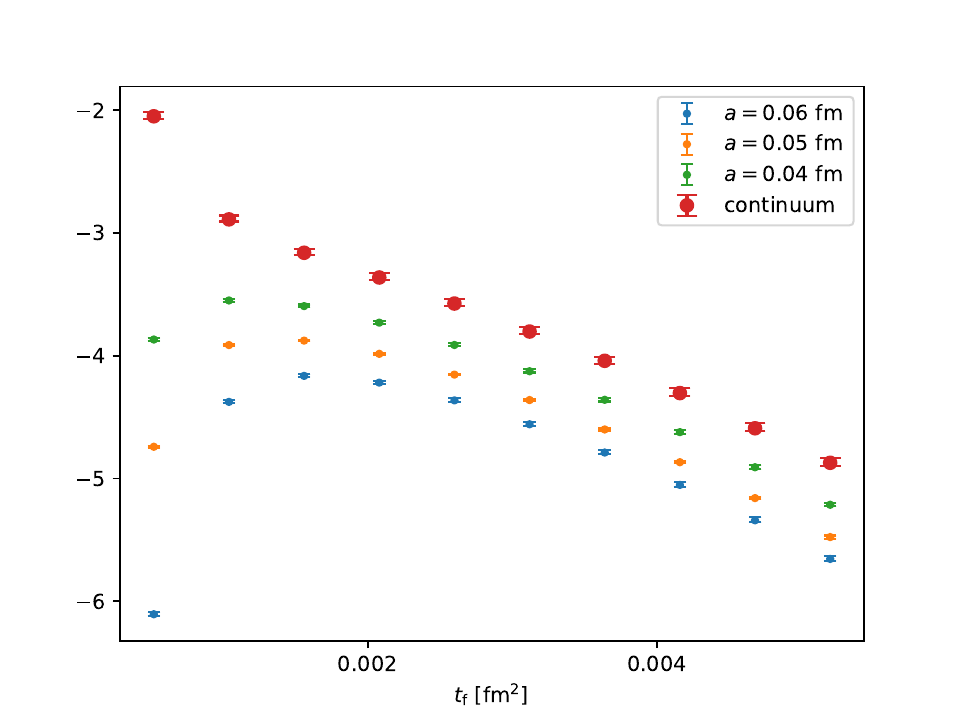}
    \includegraphics[width=0.33\textwidth]{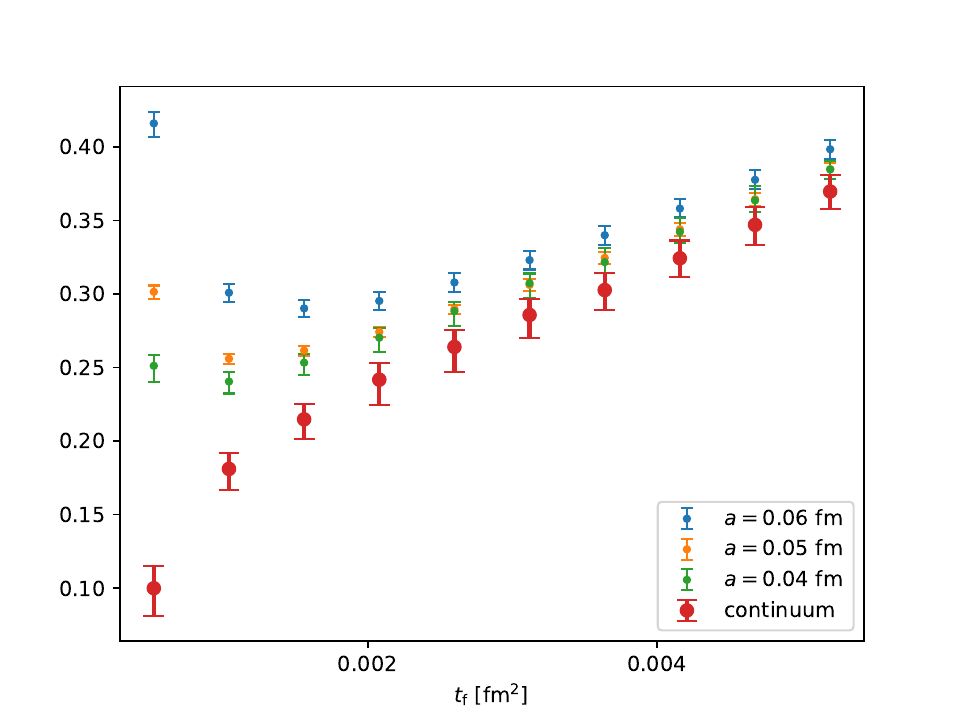}
    }
    \caption{The same as \autoref{fig:continuum_etac} but for $\frac{1}{2M}\langle J/\psi,\gamma_1|\tilO_{1,44}|J/\psi,\gamma_1\rangle$, $\frac{1}{2M}\langle J/\psi,\gamma_1|\tilO_{3,44}|J/\psi,\gamma_1\rangle$, $\frac{1}{2M}\langle J/\psi,\gamma_1|\tilO_{3,11}|J/\psi,\gamma_1\rangle$.
    }
    \label{fig:continuum_Jpsi}
\end{figure*}

The flowed matrix elements extracted from the ratio fits are still affected by lattice discretization errors~\cite{Taniguchi:2016ofw}. Based on the lattice actions and operator constructions used in this study, the dominant discretization effects are of order $O(a^2)$. To obtain the continuum limit of the flowed matrix elements, we perform extrapolations in the lattice spacing $a$ at fixed flow time $\tf$.

For each matrix elements $\langle H|\tilO_{i,\rs}|H\rangle$, and for each flow time $\tf$, we have 1800 bootstrap samples constructed from the combination of three fitting methods ($\mathrm{Fit_{str2}}$, $\mathrm{Fit_{str4}}$, and $\mathrm{SUM}$) across multiple source-sink separations and multiple ensembles.
Then at fixed flow time $\tf$, we apply the quadratic continuum extrapolation in $a^2$ sample by sample,
\begin{equation}
  \langle H|\tilO_{i,\rs}|H\rangle(\tf,a)=\langle H|\tilO_{i,\rs}|H\rangle(\tf,0)+a^2 X_{i,\rs}(\tf)\,.
\end{equation}

Examples of the continuum extrapolation behavior are illustrated in \autoref{fig:continuum_etac} for the $\eta_c$ channel and \autoref{fig:continuum_Jpsi} for the $J/\psi$ channel. The upper panels show representative examples of continuum extrapolations at $\tf = 5\ef$, where a clear linear dependence on $a^2$ is observed, supporting the validity of the extrapolation ansatz. In the lower panels, we present the continuum-extrapolated results as a function of $\tf$. As $\tf$ decreases, statistical uncertainties increase due to reduced signal and stronger sensitivity to lattice artifacts. These findings underscore the importance of multi-ensemble analysis, particularly when working with gradient-flow-smeared operators.

%
%
\section{Perturbative matching to \texorpdfstring{$\overline{\text{MS}}$}{MSbar} scheme}

The flowed EMT operator can be related to the renormalized EMT in the $\overline{\text{MS}}$ scheme through a perturbative matching procedure. Specifically, in the small flow-time limit, the flowed operator $\tilde{\mathcal{O}}_{i,\mn}(\tf)$ can be expanded in terms of renormalized local operators through,
\begin{align}
    \O^{\msbar}_i(\mu)=M_{ij}(\tf,\mu)\tilO_{j}(\tf)+O(\tf),
\end{align}
with power corrections denoted by $O(\tf)$. The matching matrix $M_{ij}(\tf,\mu)$ is given by~\cite{Harlander:2018zpi}
\begin{align}\label{eq:Mij}
    M_{ij}(\tf,\mu) = Z_{il}(\mu)\, \left[\zeta^{-1}\right]_{lj}(\tf,\mu),
\end{align}
where $\left[\zeta^{-1}\right]_{lj}(\tf,\mu)$ denotes the inverse matching from the flowed operator to the bare EMT operator in dimensional regularization and $Z_{il}(\mu)$ is the renormalization matrix in the $\overline{\text{MS}}$ scheme.

We apply the one-loop and two-loop results for $M_{ij}$ computed in Ref.~\cite{Harlander:2018zpi} to extract the renormalized operators $\mathcal{O}^{\msbar}_{i,\mn}$, $i=1,2,3,4$, in the $\overline{\text{MS}}$ scheme.

\begin{figure*}[]
\centerline{
    \includegraphics[width=0.32\textwidth]{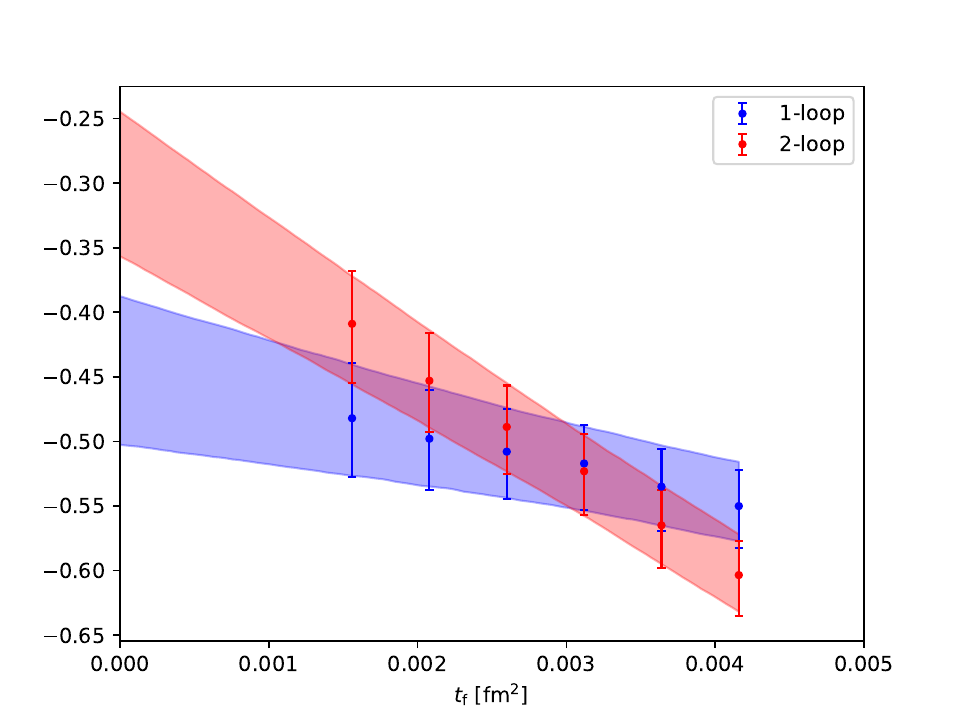}
    \includegraphics[width=0.32\textwidth]{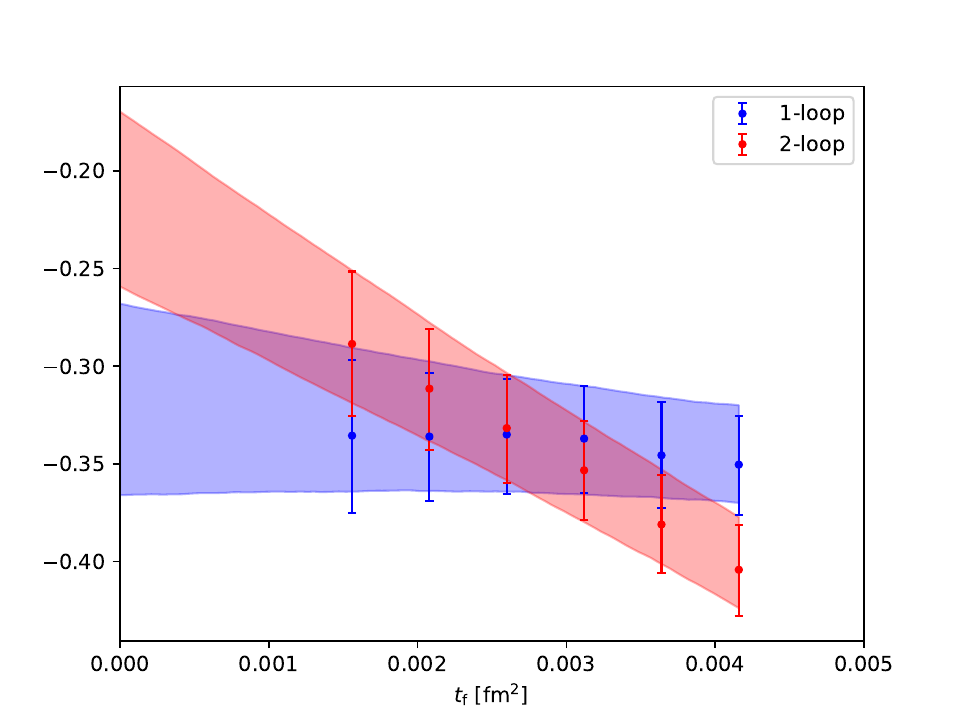}
    \includegraphics[width=0.32\textwidth]{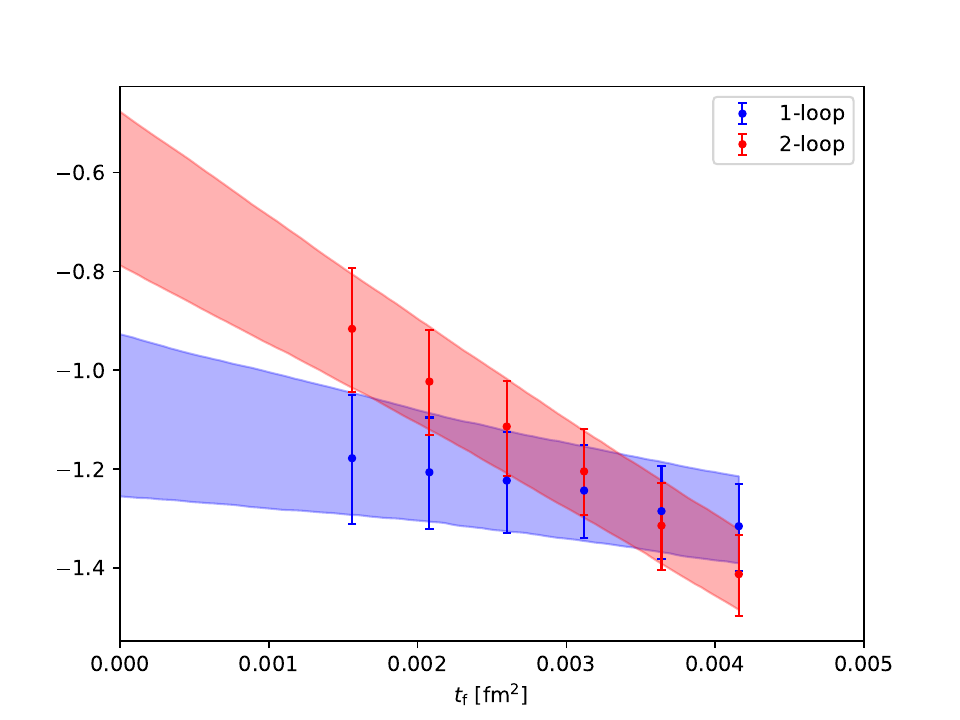}
    }
\centerline{
    \includegraphics[width=0.32\textwidth]{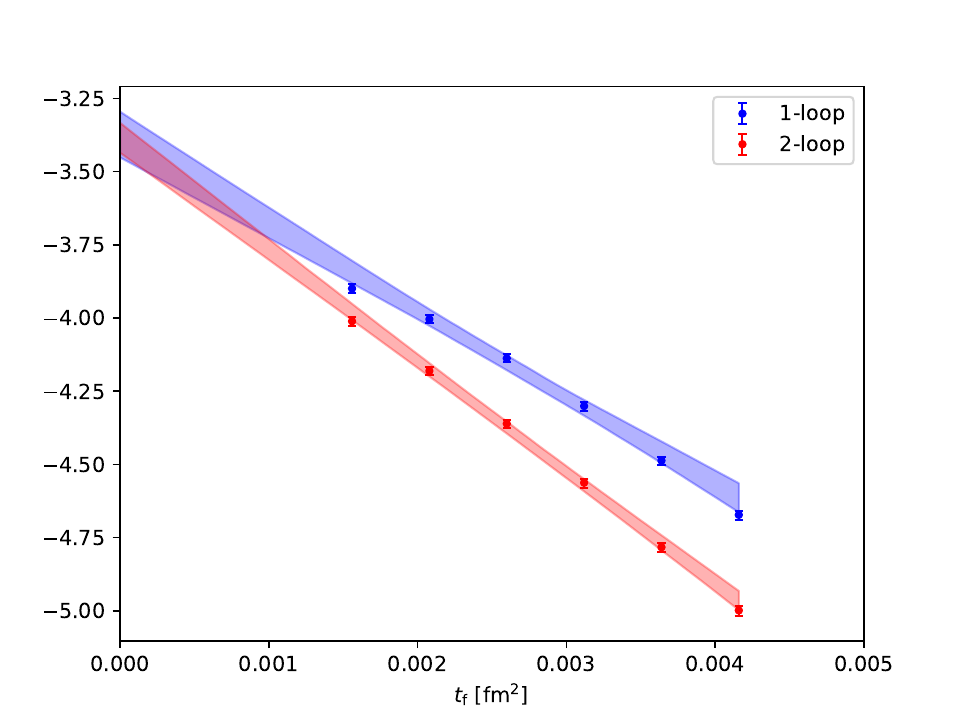}
    \includegraphics[width=0.32\textwidth]{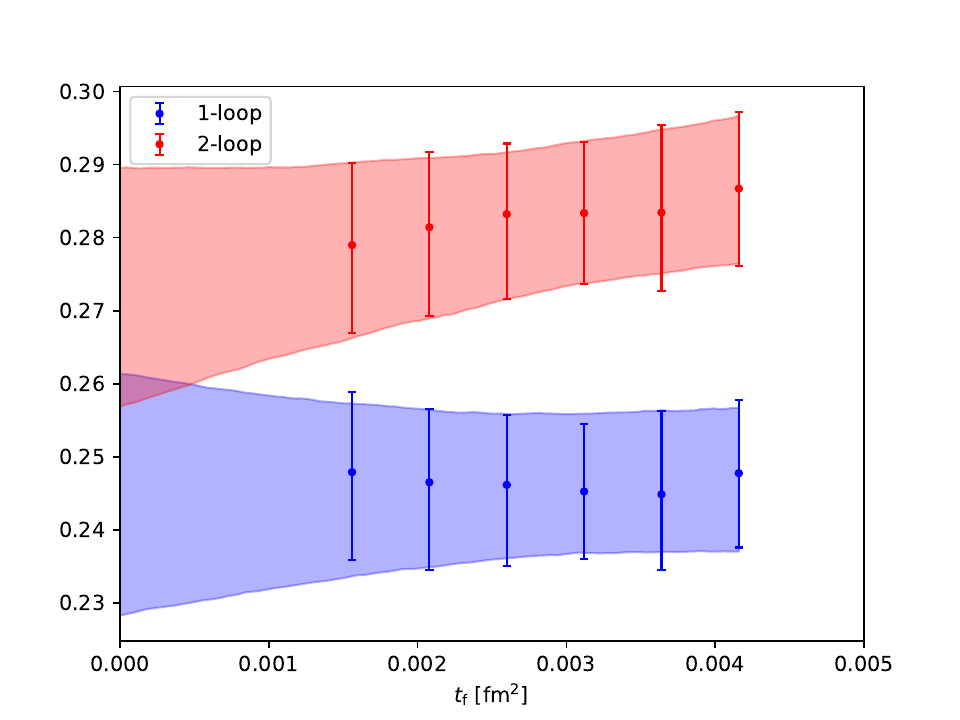}
    \includegraphics[width=0.32\textwidth]{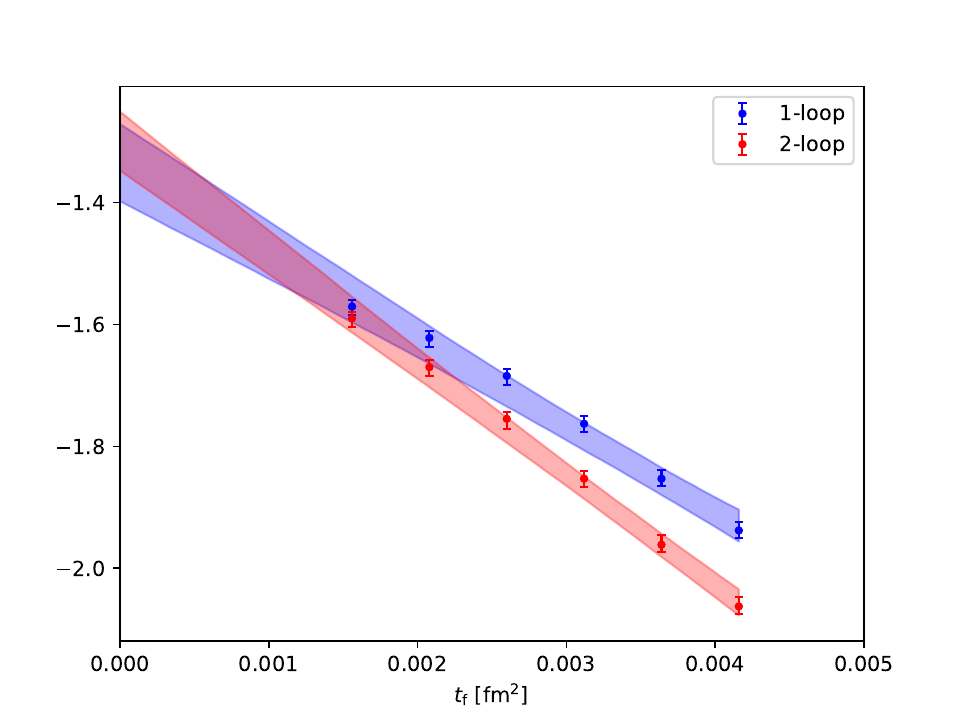}
    }
    \caption{Flow-time dependence of $\msbar$ matrix elements in the $\eta_c$ channel after continuum extrapolation and matching. Each panel compares results obtained using one-loop (blue) and two-loop (red) matching kernels. Top row (gluon sector): Left—Temporal component $\frac{1}{2M}\langle \eta_c|\O^{\msbar}_{1,44}|\eta_c\rangle$; Middle—Averaged spatial components $\frac{1}{2M}\langle \eta_c|\O^{\msbar}_{1,kk}/3|\eta_c\rangle$; Right—Trace component $\frac{1}{2M}\langle \eta_c|\O^{\msbar}_{2,44}|\eta_c\rangle$. Bottom row (quark sector):  Left—Temporal component $\frac{1}{2M}\langle \eta_c|\O^{\msbar}_{3,44}|\eta_c\rangle$; Middle—Averaged spatial components $\frac{1}{2M}\langle \eta_c|\O^{\msbar}_{3,kk}/3|\eta_c\rangle$; Right—Trace component $\frac{1}{2M}\langle \eta_c|\O^{\msbar}_{4,44}|\eta_c\rangle$.}
    \label{fig:tf_extrapolation_etac}
\end{figure*}

\autoref{fig:tf_extrapolation_etac} shows examples of the matched matrix elements in the $\eta_c$ channel after continuum extrapolation, comparing one-loop (blue) and two-loop (red) results for various operator components. The top row corresponds to the gluon sector, with the left and middle columns showing the temporal and spatial components of $\mathcal{O}^{\overline{\text{MS}}}_1$ respectively, and the right column displaying the gluon trace operator $\mathcal{O}^{\overline{\text{MS}}}_2$. The bottom row presents quark-sector results, including the temporal component $\mathcal{O}^{\overline{\text{MS}}}_{3,44}$, the averaged spatial components $\sum_{k=1}^3\mathcal{O}^{\overline{\text{MS}}}_{3,kk}/3$, and the quark trace operator. 

We observe that the results obtained from one-loop and two-loop matching are generally close to each other, with only small differences. These differences are expected, as they are of order $O(\alpha_s^2)$. The small magnitude of the deviation also suggests that higher-order corrections beyond two loops are likely subdominant relative to our current statistical precision. This observation supports the robustness of the perturbative matching procedure and indicates that the conversion to the $\overline{\text{MS}}$ scheme is under good control within our error budget.

\begin{figure*}[]
\centerline{
    \includegraphics[width=0.25\textwidth]{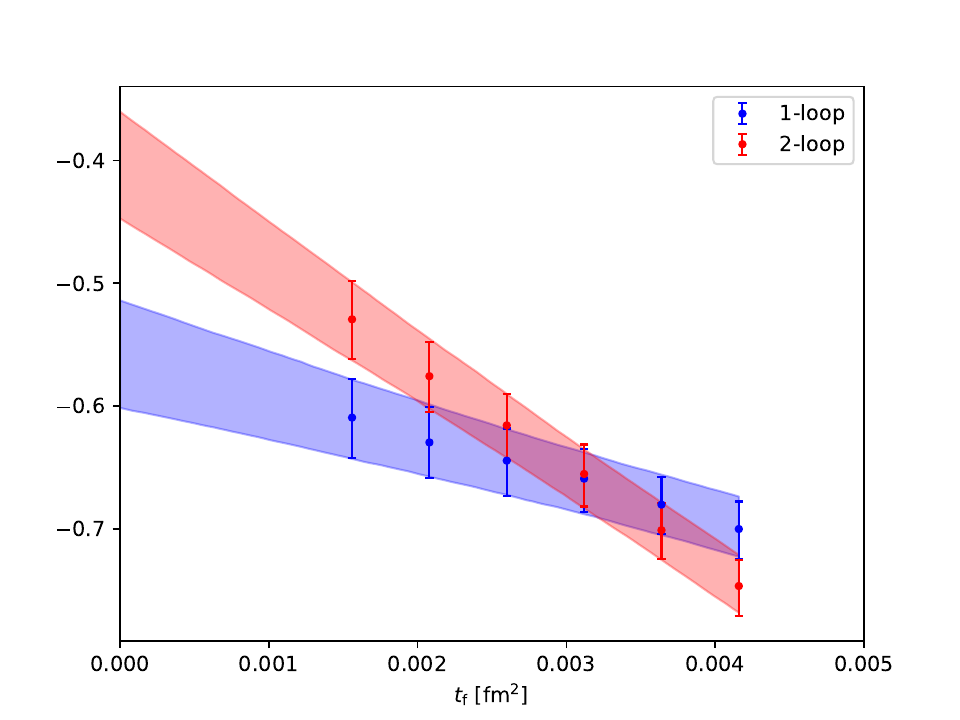}
    \includegraphics[width=0.25\textwidth]{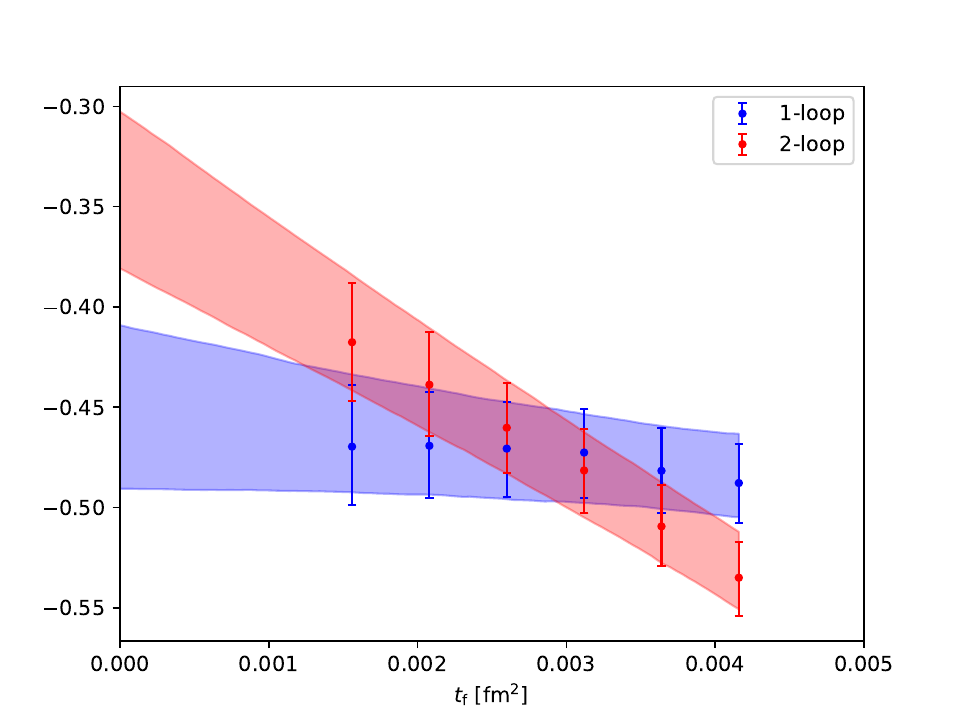}
    \includegraphics[width=0.25\textwidth]{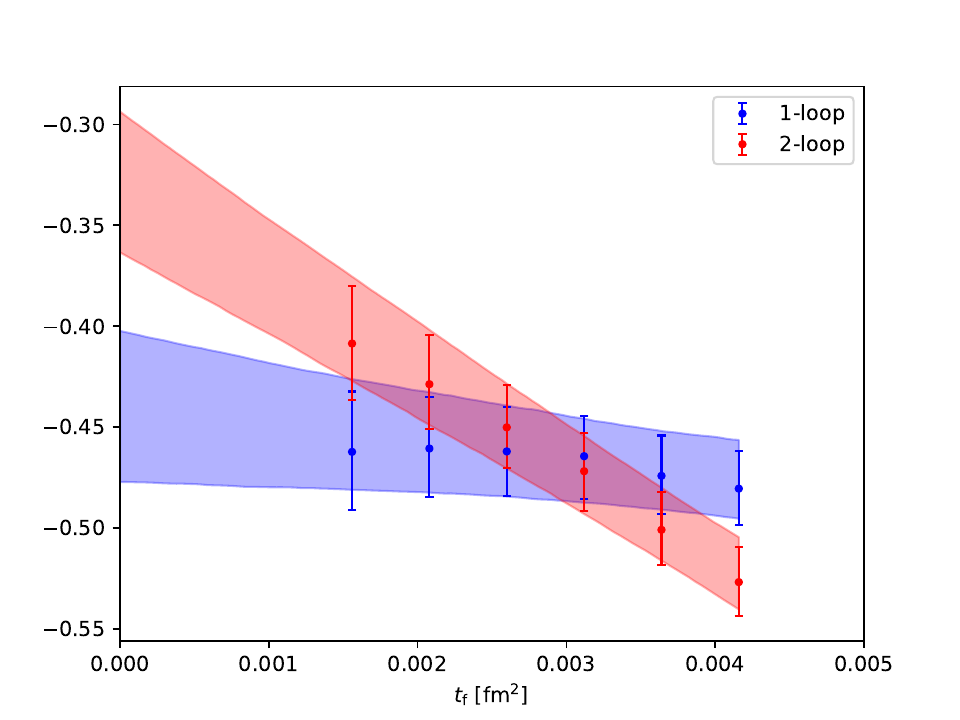}
    \includegraphics[width=0.25\textwidth]{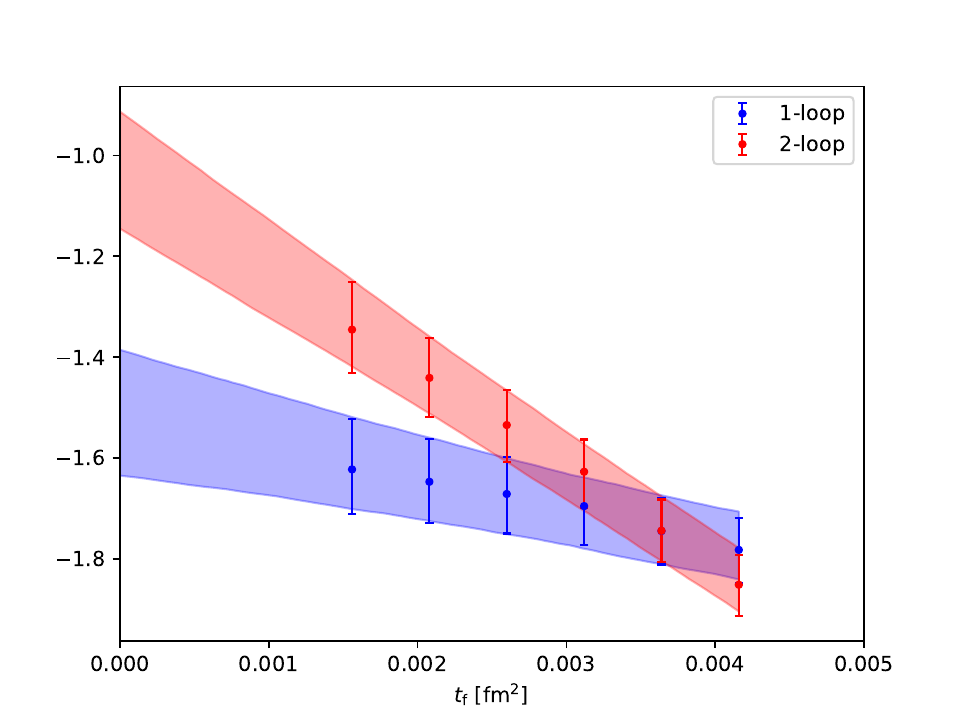}
    }
\centerline{
    \includegraphics[width=0.25\textwidth]{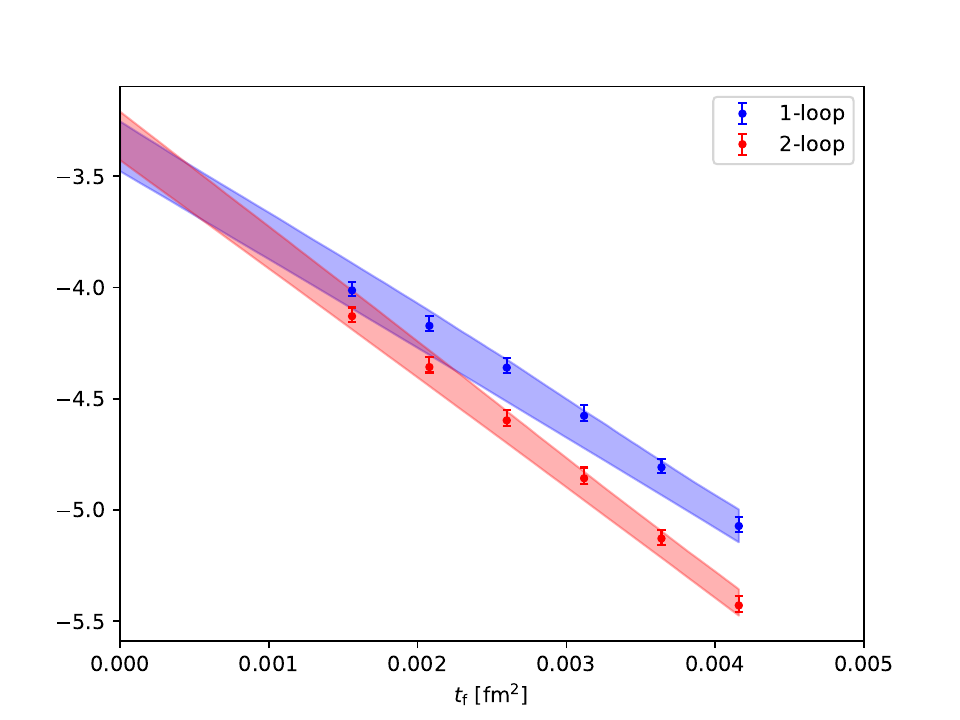}
    \includegraphics[width=0.25\textwidth]{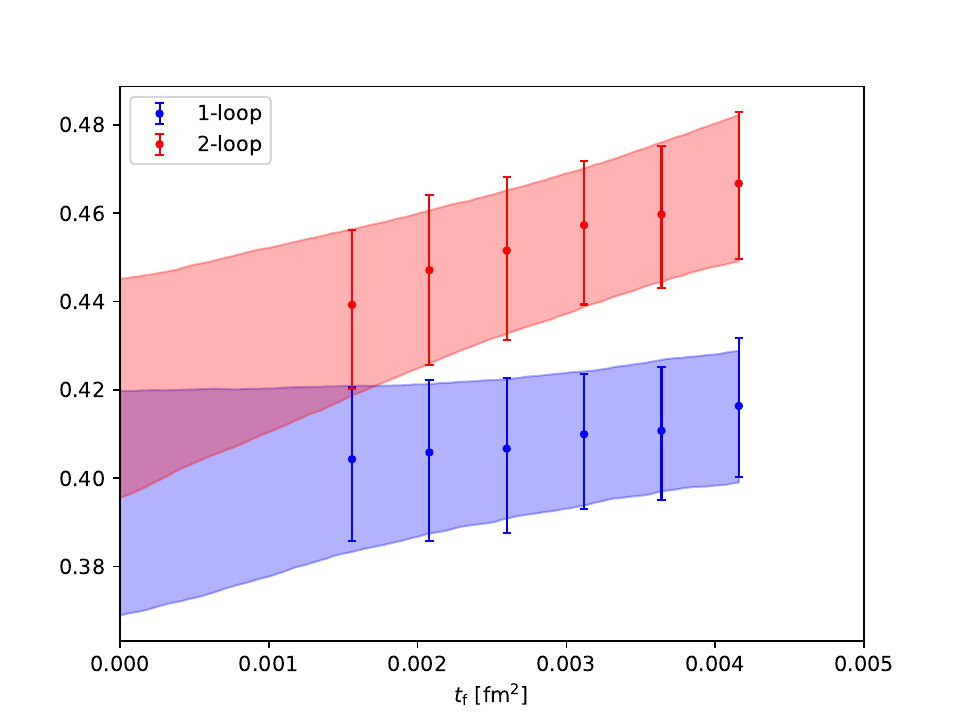}
    \includegraphics[width=0.25\textwidth]{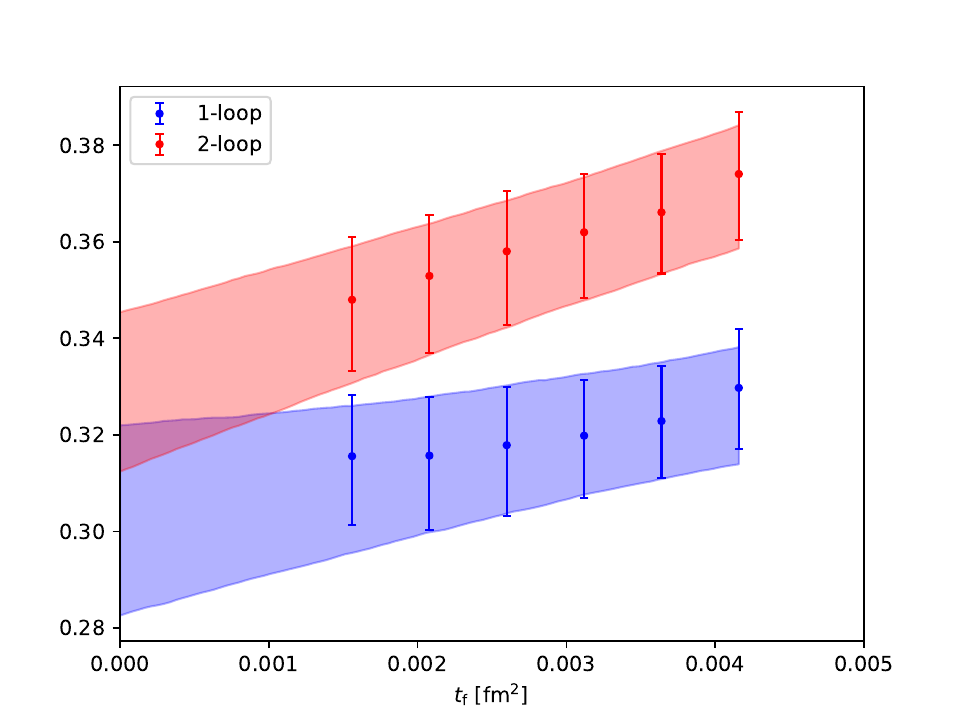}
    \includegraphics[width=0.25\textwidth]{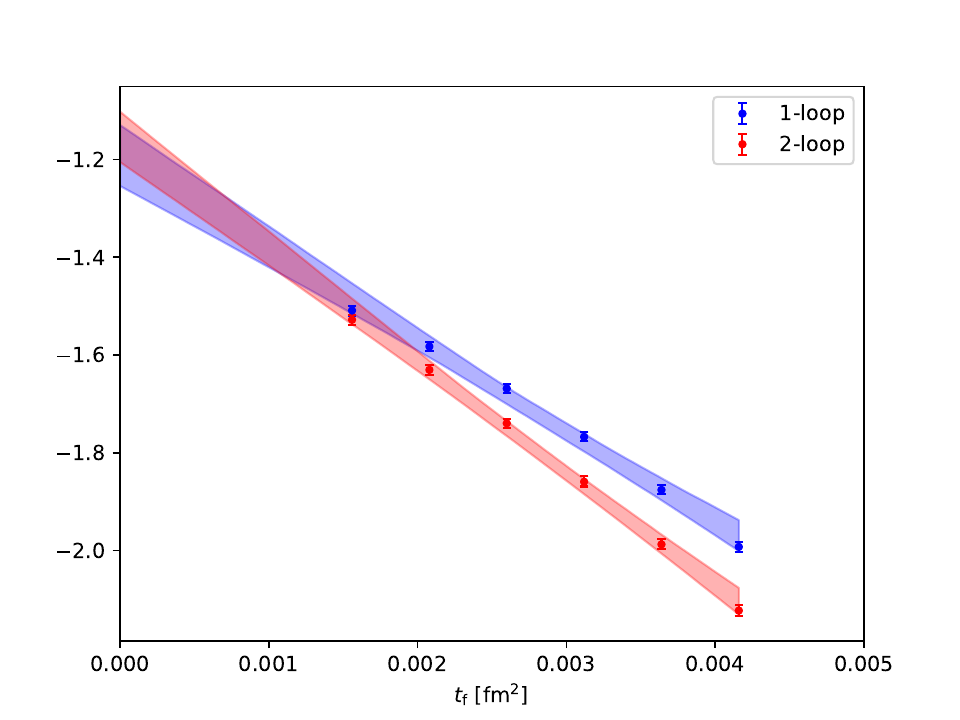}
    }
    \caption{Same as \autoref{fig:tf_extrapolation_etac} but for $J/\psi$. The upper row (left to right) shows the small-$\tf$ extrapolations of gluon sector: $\frac{1}{2M}\langle J/\psi,\gamma_1|\O^{\msbar}_{1,44}|J/\psi,\gamma_1\rangle$, $\frac{1}{2M}\langle J/\psi,\gamma_1|\O^{\msbar}_{1,11}|J/\psi,\gamma_1\rangle$, $\frac{1}{2M}\langle J/\psi,\gamma_1|(\O^{\msbar}_{1,22}+\O^{\msbar}_{1,33})/2|J/\psi,\gamma_1\rangle$, $\frac{1}{2M}\langle J/\psi,\gamma_1|\O^{\msbar}_{2,44}|J/\psi,\gamma_1\rangle$. The lower row (left to right) shows the corresponding results of quark sector: $\frac{1}{2M}\langle J/\psi,\gamma_1|\O^{\msbar}_{3,44}|J/\psi,\gamma_1\rangle$, $\frac{1}{2M}\langle J/\psi,\gamma_1|\O^{\msbar}_{3,11}|J/\psi,\gamma_1\rangle$, $\frac{1}{2M}\langle J/\psi,\gamma_1|(\O^{\msbar}_{3,22}+\O^{\msbar}_{3,33})/2|J/\psi,\gamma_1\rangle$, $\frac{1}{2M}\langle J/\psi,\gamma_1|\O^{\msbar}_{4,44}|J/\psi,\gamma_1\rangle$.}
    \label{fig:tf_extrapolation_Jpsi}
\end{figure*}

%
%
\section{Small-flow time extrapolation}

The gradient flow provides a theoretically clean method to define renormalized operators by smoothing ultraviolet fluctuations over a radius $\sqrt{8\tf}$~\cite{Luscher:2011bx}. At positive flow time $\tf>0$, composite operators constructed from flowed fields are finite and automatically renormalized. However, they are no longer identical to the original local QCD operators. Instead, they admit an operator product expansion (OPE) in flow time as indicated by \autoref{eq:Mij}. 

These $O(\tf)$ terms arise because the flowed operator smears over a finite region, and therefore mixes with a tower of operators allowed by the symmetries. Such contamination cannot be removed by continuum or perturbative matching alone and must be explicitly eliminated via an extrapolation $\tf \to 0$. Only in this limit does the flowed matrix element reliably reproduce the target renormalized operator in the $\overline{\text{MS}}$ scheme.

To perform the extrapolation, we use the form:
\begin{equation}
  \langle H| \O^{\msbar}_{i,\rs} |H\rangle (\mu,\tf) = \langle H| \O^{\msbar}_{i,\rs} |H\rangle (\mu) + Y_{i,\rs}\tf,
\end{equation}
and carry out a linear fit in $\tf$ for each matrix element. To mitigate correlations between data at different flow times and reduce the condition number of the covariance matrix, we randomly select three $\tf$ values from our available set for each of the 1800 bootstrap samples and perform an correlated linear fit. This procedure also allows us to estimate systematic uncertainties associated with the choice of fit window.

Examples of the flow-time extrapolation are shown in \autoref{fig:tf_extrapolation_etac} and \autoref{fig:tf_extrapolation_Jpsi} for the $\eta_c$ and $J/\psi$ channels, respectively. The observed linear dependence indicates that higher-order flow-time corrections are negligible within our current precision. The extrapolated values thus represent the final results for EMT matrix elements in the $\overline{\text{MS}}$ scheme, with both statistical and systematic uncertainties properly propagated through the entire analysis pipeline.

%
%
\section{Quark-line-disconnected contributions of the charm quark}\label{sec:disc}

\begin{figure}
    \centering
    \includegraphics[width=0.45\linewidth]{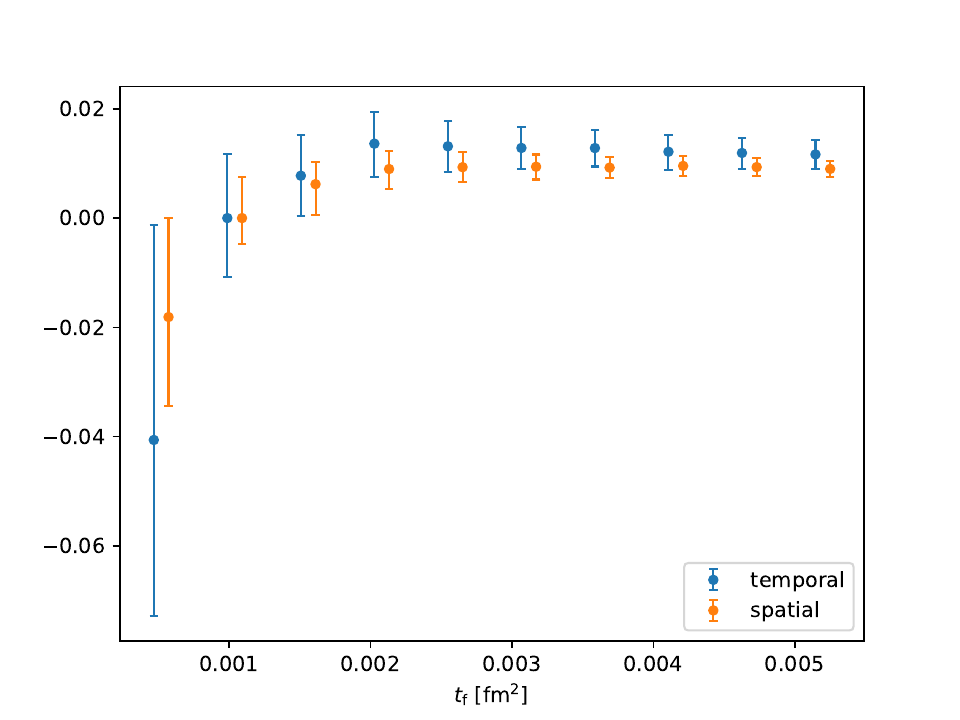}
    \caption{The quark-line disconnected contribution to $\frac{1}{2M}\langle \eta_c|\tilO_{3,44}|\eta_c\rangle$ and $\frac{1}{2M}\langle \eta_c|\tilO_{3,kk}/3|\eta_c\rangle$ on the $a=0.06$ fm ensemble.}
    \label{fig:test_dis_charm}
\end{figure}

In the main text, we computed the quark-sector matrix elements of the EMT using only the connected three-point diagrams, omitting the disconnected contributions. This choice was motivated by the expectation that, for the charm quark, disconnected diagrams are strongly suppressed and therefore make a negligible contribution to the total matrix element.

To verify this assumption, we explicitly evaluated the disconnected charm-quark contribution for $\frac{1}{2M}\langle \eta_c|\tilO_{3,\rs}|\eta_c\rangle$ on the $a=0.06$ fm ensemble. The results, shown in \autoref{fig:test_dis_charm}, indicate that while the disconnected signal is nonzero with high statistical precision, it is tiny compared to the connected contribution shown in \autoref{fig:continuum_etac}, and well below the statistical uncertainties of the connected results.

This numerical check confirms that omitting disconnected diagrams in the charm-quark sector introduces a negligible systematic uncertainty at our current level of precision. Consequently, restricting the calculation in main text to the connected contributions is fully justified for the present study.

\bibliography{ref}

\end{document}